\documentclass{ian}

\usepackage{dsfont}
\usepackage{etoolbox}
\usepackage{graphicx}
\usepackage{xcolor}

\usepackage{largearray}
\usepackage{iantheo}
\usepackage[reset_at_theo]{point}
\usepackage[problem_in_eq]{problemset}

\definecolor{Complete Equation}{HTML}{4B0082}
\definecolor{Big Equation}{HTML}{DAA520}
\definecolor{Medium Equation}{HTML}{32CD32}
\definecolor{Simple Equation}{HTML}{4169E1}
\definecolor{Quantum Monte Carlo}{HTML}{DC143C}

\def\eqformat#1{{\bf\color{#1}#1}}

\def\problemformat#1{{\bf Exercise #1}}

\begin{document}

\pagestyle{empty}

\hbox{}
\hfil{\bf\LARGE
An introduction to\par
\vskip10pt
\hfil Lieb's Simplified approach to the Bose gas
}
\vskip20pt

\hfil{\bf\large Ian Jauslin}\par
\hfil{\it Department of Mathematics, Rutgers University}\par
\hfil{\tt\color{blue}\href{mailto:ian.jauslin@rutgers.edu}{ian.jauslin@rutgers.edu}}\par
\vskip20pt

\hfil {\bf Abstract}\par
\medskip

This is a book about Lieb's Simplified approach to the Bose gas, which is a family of effective single-particle equations to study the ground state of many-body systems of interacting Bosons.
It was introduced by Lieb in 1963, and recently found to have some rather intriguing properties.
One of the equations of the approach, called the Simple equation, has been proved to make a prediction for the ground state energy that is asymptotically accurate both in the low- and the high-density regimes.
Its predictions for the condensate fraction, two-point correlation function, and momentum distribution also agree with those of Bogolyubov theory at low density, despite the fact that it is based on ideas that are very different from those of Bogolyubov theory.
In addition, another equation of the approach called the Big equation has been found to yield numerically accurate results for these observables over the entire range of densities for certain interaction potentials.
\smallskip

This book is an introduction to Lieb's Simplified approach, and little background knowledge is assumed.
We begin with a discussion of Bose gases and quantum statistical mechanics, and the notion of Bose-Einstein condensation, which is one of the main motivations for the approach.
We then move on to an abridged bibliographical overview on known theorems and conjectures about Bose gases in the thermodynamic limit.
Next, we introduce Lieb's Simplified approach, and its derivation from the many-body problem.
We then give an overview of results, both analytical and numerical, on the predictions of the approach.
We then conclude with a list of open problems.
\vskip20pt

This is a preprint of the following work: I. Jauslin, {\it An Introduction to Lieb's Simplified Approach}, 2025, Springer.
It is the version of the author's manuscript prior to acceptance for publication and has not undergone editorial and/or peer review on behalf of the Publisher (where applicable).
The final authenticated version is available online at:

\hfil{\color{blue}\href{http://dx.doi.org/10.1007/978-3-031-81393-1}{http://dx.doi.org/10.1007/978-3-031-81393-1}}

\vskip20pt

\tableofcontents

\vfill
\eject

\setcounter{page}1
\pagestyle{plain}

\section{Introduction}
\indent
This book is about interacting Bosons.
The notion of a ``Boson'' was introduced a hundred years before this book was written, in 1924\-~\cite{Bo24,Ei24}, and has garnered much attention, from physicists and mathematicians alike.
In the mathematical physics community, interest in systems of interacting Bosons ballooned at the end of the 1990's, and their study has become one of the major research areas of the field.
As such, much has been written about such systems: research and review articles as well as books.
Nevertheless, despite much work and significant advances, one of the core, foundational questions on Bose gases is still lacking a satisfactory answer: there is still no mathematical proof that interacting Bosons form condensates (at least not in realistic models).
This has been a well-known open question for several decades, and many approaches have been attempted to solve it, without success.
This book is about a new idea to study interacting Bosons: Lieb's Simplified approach.
(To be exact, the idea is old, but the realization of its significance is new.)
Now, the approach has not, as of this writing, helped solve the problem of condensation, and it may never do so.
But it is a novel approach to the study of interacting Bosons, and it has already yielded some intriguing results.
The aim of this book is to introduce Lieb's Simplified approach and the ideas behind it, in a way that is more pedagogical than the research articles currently available in the literature.
\bigskip

\indent
This book is based on a series of lectures I gave at a summer school organized by Jake Fillman, at Texas State University, in the summer of 2023.
That course was aimed at undergraduate and young graduate students, and thus required very little prior knowledge on Bose gases.
We will take a similar approach here, and start our discussion at the very beginning.
Readers who are already familiar with Bose gases may prefer to skip to Chapter\-~\ref{sec:simplified_def} where Lieb's Simplified approach is introduced.
\bigskip

\indent
And so, we start from the very beginning with what may be the most fundamental and important question about systems of interacting Bosons: why are they interesting, and why should we spend time and effort studying them?
There are many possible answers to this question; let us discuss three which I have found to be personally motivating.
First, Bosons exist, and if we want to understand the natural world, we are going to have to study them.
Second, systems of many interacting Bosons exhibit interesting and non-trivial physical behavior.
Third, they are very challenging to study mathematically, which means that new mathematical tools need to be developed.
Let us now expand on each of these points.
\bigskip

\point
One of the great successes of modern (and not-so-modern) science is the understanding that matter is made of particles, and the properties of matter can be understood from the microscopic dynamics of the underlying particles.
This is a powerful idea: the macroscopic world is very varied: on the face of it, there seems to be very little in common between a tree, a car battery, and a tectonic plate, but they are all made of the same stuff: protons, neutrons, electrons, and assorted particles.
If we can understand the behavior of protons, neutrons, and electrons, we should, at least in principle, be able to understand the properties of all matter.
There is one problem though: deriving macroscopic behaviors from microscopic properties is rather difficult to do in practice.
The branch of physics dedicated to understanding how to do so is called ``Statistical Mechanics'', and aims to compute the behavior of observable matter starting from the microscopic laws of motion guiding its constituents particles.
These laws of motion are quantum mechanical, and, in quantum mechanics there are two types of particles: Fermions (such as electrons, protons, neutrinos, etc) and Bosons (such as photons, Higgs particles, Helium-4 atoms, etc).
(At least this is the case in three dimensions: there are particles that are neither Fermions nor Bosons in two dimensions.)
As we will see, things are complicated, and whereas it would be great to study all types of particles and their mixtures, it is difficult enough to consider even the simplest case in which the particles are all identical Bosons.
And so we will focus here exclusively on systems of identical Bosons.
Such systems exist in nature, in fact, Bosons were introduced to study the photons emanating from matter that is heated (such as the filament in a light bulb)\-~\cite{Bo24,Ei24}.
Another example, which is more relevant to the discussion in this book, is a gas of Helium-4 atoms.
\bigskip

\point
Helium has many interesting properties.
One of the most intriguing is that, when the temperature is small enough (below $2\mathrm{K}$, that is, $-271^\circ\mathrm{C}$ or $-456^\circ\mathrm{F}$), it forms a superfluid phase, which looks like a liquid phase, but it can flow without any viscosity\-~\cite{Ka38}, which leads to some fairly surprising behaviors.
For instance, when a superfluid is placed inside a container, capillary forces attract it to the edges of the container, but, since there is no viscosity to compensate for them, these forces pull the superfluid all the way up the wall and trickling down the sides.
This has been observed in experiments: superfluid helium is poured into a glass container, and it can be seen to drip out from the bottom of the container\-~\cite{RS39}.
In addition, because of the lack of viscosity, superfluid helium can pass through tiny little holes in materials.
\bigskip

\indent
Understanding superfluidity from microscopic dynamics is not an easy task (see\-~\cite{LSe05} for an overview of mathematical results).
There are effective theories that are designed to explain the phenomenon (see\-~\cite{PS16} and references therein), but it is still an open problem to prove that these are linked to realistic microscopic models.
One thing that is clear, however, is that the phenomenon of superfluidity can only arise in quantum models: if the microscopic dynamics were classical, superfluidity would be impossible.
In other words, superfluidity is intrinsically a quantum phenomenon.
Another phenomenon that is intrinsically quantum and has been observed in systems of interacting Bosons is Bose-Einstein condensation\-~\cite{AEe95,DMe95}.
A Bose-Einstein condensate is a phase in which a majority of the particles are all in the same state.
In other words, most of the particles are doing the same thing at the same time.
There is an idea that Bose-Einstein condensation should be related to superfluidity: after all, if the particles are doing the same thing, there should be no friction between particles, and thus no viscosity.
However, the picture is more complicated than that, and the nature of the link between Bose-Einstein condensation and superfluidity is still unclear.
(It is worth noting, for instance, that there are cases in which superfluidity is believed to occur without Bose-Einstein condensation, for instance, in two dimensional models at positive temperature).
But Bose-Einstein condensation is interesting in its own right: it is an inherently quantum phase of matter that can be realized in experiments, in which most particles are in the same exact state.
\bigskip

\point
Bose-Einstein condensation has had a resurgence in popularity after 1995, when it was first observed experimentally\-~\cite{AEe95,DMe95}.
In particular, the mathematical physics community set off to solve an important problem: the works of Bose\-~\cite{Bo24} and Einstein\-~\cite{Ei24} that first predicted the phenomenon use a much simplified model, in which particles are assumed not to interact with each other.
This makes things much simpler: if the particles are independent, it suffices to understand how a single particle behaves, since the situation is identical for all the others.
However, in the presence of interactions, no one has yet been able to prove that Bose-Einstein condensation even occurs (and real Bosonic systems have interactions, except for photons, but for those, we don't even know how to write down an appropriate wave equation\-~\cite{Op31}).
Here, the word ``prove'' is key: most everyone accepts Bose-Einstein condensation occurs in interacting systems, but we don't have a mathematical proof yet.
In fact, before 1998, there was no complete mathematical handle on pretty much any of the properties of interacting Bosonic gases.
In 1998, Lieb and Yngvason\-~\cite{LY98} computed the asymptotic behavior at low density of the ground-state energy (see\-~(\ref{lhy_e})) and this started a flurry of activity.
Back in 1957, Lee, Huang and Yang\-~\cite{LHY57} had made a series of predictions, one of which was a low-density expansion for the ground-state energy that went one step further that what Lieb and Yngvason proved in 1998.
The mathematical physics community rose to the challenge, and embarked upon a twenty year effort to prove the Lee-Huang-Yang formula.
This was finally accomplished in 2020, after a long series of works culminating in\-~\cite{YY09,FS20,BCS21,FS23}.
\bigskip

\indent
However, these successes only concern the ground-state energy.
In particular, there still is no proof of Bose-Einstein condensation in (continuum) interacting models.
The results on the ground-state energy were obtained by finding a robust mathematical basis for {\it Boglyubov theory}.
Bogolyubov theory\-~\cite{Bo47} is an approach that was introduced by Bogolyubov in the 1940's to understand the low-density properties of Bose gases.
One can do a lot with Bogolyubov theory, and many predictions can be made from it\-~\cite{LHY57}.
However, to this day, only its approach to the ground-state energy has been used in mathematical proofs (at least for systems in the thermodynamic limit: there are proofs in other settings, as we will briefly discuss in Chapter\-~\ref{sec:bose_results}).
\bigskip

\indent
In this book, we will present a different approach, called {\it Lieb's Simplified approach}.
It was introduced by Lieb in 1963\-~\cite{Li63}, but after a short series of papers by Lieb along with Sakakura\-~\cite{LS64} and Liniger\-~\cite{LL64}, it has largely been forgotten.
We have recently had a fresh look at this approach, in a collaboration with Carlen and Lieb\-~\cite{CJL20,CJL21,CHe21,Ja22,Ja23,Ja24} and found that its predictions exceeded our expectations.
Indeed, we found that it is as good as Bogolyubov theory at low densities\-~\cite{CJL20,CJL21} (at least as far as the ground state and condensate fraction are concerned), and it also reproduces the behavior of the Bose gas at high densities\-~\cite{CJL20}.
We have even found very good agreement for some systems {\it over the entire range of densities}\-~\cite{CHe21,Ja23}.
Lieb's Simplified approach consists in replacing the original problem that involves an infinite number of particles with a non-linear, non-local equation for a single particle.
This is a familiar approach in physics, where many-body problems are replaced with non-linear effective equations.
However, in most other settings, these effective equations are only accurate in one parameter regime.
For instance Bogolyubov theory is accurate at low densities.
In contrast, Lieb's Simplified approach is asymptotically accurate at both low- and high-densities, and it does quite well at intermediate densities as well.
The approach actually gives us several levels of approximation, and defines a family of equations that are more or less accurate, and correspondingly, harder or easier to solve.
Some can be studied analytically, whereas others have only been treated numerically.
\bigskip

\indent
In short, systems of many interacting Bosons are interesting: they exist in the real world, exhibit non-trivial physical behaviors, and are mathematically challenging to study.
Lieb's Simplified approach is an approximation for the ground state of interacting Boson systems that is surprisingly accurate across densities, while being much simpler to study than the many-body system.
In this book, we will describe the derivation of Lieb's Simplified approach, and prove some properties of its solutions.
But first, let us take a step back and discuss the concept of Bose-Einstein condensation in more depth, and introduce the tools of statistical mechanics on which our entire premise is based.
\bigskip

{\bf Outline}:\par
\indent
In Chapter\-~\ref{sec:BEC}, we discuss Bose-Einstein condensation in the simplest setting in which it can be proved: for non-interacting systems.
The arguments presented there follow the original derivation by Bose and Einstein\-~\cite{Bo24,Ei24}.
We will open that section with a much abbreviated introduction to quantum statistical mechanics, for the benefit of readers with little experience in statistical mechanics.
Next, in Chapter\-~\ref{sec:bose_results} we will give a brief historical overview of some of the conjectures about the Bose gas that are most relevant to the discussion in this book.
In doing so, we will give a brief overview of scattering theory, which underlies the intuitive understanding of the low-density regime.
In Chapter\-~\ref{sec:simplified_def}, we introduce Lieb's Simplified approach.
We will discuss the approximations that are done to arrive at the various equations of the approach (in particular the Simple equation), and motivate it by discussing the accuracy of some of its predictions.
In Chapter\-~\ref{sec:existence}, we prove the existence of solutions to the Simple equation, and in Chapter\-~\ref{sec:predictions}, we prove asymptotic formulas for the prediction of the Simple equation for the energy and condensate fraction, and state similar results for the two-point correlation function and the momentum distribution.
These result use some functional and harmonic analysis, and we state the theorems that are needed without proof in Appendices\-~\ref{app:functional_analysis}, \ref{app:harmonic}, and\-~\ref{app:Ke}.
In Chapter\-~\ref{sec:numerics}, we briefly discuss the numerical computation of solutions to the Big and Medium equations.
Finally, in Chapter\-~\ref{sec:open}, we discuss some open problems.
\bigskip

{\bf Acknowledgements}:\par
\indent
This book is adapted from a course I gave at Texas State University, in the summer of 2023, organized by Jake Fillman, to whom I owe much credit!
The results described in this book are the fruit of a productive and very enjoyable collaboration with Eric A. Carlen and Elliott H. Lieb.
This book is a pedagogical summary of some of the results that we have discovered and published together (see the references below).

\indent
I wholeheartedly thank Elliott H. Lieb for introducing me to his approach, and for very useful notes and comments on this manuscript.
I am also extremely grateful for his mentorship and support, and for our many enlightening discussions over the years.
I also thank Eric A. Carlen for many useful discussions, Markus Holzmann for carrying out the Quantum Monte Carlo computations shown in this book, as well as Erik Bahnson who read an early draft of this manuscript and gave very valuable feedback.
This work was partially supported by the Simons Foundation grant number 825876, and the National Science Foundation grant number DMS-2349077.

\section{Bose-Einstein condensation}\label{sec:BEC}
\indent
Bose-Einstein condensation serves as the main motivation for Lieb's Simplified approach.
Proving the existence of Bose-Einstein condensation in interacting systems is one of the main open problems in the field.
In this chapter, we will discuss what Bose-Einstein condensation is in more detail, and prove that it occurs in a much simplified model: non-interacting Bosons.
But first, we must define what a ``Boson'' is, as well as the model we will be using.
\bigskip

\subsection{Elements of quantum statistical mechanics}
\indent
One of the great achievements of modern science is the understanding that macroscopic matter is made of microscopic elementary particles, and that the properties of the whole stem from the interaction between these elementary particles.
The goal of statistical mechanics is to understand how this works, that is, the relation between the microscopic and the macroscopic worlds.
In order to motivate the main ideas of quantum statistical mechanics, let us begin with a short primer on classical systems.
\bigskip

\subsubsection{Classical statistical mechanics}
\indent
Classical particles are governed by Newton's second law of motion: $F=ma$.
In the case of conservative systems (at their deepest level, all systems are actually conservative: friction is an effective description of a complicated, conservative behavior), this can be restated in the Hamiltonian formalism.
A system of $N$ particles is described by a Hamiltonian, which is a function on $\mathbb R^{6N}$ that may look something like this:
\begin{equation}
  H_N(x_1,\cdots,x_n;p_1,\cdots,p_N)=\sum_{i=1}^N\frac1{2m}p_i^2+\sum_{1\leqslant i<j\leqslant N}v(x_i-x_j)
  \label{H_classical}
\end{equation}
in which $x_i\in\mathbb R^3$ and $p_i\in\mathbb R^3$ are the positions and momenta of the particles, and $v$ is a pair interaction potential.
This dictates the {\it microscopic} dynamics.
But what if we were to use this formalism to understand the properties of a gas of particles?
Say we want to compute the pressure and temperature in a room full of air.
In an average room, there may be somewhere around $N=10^{24}$ particles, that's a trillion trillion particles.
The Hamiltonian picture gives us a trajectory for each of these particles.
This is an incredibly and uselessly large amount of information, if all we were after in the first place was pressure and temperature.
\bigskip

\indent
Instead, we will approach the problem statistically: instead of following each individual particle, we compute average behaviors.
The question is: average with respect to what probability measure?
Here, things get complicated, and, to make a long story short, there are three satisfactory measures to consider, that are called the {\it microcanonical}, {\it canonical} and {\it grand-canonical} measures.
In this section we'll focus on the grand-canonical measure.
In this setting, there are two parameters: the {\it temperature} $T>0$ (in this formalism, temperature is a parameter) and the {\it chemical potential} $\mu\in\mathbb R$.
Given these two, the probability of a given configuration $(x_1,\cdots,x_N;p_1,\cdots,p_N)$ with $N$ particles is
\begin{equation}
  \frac1{\Xi}e^{-\beta H_N(x_1,\cdots,x_N;p_1,\cdots,p_N)+\beta\mu N}
  \label{gc}
\end{equation}
where
\begin{equation}
  \beta=\frac1{k_BT}
\end{equation}
in which $k_B$ is the {\it Boltzmann constant} and $\Xi$ is a normalization:
\begin{equation}
  \Xi:=1+\sum_{N=1}^\infty\frac1{N!}\int dx_1\cdots dx_Ndp_1\cdots dp_N\ e^{-\beta H_N(x_1,\cdots,x_N;p_1,\cdots,p_N)+\beta\mu N}
  .
\end{equation}
We immediately notice an issue: if the $x_i$ take values anywhere in $\mathbb R^3$, $\Xi$ will come out infinite!
To avoid this, let us restrict $x_i\in[0,L]^3$ for some finite $L$, and, later on, we will take $L\to\infty$.
Note that $\Xi$ involves a sum over $N$: in the grand-canonical picture, we consider all possible numbers of particles.
In addition, note the $1/N!$ factor, which indicates that all the particles are identical, and the labelling $i=1,\cdots,N$ does not affect the statistics.
\bigskip

\indent
The precise reason why the probability is taken to be\-~(\ref{gc}) is not so simple.
There are several approaches to justifying it: by arguments based on finding {\it typical} configuration (see, for instance, \cite{GHe17} and references therein), or simply by arguing that this is the only way to reproduce the laws of thermodynamics (see, for instance,\cite{Ru99,Ga99}).
We will not enter in a profound discussion about this here, but instead, let us have check that this distribution is plausible, using some physical intuition.
The first term in the exponential is $e^{-\beta H}$, and, since $\beta>0$, it says that configurations with small energies will be more likely than those with large energies.
The parameter $\beta$ (the inverse of the temperature) quantifies how much more likely these are.
This is rather sensible: we know that energy tends to be minimized, and that this is more true at low temperatures than higher ones.
The second term is $e^{\beta\mu N}$ controls the number of particles.
If $\mu>0$, then configurations with more particles are more likely, whereas $\mu<0$ makes more particles unlikely.
Thus the chemical potential $\mu$ controls the {\it density} of particles.
\bigskip

\indent
Why do we use the chemical potential instead of simply fixing the number of particles (which is, after all, the most direct way of setting the density)?
This can be done: in statistical mechanics, the situation in which the number of particles is fixed is called the {\it canonical formalism}.
The parameters are then the temperature and the number of particles.
One can show that the two formalisms are equivalent: there exists a mapping between the density and chemical potential that makes all average observables identical in both formalisms\-~\cite{Ru99}.
However, in certain situations, it may be easier to compute these averages in one formalism rather than the other.
In the case of Bose-Einstein condensation, the computation is easy in the grand-canonical, and very difficult in the canonical, which is why we use the chemical potential instead of the density.
\bigskip

\indent
Thinking probabilistically in this manner, we can then compute the average behavior of these particles, and compute properties for the macroscopic system.
More specifically, we define an {\it observable} $A$ as a map from $\mathbb N$ (the number of particles) to functions on the phase space with $N$ particles: $A_N(x_1,\cdots,x_N;p_1,\cdots,p_N)$.
The average of an observable $A$ is
\begin{equation}
  \begin{largearray}
    \left<A\right>
    =
    \frac1\Xi
    \Bigg(
      A_0
      +
      \\\hfill+
      \sum_{N=1}^\infty\frac1{N!}\int dx_1\cdots dx_Ndp_1\cdots dp_N\ A_N(x_1,\cdots,x_N;p_1,\cdots,p_N)e^{-\beta H_N(x_1,\cdots,x_N;p_1,\cdots,p_N)+\beta\mu N}
    \Bigg)
    .
  \end{largearray}
\end{equation}
An example of an observable could be the number of particles itself:
\begin{equation}
  \mathcal N_N:=N
\end{equation}
or the total energy
\begin{equation}
  \mathcal E_N(x_1,\cdots,x_N;p_1,\cdots,p_N):=H_N(x_1,\cdots,x_N;p_1,\cdots,p_N)
  .
\end{equation}

\subsubsection{Quantum statistical mechanics}
\indent
We have thus defined the grand-canonical ensemble for systems of classical particles.
What about quantum particles?
The microscopic state of a system of quantum particles is given by a wavefunction, for instance, $\psi\in L_2(\mathbb R^3)^{\otimes N}$, and the dynamics is given by a Hamiltonian operator, for instance,
\begin{equation}
  H_N=-\sum_{i=1}^N\frac1{2m}\Delta_{i}+\sum_{1\leqslant i<j\leqslant N}v(x_i-x_j)
  \label{H}
\end{equation}
where $\Delta_i$ is the Laplacian with respect to $x_i$: $\Delta_i:=\partial_{x_i}^2$.
(Throughout this document, we choose units so that $\hbar=1$; we keep track of the mass because different authors use $m=1$ or $m=1/2$, so we keep it to avoid confusion.)
This is the microscopic description of the system.
Similarly to the classical case, to treat situations with a huge number of particles, we will approach the system statistically.
In the classical case, we considered a probability distribution over all possible configurations, without fixing the number of particles.
Here, we wish to do the same.
The configurations are wavefunctions, which, again, are in $L_2(\mathbb R^3)^{\otimes N}$.
Instead of integrating over configurations as we did in the classical case, we will take traces over $L_2(\mathbb R^3)^{\otimes N}$.
\bigskip

\indent
There is just one subtlety here: the particles are indistinguishable.
In the classical setting, this is dealt with simply by adding a $1/N!$ that accounts for the fact that configurations that are permutations of each other are really identical.
In the quantum mechanical case, however, the notion of ``identical'' is more complicated: two wavefunctions that have the same absolute value, but a different phase, are indistinguishable.
Therefore, to enforce the fact that particles are indistinguishable, it suffices to consider wavefunctions whose absolute value does not change under exchanges of particles, but these permutations of particles may lead to a different phase.
As it turns out, there are only two possible phase shifts that are consistent in three dimensions: one possibility is the trivial one, in which wavefunctions are symmetric under exchanges of particles: for any permutation $\sigma\in\mathcal S_N$,
\begin{equation}
  \psi_{\mathrm{sym}}(x_1,\cdots,x_N)
  =
  \psi_{\mathrm{sym}}(x_{\sigma(1)},\cdots,x_{\sigma(N)})
\end{equation}
and the other is that exchanging two particles leads to the wavefunction changing sign:
\begin{equation}
  \psi_{\mathrm{antisym}}(x_1,\cdots,x_N)
  =
  (-1)^\sigma\psi_{\mathrm{antisym}}(x_{\sigma(1)},\cdots,x_{\sigma(N)})
  .
\end{equation}
Why are these the only two possibilities (in three dimensions)?
This comes from the fact that if two particles are exchanged, the wavefunction should pick up a phase $e^{i\theta}$, but if I exchange the particles back, we should return to the original wavefunction, and thus $e^{2i\theta}=1$ and so $\theta$ is either $0$ or $\pi$.
(This is not the case in two dimensions, for reasons we will not elaborate here.)
\bigskip

\indent
We therefore define the symmetric subspace:
\begin{equation}
  L_{2,\mathrm{sym}}^{(N)}(\mathbb R^3)
  :=\{\psi\in L_2(\mathbb R^3)^{\otimes N}:\ 
  \forall \sigma\in\mathcal S_N,\ \psi(x_1,\cdots,x_N)=\psi(x_{\sigma(1)},\cdots,x_{\sigma(N)})
  \}
\end{equation}
and the antisymmetric subspace:
\begin{equation}
  L_{2,\mathrm{antisym}}^{(N)}(\mathbb R^3)
  :=\{\psi\in L_2(\mathbb R^3)^{\otimes N}:\ 
  \forall \sigma\in\mathcal S_N,\ \psi(x_1,\cdots,x_N)=(-1)^\sigma\psi(x_{\sigma(1)},\cdots,x_{\sigma(N)})
  \}
  .
\end{equation}
Note that the symmetric and antisymmetric subspaces span all of $L_2$:
\begin{equation}
  L_2(\mathbb R^3)^{\otimes N}
  =
  L_{2,\mathrm{sym}}^{(N)}(\mathbb R^3)
  \oplus
  L_{2,\mathrm{antisym}}^{(N)}(\mathbb R^3)
  .
\end{equation}
Now, from a physical point of view, systems of many indistinguishable particles come in two flavors: those with symmetric wavefunctions and those with antisymmetric wavefunctions.
Symmetric ones are called {\it Bosons} and anstisymmetric ones are called {\it Fermions}.
(Another way of seeing this is to think of particles as corresponding to irreducible representations of symmetry groups; the Boson/Fermion distinction corresponds to the two irreducible representations of the permutation group.)
Electrons, protons, quarks are examples of Fermions.
Photons, Helium atoms, Higgs particles are examples of Bosons.
Here, we will consider systems of many Bosons, which is to say that the configuration space we will consider is $L_{2,\mathrm{sym}}^{(N)}(\mathbb R^3)$.
\bigskip

\indent
An observable in the quantum setting is a map $A$ from $\mathbb N$ (the number of particles) to a self-adjoint operator $A_N$ on $L_{2,\mathrm{sym}}^{(N)}(\mathbb R^3)$.
An example of an observable is the number of particles operator:
\begin{equation}
  \mathcal N_N\psi:=N\psi
\end{equation}
another is the total energy: $H_N$.
The average of an observable is then defined as
\begin{equation}
  \left<A\right>
  :=\frac1\Xi
  \left(
    A_0+
    \sum_{N=1}^\infty
    \mathrm{Tr}_{L_{2,\mathrm{sym}}^{(N)}(\mathbb R^3)}(A_Ne^{-\beta H_N+\beta \mu\mathcal N_N})
  \right)
\end{equation}
with
\begin{equation}
  \Xi:=
  1+
  \sum_{N=1}^\infty
  \mathrm{Tr}_{L_{2,\mathrm{sym}}^{(N)}(\mathbb R^3)}(e^{-\beta H_N+\beta \mu\mathcal N_N})
  .
\end{equation}
\bigskip

\indent
As in the classical case, we immediately notice an issue: the trace in $\Xi$ is infinite!
To prevent this we replace $\mathbb R^3$ with $[0,L)^3$:
\begin{equation}
  \Xi:=
  1+
  \sum_{N=1}^\infty
  \mathrm{Tr}_{L_{2,\mathrm{sym}}^{(N)}([0,L)^3)}(e^{-\beta H_N+\beta \mu\mathcal N_N})
  .
  \label{Xi_box}
\end{equation}

\subsection{The ideal Bose gas}
\indent
Let us now apply these concepts to the non-interacting (a.k.a. the ideal) Bose gas.
The Hamiltonian is
\begin{equation}
  H_N=-\sum_{i=1}^N\frac1{2m}\Delta_{i}
  .
\end{equation}
We will consider this system on the cube $[0,L]^3$ with periodic boundary conditions (other boundary conditions can be treated similarly).
In other words, we replace $\mathbb R^3$ with $\mathbb T^3:=(\mathbb R/(L\mathbb Z))^3$.
This Hamiltonian is easily diagonalized: the eigenvectors of $\frac1{2m}\Delta$ are plane waves:
\begin{equation}
  e^{ikx}
  ,\quad
  k\in\frac{2\pi}L\mathbb Z^3
\end{equation}
whose eigenvalue is
\begin{equation}
  \epsilon_k:=\frac{k^2}{2m}
  .
\end{equation}
We therefore have a (Hilbert) basis, in which each element with $N$ particles is indexed by an unordered sequence of $N$ wavevectors $k\in\frac{2\pi}L\mathbb Z^3$ (the vectors need not be different).
Equivalently, a basis vector can be written as a function that takes a wavevector $k$ and returns the number of particles of that wavevector.
This allows us to compute $\Xi$ in the case $\mu<0$: we write the sum over $N$ of the trace in\-~(\ref{Xi_box}) as
\begin{equation}
  \Xi=
  \prod_{k\in\frac{2\pi}L\mathbb Z^3}
  \sum_{N_k=0}^\infty
  e^{-\beta(\frac{k^2}{2m}-\mu)N_k}
\end{equation}
In the case $\mu<0$, this series is absolutely convergent:
\begin{equation}
  \Xi=
  \prod_{k\in\frac{2\pi}L\mathbb Z^3}
  \frac1{1-e^{-\beta(\frac{k^2}{2m}-\mu)}}
  .
  \label{Xi_ideal}
\end{equation}
\bigskip

\indent
Let us now compute some observables.
Let $\mathcal B_q$ be the number of particles in the state $q$ whose average is thus easily computed:
\begin{equation}
  \left<\mathcal B_q\right>=
  \frac1\Xi
  \left(
    \sum_{N_q=0}^\infty
    N_qe^{-\beta(\frac{q^2}{2m}-\mu)N_q}
  \right)
  \prod_{k\in\frac{2\pi}L\mathbb Z^3\setminus\{q\}}
  \sum_{N_k=0}^\infty
  e^{-\beta(\frac{k^2}{2m}-\mu)N_k}
  \label{BE_pre}
\end{equation}
and so (see Exercise\-~\ref{ex:BE_distr})
\begin{equation}
  \left<\mathcal B_q\right>=
  \frac1{e^{\beta(\frac{q^2}{2m}-\mu)}-1}
  \label{BE}
\end{equation}
which is called the {\it Bose-Einstein distribution}.
It is the distribution of particles as a function of the wavevector $q$.

\subsection{Bose-Einstein condensation in the ideal Bose gas}
\indent
By summing the number of particles of momentum $q$, we can compute the average total number of particles:
\begin{equation}
  \left<\mathcal N\right>
  =\sum_{q\in\frac{2\pi}L\mathbb Z^3}\left<\mathcal B_q\right>
  =\sum_{q\in\frac{2\pi}L\mathbb Z^3}
  \frac1{e^{\beta(\frac{q^2}{2m}-\mu)}-1}
\end{equation}
and the average density:
\begin{equation}
  \rho_L(\mu):=\frac{\left<\mathcal N\right>}{L^3}
  =
  \frac1{L^3}
  \sum_{q\in\frac{2\pi}L\mathbb Z^3}
  \frac1{e^{\beta(\frac{q^2}{2m}-\mu)}-1}
  .
  \label{rhomu_pre}
\end{equation}
The density is an important quantity.
As was discussed above, the mapping between density and chemical potential is what establishes the equivalence between the grand-canonical and canonical formalisms (which are the points of view with fixed chemical potential and fixed density respectively).
Therefore, we will prove properties of the mapping $\mu\mapsto\rho_L(\mu)$.
But first, let us simplify the expression by taking the $L\to\infty$ limit.
\bigskip

\indent
For fixed $\mu<0$, this is a Riemann sum, so the limit $L\to\infty$ turns the sum to an integral (see Exercise\-~\ref{ex:bec}):
\begin{equation}
  \rho(\mu):=\lim_{L\to\infty}\rho_L(\mu)
  =
  \int_{\mathbb R^3}\frac{dq}{(2\pi)^3}\ 
  \frac1{e^{\beta(\frac{q^2}{2m}-\mu)}-1}
  .
  \label{rhomu}
\end{equation}
By dominated convergence, one easily checks (see Exercise\-~\ref{ex:bec}) that this is a strictly increasing function of $\mu$.
Furthermore, by dominated convergence, (see Exercise\-~\ref{ex:bec})
\begin{equation}
  \lim_{\mu\to-\infty}\rho(\mu)=0
  \label{lim_rhomu_infty}
\end{equation}
and
\begin{equation}
  \lim_{\mu\to0}\rho(\mu)=
  \int_{\mathbb R^3}\frac{dq}{(2\pi)^3}\ 
  \frac1{e^{\beta\frac{q^2}{2m}}-1}
  =
  \left(\frac{2m}\beta\right)^{\frac32}
  \int_{\mathbb R^3}\frac{dq}{(2\pi)^3}\ 
  \frac1{e^{q^2}-1}
  =:\rho_c(\beta)
  \label{lim_rhomu_0}
\end{equation}
Thus, for $\mu\in(-\infty,0)$, $\rho(\mu)$ increases from $0$ to $\rho_c(\beta)$, which is finite.
For $\mu>0$, the right side of\-~(\ref{rhomu}) is infinite.
In particular, proceeding in this way, we only cover a finite range of densities.
What of the densities with $\rho>\rho_c(\beta)$?
In other words, can we set the chemical potential $\mu$ in such a way that $\rho>\rho_c(\beta)$?
\bigskip

\indent
To achieve this, instead of fixing $\mu$ and taking $L\to\infty$, we allow $\mu$ to depend on $L$.
To make this clear, we denote $\mu$ by $\mu_L$.
We will choose $\mu_L$ such that $\mu_L\to0$ as $L\to\infty$ (if $\mu_L$ converged to a negative value, then we would find the same result as above).
Doing so changes the computation in one way: when passing to the $L\to\infty$ limit, we must pay attention to the $q=0$ term in the sum.
Let us separate it out:
\begin{equation}
  \rho_L(\mu_L)
  =
  \frac1{L^3}
  \frac1{e^{-\beta\mu_L}-1}
  +
  \frac1{L^3}
  \sum_{q\in\frac{2\pi}L\mathbb Z^3\setminus\{0\}}
  \frac1{e^{\beta(\frac{q^2}{2m}-\mu_L)}-1}
  .
\end{equation}
The second term converges uniformly in $\mu$:
\begin{equation}
  \frac1{L^3}\frac1{e^{\beta(\frac{q^2}{2m}-\mu_L)}-1}
  \leqslant
  \frac1{L^3}
  \frac1{e^{\beta\frac{2\pi^2}{mL^2}}-1}
\end{equation}
which is bounded.
Thus, taking the limit $L\to\infty$,
\begin{equation}
  \lim_{L\to\infty}\rho_L(\mu_L)
  =
  \lim_{L\to\infty}
  \frac1{L^3}
  \frac1{e^{-\beta\mu_L}-1}
  +
  \int_{\mathbb R^3}\frac{dq}{(2\pi)^3}\ 
  \frac1{e^{\beta\frac{q^2}{2m}}-1}
\end{equation}
that is
\begin{equation}
  \lim_{L\to\infty}\rho_L(\mu_L)
  =
  \lim_{L\to\infty}
  \frac1{L^3}
  \frac1{e^{-\beta\mu_L}-1}
  +
  \rho_c(\beta)
  .
\end{equation}
Thus, any density $\rho>\rho_c(\beta)$ can be attained by taking $\mu_L$ such that
\begin{equation}
  \frac1{L^3}
  \frac1{e^{-\beta\mu_L}-1}
  =\rho-\rho_c(\beta)
\end{equation}
that is
\begin{equation}
  \mu_L=-\frac1\beta\log\left(1+\frac1{L^3(\rho-\rho_c(\beta))}\right)
  .
\end{equation}
In this case,
\begin{equation}
  \lim_{L\to\infty}\frac1{L^3}\left<\mathcal B_0\right>
  =\lim_{L\to\infty}\frac1{L^3}\frac1{e^{-\beta\mu_L}-1}
  =\rho-\rho_c(\beta)
\end{equation}
and for $q\neq 0$,
\begin{equation}
  \lim_{L\to\infty}\frac1{L^3}\left<\mathcal B_q\right>
  =\lim_{L\to\infty}\frac1{L^3}\frac1{e^{\beta(\frac{q^2}{2m}-\mu_L)}-1}
  =0
  .
\end{equation}
Thus, choosing $\mu_L$ in this way, we find that the number of particles in the state $q=0$ is $L^3(\rho-\rho_c)$, that is, the number of particles in this state is proportional to $N$.
There is a {\it macroscopic} proportion of particles that are all in this one state.
Whereas in every other state $q\neq0$, the fraction of particles is 0.
\bigskip

\begin{figure}
  \hfil\includegraphics[width=8cm]{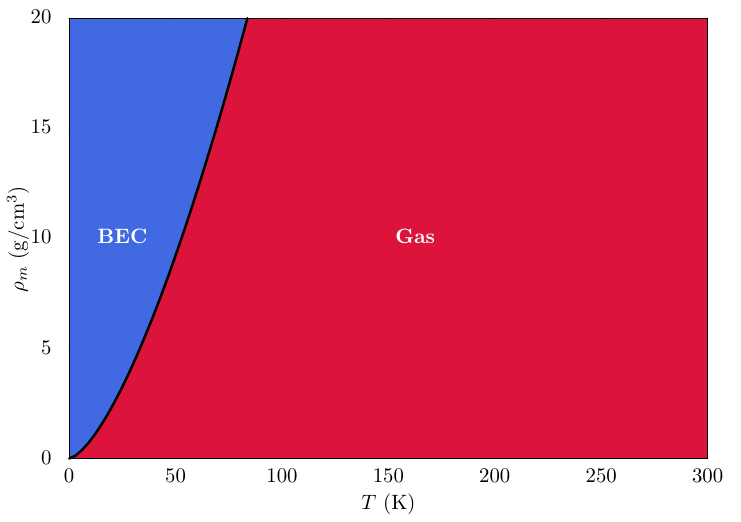}
  \caption{The phase diagram of the ideal gas of Helium-4 atoms.
  The $y$-axis is the mass density.}
  \label{fig:phase}
\end{figure}

\indent
Let's take a step back.
We have been working in the grand-canonical ensemble, that is, with a fixed chemical potential and a variable number of particles.
Proceeding in this way, we have computed the average density, and found a bijection between the chemical potential and the density, provided we allow the chemical potential to depend on the size of the system $L$.
This allows us to switch the point of view and consider the system at fixed density and temperature.
Thinking in this way, we find that the behavior of the system depends in a non-trivial way on the density.
If $\rho<\rho_c(\beta)$, then $\left<\mathcal B_0\right>/L^3\to0$, but if $\rho>\rho_c(\beta)$, then $\left<\mathcal B_0\right>/L^3\to c_0>0$, see Figure\-~\ref{fig:phase}.
Or, to put it into words, for small densities, the system behaves like a regular, classical gas, but when the density exceeds $\rho_c(\beta)$, a huge number of particles all occupy the same state.
This phenomenon is called Bose-Einstein condensation.

\bigskip

\indent
Bose-Einstein condensation was first predicted by Bose\-~\cite{Bo24} and Einstein\-~\cite{Ei24} in the 1920's (following roughly the argument given above).
It took until 1995 before it was finally observed experimentally\-~\cite{AEe95,DMe95}.
The experiments consist in trapping very cold Bosonic atoms (rubidium and sodium) at a high enough density so as to be in the Bose-Einstein condensate phase.
The trap is then released, and the momentum of the atoms measured.
The result is a sharply peaked distribution: many atoms were in the same state!

\section{Knowns and unknowns of the interacting Bose gas}\label{sec:bose_results}
\subsection{Interacting Bose gases}
\indent
We have proved that the ideal Bose gas exhibits Bose-Einstein condensation.
The approach we took consisted in computing exact expressions for the partition function $\Xi$ and the average occupation numbers $\left<\mathcal B_q\right>$.
We were able to do this because the ideal Bose gas is an ``exactly solvable model'', but this is a very special feature of that specific model.
Furthermore, the ideal gas is not very realistic: in that model, there are no interactions between particles, which is not how real matter behaves.
Let us consider a more realistic Hamiltonian:
\begin{equation}
  H_N=-\sum_{i=1}^N\frac1{2m}\Delta_{i}+\sum_{1\leqslant i<j\leqslant N}v(x_i-x_j)
  \label{Ham}
\end{equation}
where $v$ is the potential that accounts for the interaction between pairs of particles.
(We choose units so that $\hbar=1$; we keep track of the mass because different authors use $m=1$ or $m=1/2$, so we keep it to avoid confusion.)
This potential could be the Coulomb potential $1/|x|$ for charged particles, or the Lennard-Jones potential for interacting atoms, or something more realistic that accounts for the fine structure of atoms.
Here, we will restrict our focus a bit and assume that $v$ is integrable ($v\in L_1(\mathbb R^3)$) and non-negative: $v(x)\geqslant 0$.
In addition, we will only consider the zero-temperature state ($T=0$ that is $\beta=\infty$), in other words, we will be considering only the ground state of $H_N$, which is the eigenstate with lowest eigenvalue.
Ultimately, we would like to study the system at positive temperatures, but, as we will see, the zero-temperature analysis is difficult enough.
\bigskip

\indent
Adding the potential breaks the exact solvability of the model: there no longer is an exact expression for the observables associated with this Hamiltonian, so we will have to proceed differently.
However, as of this writing, there are no tools to compute very much for this model.
The main rigorous results that have been proved so far are computations of the average energy for small\-~\cite{YY09,FS20} or large\-~\cite{Li63} densities.
In particular, there is no proof of Bose-Einstein condensation.
\bigskip

\indent
If we can't solve this model, what can we do?
One approach is to change the model.
For instance, we could make the potential depend on the number of particles.
This is not very realistic for a microscopic model, but it can accurately capture the behavior of the system in certain asymptotic regimes.
For instance, the Gross-Pitaevskii scaling, which consists in taking the potential to be
\begin{equation}
  v(x)\mapsto N^{\frac23}v(N^{\frac13}x)
\end{equation}
corresponds to an infinitely dilute gas $\rho=0$, and the mean-field scaling
\begin{equation}
  v(x)\mapsto \frac1{N^{\frac53}}v(N^{-\frac13}x)
\end{equation}
corresponds to an infinitely dense system $\rho=\infty$.
Both regimes have been studied extensively (see\-~\cite{Sc22,Se11} and the references therein) and much has been proved about these.
However, these scaling regimes do not capture the physical behavior of the system at finite density.
\bigskip

\indent
Another approach is to make some approximations that, while not being exact, capture some of the physical behavior of the system.
In a sense, the ideal Bose gas is such an approximation.
However, the potential cannot be ignored in practice, so a more refined approximation is in order.
One such approximation was introduced by Bogolyubov\-~\cite{Bo47} and has been studied extensively since then.
It has since become known as {\it Bogolyubov theory}.
The key observation made by Bogolyubov is that the behavior of the particles in the condensate is simple, we can deal with them in a simplified way, and that what is left over does not interact too much so we can drop all the interaction terms in the Hamiltonian that we don't know how to deal with.
This idea seems to work very well as long as the density is small enough.
One distinct advantage is that, after having ignored certain terms, the model becomes exactly solvable like the ideal Bose gas, so everything can be computed.
This is what was done by Lee, Huang and Yang\-~\cite{LHY57}, who thus made a number of predictions about the interacting Bose gas, concerning the energy, condensate fraction and two point correlation function.
In all of these, the effect of the interaction appears through a quantity called the {\it scattering length}.
Before stating the conjectures of Lee, Huang and Yang, let us discuss the scattering length.

\subsection{Scattering theory}
\indent
The physical picture that underlies Bogolyubov theory is that, at low densities, the behavior of the system should be dominated by single body dynamics, and the interaction only comes in in the rare events when particles get close to each other.
Because these are rare, these interaction events will only involve single pairs of particles.
These pair interaction events are called {\it scattering events}.
The physical picture to have in mind is two particles coming closer together in straight lines (free states), interacting, and leaving along different lines from the ones they started in (different free states).
By translation invariance, we can work in the center of mass reference frame, in which the scattering event can be seen as a single particle flying through a fixed potential.
The Schr\"odinger equation for this process is
\begin{equation}
  -\frac1m\Delta\psi+v(x)\psi=i\partial_t\psi
\end{equation}
(note the absence of a $1/2$ factor in front of the Laplacian which comes from the fact that the effective mass in the center of mass frame is $\frac{m_1m_2}{m_1+m_2}=m/2$.)
(Recall that we chose units so that $\hbar=1$.)
\bigskip

\indent
We will focus on stationary solutions of the equation:
\begin{equation}
  -\frac1m\Delta\psi+v(x)\psi=E\psi
\end{equation}
and will study spherically symmetric solutions, which must have $E=0$ (this is an {\it s-wave}).
That is, we will solve the {\it scattering equation}
\begin{equation}
  -\frac1m\Delta\psi+v(x)\psi=0
  .
  \label{scateq}
\end{equation}
We impose the boundary condition
\begin{equation}
  \lim_{|x|\to\infty} \psi(x)=1
  .
\end{equation}
\bigskip

\indent
Let us consider a simple example: the hard core potential:
\begin{equation}
  v(x)=\left\{\begin{array}{>\displaystyle ll}
    0&\mathrm{if\ }|x|>R\\
    \infty&\mathrm{otherwise}
    .
  \end{array}\right.
\end{equation}
In other words, the scattering equation is
\begin{equation}
  \left\{\begin{array}{>\displaystyle ll}
    \psi(x)=0&\mathrm{if\ }|x|\leqslant R\\[0.3cm]
    -\frac1m\Delta\psi=0&\mathrm{otherwise}
    .
  \end{array}\right.
\end{equation}
In spherical coordinates, assuming $\psi$ is spherically symmetric, that is, it only depends on $r\equiv|x|$,
\begin{equation}
  \Delta\psi=\frac1r\partial_r^2(r\psi)
\end{equation}
so $\Delta\psi=0$ yields the solution, for $r>R$,
\begin{equation}
  \psi=c-\frac ar
\end{equation}
and, using $\psi(R)=0$ and $\psi\to1$,
\begin{equation}
  c=1
  ,\quad
  a=R
\end{equation}
so
\begin{equation}
  1-\psi=\frac Rr
  .
\end{equation}
The constant $c$ is dimensionless, and the constant $a$ has the dimension of a length.
In this example, we see that $a$ is the radius of the potential.
\bigskip

\indent
Let us now turn to a different potential: the soft core potential:
\begin{equation}
  v(x)=\left\{\begin{array}{>\displaystyle ll}
    0&\mathrm{if\ }|x|>R\\
    U&\mathrm{otherwise}
  \end{array}\right.
\end{equation}
with $U>0$.
The scattering equation is thus
\begin{equation}
  \left\{\begin{array}{>\displaystyle ll}
    -\frac1m\Delta\psi+U\psi=0&\mathrm{if\ }|x|\leqslant R\\[0.3cm]
    -\frac1m\Delta\psi=0&\mathrm{otherwise}
  \end{array}\right.
\end{equation}
which, in spherical coordinates is
\begin{equation}
  \left\{\begin{array}{>\displaystyle ll}
    -\frac1{mr}\partial_r^2(r\psi)+U\psi=0&\mathrm{if\ }|x|\leqslant R\\[0.3cm]
    -\frac1{mr}\partial_r^2(r\psi)=0&\mathrm{otherwise}
    .
    \label{scat_softcore_spherical}
  \end{array}\right.
\end{equation}
The solution is (see Exercise\-~\ref{ex:scattering_softcore}), using the fact that $r\psi$ should equal $0$ at $r=0$ and that $r\psi$ is continuous at $r=R$,
\begin{equation}
  r\psi=
  \left\{\begin{array}{>\displaystyle ll}
    (R-a)\frac{\sinh(\sqrt{Um}r)}{\sinh(\sqrt{Um}R)}&\mathrm{if\ }r\leqslant R\\[0.3cm]
    r-a&\mathrm{otherwise}
  \end{array}\right.
  \label{sol_softcore}
\end{equation}
with
\begin{equation}
  a=R-\frac{\tanh(\sqrt{Um}R)}{\sqrt{Um}}
  .
  \label{scat_softcore}
\end{equation}
Note that $a$ is an increasing function of $U$, ranging from $0$ for $U=0$ to $R$ at infinity.
Therefore, the solution of the scattering equation for large $r$ looks like that of a hard core, but with radius $a$ instead of $R$.
\bigskip

\indent
This length scale $a$ is called the {\it scattering length}.
We can define it for potentials that are compactly supported, that is, for which there exists $R$ such that $v(x)=0$ for $|x|>R$.
In this case, for $|x|>R$,
\begin{equation}
  \psi(x)=1-\frac a{|x|}
\end{equation}
where $a$ is the scattering length.
Alternatively, the scattering length can be computed as follows.
\bigskip

\theo{Lemma}\label{lemma:scattering}
  If $v$ is spherically symmetric, compactly supported and integrable, then
  \nopagebreakaftereq
  \begin{equation}
    a=\frac m{4\pi}\int dx\ v(x)\psi(x)
    .
  \end{equation}
\endtheo
\restorepagebreakaftereq
\bigskip

\indent\underline{Proof}:
  By the scattering equation\-~(\ref{scateq}),
  \begin{equation}
    \frac m{4\pi}\int dx\ v(x)\psi(x)
    =
    \frac1{4\pi}\int dx\ \Delta\psi(x)
  \end{equation}
  which we rewrite in spherical coordinates, using $\Delta \psi=\frac1{r^2}\partial_r(r^2\partial_r\psi)$:
  \begin{equation}
    \frac m{4\pi}\int dx\ v(x)\psi(x)
    =
    \int_0^\infty dr\ \partial_r(r^2\partial_r\psi(r))
    =
    \lim_{R\to\infty}\int_0^R dr\ \partial_r(r^2\partial_r\psi(r))
    =
    \lim_{R\to\infty}R^2\partial_r\psi(R)
    .
  \end{equation}
  Since $v$ has compact support, $v(x)=0$ for $|x|>R$, so
  \begin{equation}
    \psi=1-\frac ar
  \end{equation}
  and
  \begin{equation}
    \partial_r\psi=\frac a{r^2}
  \end{equation}
  so
  \begin{equation}
    \lim_{R\to\infty}R^2\partial_r\psi(R)
    =a
  \end{equation}
  which proves the lemma.
\qed
\bigskip

\indent
This lemma gives us an alternative definition of the scattering length, which does not require $v$ to be compactly supported, just that it be in $L_1(\mathbb R^3)$ so that the integral is finite.

\subsection{The Lee-Huang-Yang predictions}\label{sec:lhy}
\indent
By following the prescriptions of Bogolyubov theory, Lee, Huang and Yang\-~\cite{LHY57} made a number of predictions about the ground state (zero-temperature state) of the interacting Bose gas.
Here, we will focus on their predictions for the energy per particle and the condensate fraction.
\bigskip

\indent
As was mentioned above, we study the ground state of the Hamiltonian\-~(\ref{Ham}).
Let us be more specific.
As we did in Chapter\-~\ref{sec:BEC}, we will consider the system to be in a finite volume with periodic boundary conditions.
We denote the volume by $V$, and place the particles on the torus $\mathbb R^3/(V^{\frac13}\mathbb Z)^3$.
Unlike Chapter\-~\ref{sec:BEC}, we will work in the canonical ensemble, which means that we fix the number of particles to some $N\in\mathbb N$.
We will be interested in the {\it thermodynamic limit}, in which
\begin{equation}
  N\to\infty
  ,\quad
  V\to\infty
  ,\quad
  \frac NV=\rho
\end{equation}
with $\rho$ (the density) fixed.
In this case, the Hamiltonian\-~(\ref{Ham}) has pure-point spectrum (by Theorem\-~\ref{theo:compact_schrodinger}, $e^{-H_N}$ is compact, so it has discrete spectrum, and therefore so does $H_N$).
Its lowest eigenvalue is denoted by $E_0$, and is called the {\it ground-state energy}.
The corresponding eigenvector is denoted by $\psi_0$:
\begin{equation}
  H_N\psi_0=E_0\psi_0
  .
  \label{eigenvalue}
\end{equation}
(Note that the ground state is unique, see Exercise\-~\ref{ex:perron_frobenius}.)
\bigskip

\indent
The ground-state energy per particle is defined as
\begin{equation}
  e_0:=\lim_{\displaystyle\mathop{\scriptstyle N,V\to\infty}_{\frac NV=\rho}}\frac{E_0}N
  .
\end{equation}
Lee, Huang and Yang predicted the following low-density expansion for the ground state energy.
\bigskip

\theoname{Conjecture}{{\rm\cite[(25)]{LHY57}}}\label{conjecture:lhy_e}
  For any potential $v\in L_1(\mathbb R^3)$ with $v\geqslant 0$ (actually, the potential may include a hard-core component), as $\rho\to 0$,
  \begin{equation}
    e_0=\frac{2\pi}m\rho a\left(1+\frac{128}{15\sqrt\pi}\sqrt{\rho a^3}+o(\sqrt{\rho})\right)
    \label{lhy_e}
  \end{equation}
  where $a$ is the scattering length of the potential, and
  \nopagebreakaftereq
  \begin{equation}
    \lim_{\rho\to 0}\frac{o(\sqrt{\rho})}{\sqrt{\rho}}=0
    .
  \end{equation}
\endtheo
\restorepagebreakaftereq
\bigskip

The first two orders of this expansion thus only depend on the potential through its scattering length.
This is a universality result: any two potentials that share the same scattering length, even if they are radically different, will yield the same first two orders of the low-density expansion of the ground state energy.
\bigskip

\indent
The condensate fraction is the fraction of particles that are in the Bose-Einstein condensate.
In the case of the ideal Bose gas, the particles condensed in the $k=0$ state, which is the constant wavefunction.
In the interacting case, the condensate state will still be the constant state, simply because the condensate state is unique, and the system is invariant under translations.
Therefore, the condensate state must be invariant under translations, and so must be the constant state
\begin{equation}
  \varphi_0=\frac1{\sqrt V}
  .
\end{equation}
To compute the average number of particles in the condensate state, we project the wavefunction onto the condensate state: let
\begin{equation}
  P_i:=\mathds 1^{\otimes(i-1)}\otimes\left|\varphi_0\right>\left<\varphi_0\right|\otimes \mathds 1^{\otimes(N-i)}
  \label{Pi}
\end{equation}
be the projector onto the subspace where the $i$-th particle is in the constant state (we use the ``bra-ket'' notation of quantum mechanics: $\left|\varphi_0\right>\left<\varphi_0\right|f\equiv\left<\varphi_0,f\right>\varphi_0$).
The average number of particles in the condensate is then
\begin{equation}
  \sum_{i=1}^N\left<\psi_0\right|P_i\left|\psi_0\right>
  .
\end{equation}
The condensate fraction is then
\begin{equation}
  \eta_0:=
  \lim_{\displaystyle\mathop{\scriptstyle N,V\to\infty}_{\frac NV=\rho}}
  \frac1N
  \sum_{i=1}^N\left<\psi_0\right|P_i\left|\psi_0\right>
  .
  \label{eta_Pi}
\end{equation}
The Lee-Huang-Yang prediction of the condensation fraction for low densities is the following.
\bigskip

\theoname{Conjecture}{{\rm\cite[(41)]{LHY57}}}\label{conjecture:lhy_h}
  For any potential $v\in L_1(\mathbb R^3)$ with $v\geqslant 0$ (actually, the potential may include a hard-core component), as $\rho\to 0$,
  \begin{equation}
    1-\eta_0\sim\frac{8\sqrt{\rho a^3}}{3\sqrt\pi}
    \label{lhy_h}
  \end{equation}
  where $a$ is the scattering length of the potential.
\endtheo
\bigskip

In particular, this conjecture would imply Bose-Einstein condensation, which means that
\begin{equation}
  \eta_0>0
  .
\end{equation}
\bigskip

\indent
Only the first of these two conjectures has been proved, under some mild assumptions on $v$:
\bigskip

\theoname{Theorem}{{\rm\cite{YY09,FS20,BCS21}}}\label{theo:lhy_e}
  For any compactly supported potential $v\in L_3(\mathbb R^3)$ with $v\geqslant 0$, as $\rho\to 0$,
  \begin{equation}
    e_0=\frac{2\pi}m\rho a\left(1+\frac{128}{15\sqrt\pi}\sqrt{\rho a^3}+o(\sqrt{\rho})\right)
    \label{lhy}
  \end{equation}
  where $a$ is the scattering length of the potential, and
  \nopagebreakaftereq
  \begin{equation}
    \lim_{\rho\to 0}\frac{o(\sqrt{\rho})}{\sqrt{\rho}}=0
    .
  \end{equation}
\endtheo
\restorepagebreakaftereq
\bigskip

The first order term, $2\pi\rho a/m$ was proved by Lieb and Yngvason\-~\cite{LY98}, and the second order term was proved by\-~\cite{FS20} and\-~\cite{YY09,BCS21}.
\cite{FS20} proved that\-~(\ref{lhy_e}) holds as a lower bound for soft-core potentials of finite range (they also extended their result to hard core potentials in\-~\cite{FS23}).
\cite{YY09}\-~proved the corresponding upper bound for small potentials, which was generalized to $L_3$ finite-range potentials by\-~\cite{BCS21} (the upper bound in the case of a hard-core potential is still open).
Conjecture\-~\ref{conjecture:lhy_h} is still wide open.
In particular, there is still no proof of Bose-Einstein condensation for interacting models.

\section{Definition of Lieb's Simplified approach}\label{sec:simplified_def}
\indent
Proving properties of interacting Bose gases is a challenging task, and, so far, some major and natural questions are still open, such as proving Bose-Einstein condensation.
In this Chapter, we will introduce Lieb's Simplified approach, which is an approximation scheme that reproduces some of the physical behavior of the interacting Bose gas, while being much easier to study.
It is worth stressing that, at least as far as we understand, Lieb's Simplified approach is quite different from Bogolyubov theory, or from the rigorous methods that lead to\-~\cite{FS20}.
At the same time, the evidence discussed below shows that this approach reproduces several physical predictions for interacting Bose gases.
This makes it a promising candidate to study the mathematics and physics of interacting Bose gases.
\bigskip

\subsection{Overview}
\indent
Lieb's Simplified approach is what its name suggests: a simplified way of studying the interacting Bose gas that is considerably simpler than dealing with the Hamiltonian\-~(\ref{Ham}).
By using the term ``simplified'', we mean to imply that the results found using the approach are not (at least not yet) proven to be valid for the original Hamiltonian\-~(\ref{Ham}).
In other words, it has (so far) not been used to prove any properties of\-~(\ref{Ham}) (though there is ongoing work in that direction).
However, we have ample numerical and analytical evidence that the Simplified approach captures a lot of the physics of the Bose gas.
In fact, it captures significantly more than Bogolyubov theory.
This makes it quite an intriguing theory, and seems to be worth studying in more detail.
We will define Lieb's Simplified approach in the next section; for now, let us discuss the results predicted by the approach, to motivate the more detailed study that will follow.
\bigskip

\indent
Lieb's Simplified approach reduces the computation of the ground state energy to solving a partial differential equation on $\mathbb R^3$ which is non-linear and non-local.
Actually, it contains a family of equations.
Some are more accurate and harder to study, while others are easier to study but less accurate.
Here we will discuss three of these equations, called the Big\-~(\ref{bigeq}), Medium\-~(\ref{medeq}) and Simple\-~(\ref{simpleq}) Equations.
Solving these equations is not, strictly speaking, easy, but it is much more so than solving the equation\-~(\ref{eigenvalue}), which involves an infinite number of particles, whereas Lieb's Simplified approach provides single-particle equations.
\bigskip

\indent
Within the Simplified approach, we have proved analogous statements to Theorem\-~\ref{theo:lhy_e}\-~\cite{CJL20}, and Conjecture\-~\ref{conjecture:lhy_h}\-~\cite{CJL21}.
In particular, we have proved that Bose-Einstein condensation occurs in Lieb's Simplified approach.
Thus, it tells us at least as much as Bogolyubov theory (at least for the ground state of\-~(\ref{Ham}): it does not have anything to say about the thermal state (at least for now, see Chapter\-~\ref{sec:open})).
But it actually goes quite a bit further.
The large density expansion of the ground state energy per particle for the interacting Bose gas had been proved to be\-~\cite{Li63}, at least in the case in which the Fourier transform of the potential is non-negative: $\hat v\geqslant 0$,
\begin{equation}
  e_0\sim_{\rho\to\infty}\frac\rho2\int dx\ v(x)
  .
  \label{e_highrho}
\end{equation}
Within the Simplified approach, this also holds.
This is the first indication that something rather significant is going on: Lieb's Simplified approach gives accurate results for {\it both low and high densities}.
This is rather unusual: there are many effective theories in physics (Bogolyubov theory is one of them), but in almost all such cases, the effective theory works in just one parameter regime.
Lieb's Simplified approach works in two opposite ranges of density.
\bigskip

\begin{figure}
  \hfil\includegraphics[width=8cm]{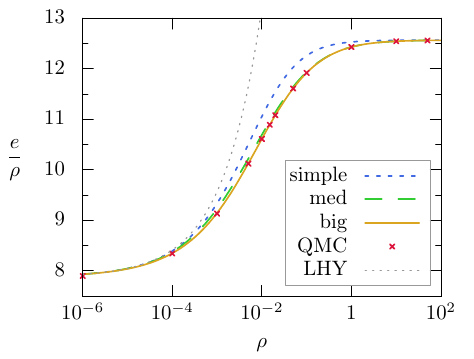}
  \caption{%
    \cite[Fig 1]{CHe21} The predictions of the energy per particle as a function of the density for the \eqformat{Simple Equation}, \eqformat{Medium Equation}, and \eqformat{Big Equation}, compared to a \eqformat{Quantum Monte Carlo} (QMC) simulation (computed by M.\-~Holzmann) and the Lee-Huang-Yang (LHY) formula, for $v=e^{-|x|}$.
  }
  \label{fig:energy}
\end{figure}

\indent
A natural question that arises is: if Lieb's Simplified approach is accurate for small and large densities, how well does it do for intermediate densities?
The energy per particle is plotted in Figure\-~\ref{fig:energy}.
The red crosses come from a Quantum Monte Carlo computation, which is the gold standard for quantum mechanical predictions (even though the theoretical control over the accuracy of this method is relatively incomplete; quantum mechanics is hard...).
The gray dotted line is the Bogolyubov prediction, which we see is good for small densities, but quickly falls off the mark.
The predictions for the Big, Medium and Simple Equations are plotted as well.
The Simple Equation is accurate at small and large densities, as was mentioned above, but not great at intermediate densities.
The Medium and Big Equations are quite accurate at {\it all} densities, especially the Big Equation which visually looks like a fit of the Quantum Monte Carlo data.
This means that the Big Equation is capturing at least some of the physics of the many-body Bose gas {\it at all densities}!
In addition, the Big Equation has a distinct advantage over Quantum Monte Carlo: it is much easier to solve numerically.
Quantum Monte Carlo requires a significant amount of artistry: it is too costly to simulate very many particles, so the results need to be inferred from computations with a few dozen particles at most.
On the other hand, the Big Equation is a partial differential equation that can be solved numerically with any of the usual tools.
\bigskip

\begin{figure}
  \hfil\includegraphics[width=8cm]{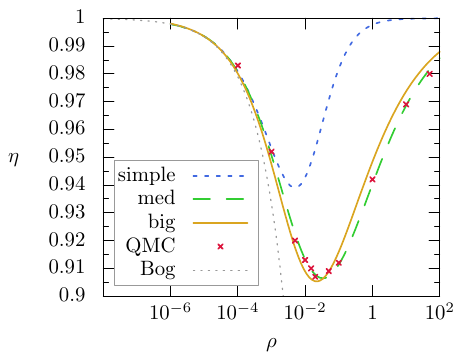}
  \caption{%
    The predictions of the condensate fraction as a function of the density for the \eqformat{Simple Equation}, \eqformat{Medium Equation}, and \eqformat{Big Equation}, compared to a \eqformat{Quantum Monte Carlo} (QMC) simulation (computed by M.\-~Holzmann) and the Bogolyubov prediction (Bog), for $v=e^{-|x|}$.
  }
  \label{fig:condensate}
\end{figure}

\indent
The condensate fraction is plotted in Figure\-~\ref{fig:condensate}.
Again, we see that the Bogolyubov prediction is good for very small densities, but quickly goes wrong.
The Big and Medium equations, on the other hand, seem to be very accurate throughout the range of densities.
\bigskip

\indent
Thus, Lieb's Simplified approach provides us with a simple tool to study systems of interacting Bosons.
The simplest equation can be studied analytically, and reproduces known and conjectured results\-~\cite{CJL20,CJL21,Ja22}.
The more complicated equations are surprisingly accurate, while being much simpler to handle than traditional methods\-~\cite{CHe21,Ja23}.
In particular, the Big Equation has been used\-~\cite{Ja23} to find evidence for a previously unknown phase in Bosons at intermediate densities.

\subsection{Lieb's Simplified approach and its approximations}\label{sec:simplified_construction}
\indent
The Simplified approach dates back to a paper by Lieb from 1963\-~\cite{Li63}.
The discussion below follows\-~\cite{Li63} closely (though with more detail).
The starting point is\-~(\ref{eigenvalue}), which we write out using\-~(\ref{Ham}):
\begin{equation}
  -\sum_{i=1}^N\frac1{2m}\Delta_{i}\psi_0+\sum_{1\leqslant i<j\leqslant N}v(x_i-x_j)\psi_0
  =E_0\psi_0
  .
  \label{starteq}
\end{equation}
\bigskip

\subsubsection{Expression of the energy}
In quantum mechanics, the probability distribution of the positions of the particles is $|\psi_0|^2$, so a natural approach to computing the energy from\-~(\ref{starteq}) would be to multiply both sides by $\psi_0^*$ and integrate.
This approach would lead to an expression of the energy as a functional to be minimized.
However, this is quite difficult (if not impossible) to do when there are so many particles.
Instead, we will simply integrate both sides of the equation (which is equivalent to taking a scalar product on both sides with the constant function (which, we recall, is the condensate wavefunction)).
In doing so, we find the integral of the Laplacian of $\psi_0$, which, since we took periodic boundary condition, is identically zero.
We are then left with
\begin{equation}
  \sum_{1\leqslant i<j\leqslant N}\int dx_1\cdots dx_N\ v(x_i-x_j)\psi_0(x_1,\cdots,x_N)=E_0\int dx_1\cdots dx_N\ \psi_0(x_1,\cdots,x_N)
\end{equation}
which we can rewrite using the symmetry under particle exchanges as
\begin{equation}
  E_0=\frac{N(N-1)}{2V^2}\int dx_1dx_2\ v(x_1-x_2)
  \frac{\int\frac{dx_3}V\cdots\frac{dx_N}V\ \psi_0(x_1,\cdots,x_N)}{\int\frac{dy_1}V\cdots\frac{dy_N}V\ \psi_0(y_1,\cdots,y_N)}
  .
\end{equation}
This is written rather suggestively: if $\psi_0/\int\psi_0$ were a probability distribution, then
\begin{equation}
  g_2(x_1-x_2):=\frac{\int\frac{dx_3}V\cdots\frac{dx_N}V\ \psi_0(x_1,\cdots,x_N)}{\int\frac{dy_1}V\cdots\frac{dy_N}V\ \psi_0(y_1,\cdots,y_N)}
\end{equation}
would be its {\it two-point correlation function} (which is different from the quantum mechanical two-point correlation function in\-~(\ref{C2})).
But, $\psi_0/\int\psi_0$ {\it is} a probability distribution: by an argument based on the Perron-Frobenius theorem, see Exercise\-~\ref{ex:perron_frobenius}, $\psi_0\geqslant 0$.
Thus, we can rewrite $E_0$ in terms of the two-point correlation function $g_2$:
\begin{equation}
  \frac{E_0}N=\frac{(N-1)}{2V}\int dx\ v(x)g_2(x)
  .
  \label{E0}
\end{equation}

\subsubsection{Hierarchy of equations}
\indent
To compute $g_2$, we follow the same idea, except that instead of integrating\-~(\ref{starteq}) with respect to all variables, we do so with respect to all but the first two: $\int dx_3\cdots dx_N$:
\begin{equation}
  \begin{largearray}
    -\frac1{2m}\int dx_3\cdots dx_N\ (\Delta_1\psi_0+\Delta_2\psi_0)
    +v(x_1-x_2)\int dx_3\cdots dx_N\ \psi_0
    +\\\indent
    +\sum_{i=3}^\infty \int dx_3\cdots dx_N\ (v(x_1-x_i)+v(x_2-x_i))\psi_0
    +\sum_{3\leqslant i<j\leqslant N}
    \int dx_3\cdots dx_N\ v(x_i-x_j)\psi_0
    =\\\hfill
    =E_0\int dx_3\cdots dx_N\ \psi_0
  \end{largearray}
\end{equation}
which, again using the invariance under particle exchanges, reads
\begin{equation}
  \begin{largearray}
    -\frac1{2m}\int dx_3\cdots dx_N\ (\Delta_1\psi_0+\Delta_2\psi_0)
    +v(x_1-x_2)\int dx_3\cdots dx_N\ \psi_0
    +\\\indent
    +(N-2) \int dx_3\ (v(x_1-x_3)+v(x_2-x_3))\int dx_4\cdots dx_N\ \psi_0
    +\\\hfill
    +\frac{(N-2)(N-3)}2
    \int dx_3dx_4\ v(x_3-x_4)
    \int dx_5\cdots dx_N\ \psi_0
    =E_0\int dx_3\cdots dx_N\ \psi_0
    .
  \end{largearray}
\end{equation}
Defining the $n$-point correlation function of $\psi_0/\int\psi_0$ as
\begin{equation}
  g_n(x_1,\cdots,x_n):=\frac{\int\frac{dx_{n+1}}V\cdots\frac{dx_N}V\ \psi_0(x_1,\cdots,x_N)}{\int\frac{dy_1}V\cdots\frac{dy_N}V\ \psi_0(y_1,\cdots,y_N)}
  \label{gn}
\end{equation}
we have
\begin{equation}
  \begin{largearray}
    -\frac1{m}\Delta g_2(x_1-x_2)
    +v(x_1-x_2)g_2(x_1-x_2)
    +\\\indent
    +\frac{N-2}V \int dx_3\ (v(x_1-x_3)+v(x_2-x_3))g_3(x_1,x_2,x_3)
    +\\\hfill
    +\frac{(N-2)(N-3)}{2V^2}
    \int dx_3dx_4\ v(x_3-x_4)
    g_4(x_1,x_2,x_3,x_4)
    =
    E_0g_2(x_1-x_2)
    .
    \label{g2_hierarchy}
  \end{largearray}
\end{equation}
We have thus obtained an equation for $g_2$ that involves $g_3$ and $g_4$.
We could keep on going, and we would find a full hierarchy of equations involving all the $g_n$.
\bigskip

\subsubsection{Main approximation of Lieb's Simplified approach}
\indent
So far, we have done nothing fishy, and solving this hierarchy is equivalent to solving the original eigenvalue problem\-~(\ref{eigenvalue}).
However, this would be too difficult.
Instead, we will make an approximation, which consists in expressing $g_3$ and $g_4$ in terms of $g_2$.
This will turn\-~(\ref{g2_hierarchy}) into an equation for $g_2$ alone.
\bigskip

\theoname{Assumption}{The factorization assumption}\label{assum:factorization}
  For $i=3,4$,
  \begin{equation}
    g_i(x_1,\cdots,x_i)
    =
    \prod_{1\leqslant j<l\leqslant i}
    (1-u_i(x_j-x_l))
    \label{g_factorized}
  \end{equation}
  in which $u_i$ is bounded independently of $V$ and is uniformly integrable in $V$:
  \begin{equation}
    |u_i(x)|\leqslant\bar u_i(x)
    ,\quad
    \int dx\ \bar u_i(x)\leqslant c_i
    \label{assum_bound}
  \end{equation}
  with $c_i$ independent of $V$.
\endtheo
\bigskip

In other words, $g_3$ and $g_4$ factorize exactly into a product of pair terms.
Here, we use the term ``assumption'' because it leads to Lieb's Simplified approach.
However, it is really an {\it approximation} rather than an assumption: this factorization will certainly not hold true exactly (the only wavefunction for which all the correlation functions factorize is the constant function).
At best, one might expect that the assumption holds approximately in the limit of small and large $\rho$, and for distant points, as numerical evidence suggests in the translation invariant case.
We will not attempt an argument here that this approximation is accurate.
Suffice it to say that this approximation is one of {\it statistical independence} that is reminiscent of phenomena arising in statistical mechanics when the density is low, that is, when the interparticle distances are large.
In the current state of the art, we do not have much in the way of an explanation for why this statistical independence should hold; instead, we have extensive evidence, both numerical\-~\cite{CHe21} and analytical\-~\cite{CJL20,CJL21}, that this approximation leads to very accurate predictions.
Understanding why this approximation is accurate is a major open problem, see Chapter\-~\ref{sec:open}
\bigskip

\indent
In order to derive a relation between $g_3,g_4$ and $g_2$, we will enforce a condition that holds true for the real $g_3,g_4$.
By definition\-~(\ref{gn}),
\begin{equation}
  \int\frac{dx_n}V\ g_n(x_1,\cdots,x_n)
  =g_{n-1}(x_1,\cdots,x_{n-1})
  .
\end{equation}
We will enforce that this condition holds in order to compute values for $u_3$ and $u_4$: specifically,
\begin{equation}
  \int\frac{dx}V\ g_2(x)=1
  ,\quad
  \int\frac{dx_3}V\ g_3(x_1,x_2,x_3)
  =g_{2}(x_1-x_2)
  ,\quad
  \int\frac{dx_3}V\frac{dx_4}V\ g_4(x_1,x_2,x_3,x_4)
  =g_{2}(x_1-x_2)
  .
  \label{cd_g3g4}
\end{equation}
Note that the only way for Assumption\-~(\ref{assum:factorization}) to hold under the constraint that
\begin{equation}
  \int\frac{dx_4}V\ g_4(x_1,x_2,x_3,x_4)
  =g_{3}(x_1,x_2,x_3)
\end{equation}
is that $g_4=g_3=g_2$ is equal to a constant, which is one reason why Assumption\-~\ref{assum:factorization} cannot hold exactly for non-constant functions.
As we will now see, (\ref{eigenvalue}), Assumption\-~\ref{assum:factorization} and (\ref{cd_g3g4}) together fix a definite value for $u_3$ and $u_4$.
\bigskip

\theo{Theorem}\label{theo:compleq}
  If $g_3,g_4$ satisfies Assumption\-~\ref{assum:factorization}, the eigenvalue equation\-~(\ref{eigenvalue}) and\-~(\ref{cd_g3g4}), then
  \begin{equation}
    u(x):=1-g_2(x)
  \end{equation}
  satisfies, in the limit $V\to\infty$,
  \begin{equation}
    -\frac1m\Delta u(x)
    =
    (1-u(x))v(x)
    +
    (1-u(x))(-2\rho K(x)+\rho^2L(x))
    \label{compleq}
  \end{equation}
  where
  \begin{equation}
    S(x):=v(x)(1-u(x))
    ,\quad
    K(x)
    :=
    S\ast u(x)
    ,\quad
    f_1\ast f_2(x):=\int dz\ f_1(x-z)f_2(z)
  \end{equation}
  \begin{equation}
    L(x)
    :=
    S\ast u\ast u(x)
    -2u\ast(u(u\ast S))(x)
    +\frac12\int dzdt\ S(z-t) u(x-z)u(x-t)u(z)u(t)
    .
    \label{compleqL}
  \end{equation}
  This equation is called the {\it Complete Equation}.
\endtheo
\bigskip

The proof of this theorem is simple in principle, but somewhat lengthy, and is deferred to Appendix\-~\ref{app:proof_factorization}.

\subsection{The Equations of Lieb's Simplified approach}\label{sec:equations}
\indent
We have thus introduced the first, and most difficult equation of Lieb's Simplified approach: the {\it Complete Equation}, see Theorem\-~\ref{theo:compleq}.
However, this equation is rather difficult to deal with, both analytically and numerically.
From a numerical point of view, it is costly to compute nested integrals: a very accurate method to compute integrals numerically is a ``Gauss quadrature'', in which each integral is replaced by a sum of $N$ terms, and taking an integral in $n$ variables requires $N^n$ points.
In practice, a quintuple integral is near the upper limit of what is computable quickly on ordinary hardware.
Several tricks are at our disposal to reduce the number of variables: in the case of a convolution, the integral is three dimensional, but by changing to {\it two-center bipolar coordinates}, we can reduce it to a double integral.
In addition, (\ref{compleq}) has fewer integrations in Fourier space, in which convolutions become products and vice versa.
These allow us to reduce the computation to a sextuple integral (for details, see the documentation of the {\tt simplesolv} software package\-~\cite{ss}), which is still rather heavy to compute.
However, the sextuple integral appears only in the last term in\-~(\ref{compleqL}), and if this term is dropped, we are left with only double integrals.
We thus define an equation obtained from the Complete Equation in which we drop the last term in\-~(\ref{compleqL}), which we will call the ``{\it Big Equation}''.
\bigskip

\theoname{Definition}{Big Equation}
  \begin{equation}
    -\frac1m\Delta u(x)
    =
    (1-u(x))v(x)+(1-u(x))\left(-2\rho K_{\mathrm{bigeq}}(x)+\rho^2 L_{\mathrm{bigeq}}(x)\right)
    \label{bigeq}
  \end{equation}
  \begin{equation}
    K_{\mathrm{bigeq}}:=
    u\ast S
    ,\quad
    S(y):=(1-u(y))v(y)
    \label{bigeqK}
  \end{equation}
  \nopagebreakaftereq
  \begin{equation}
    L_{\mathrm{bigeq}}:=
    u\ast u\ast S
    -2u\ast(u(u\ast S))
    .
  \label{bigeqL}
  \end{equation}
\endtheo
\restorepagebreakaftereq
\bigskip

In practice, we have found that the predictions of the Big Equation are extremely close to those of the Complete Equation.
\bigskip

\indent
Whereas the Big Equation is (somewhat) easily computable numerically (see the documentation of the {\tt simplesolv} software package\-~\cite{ss}), it remains rather difficult to study analytically.
To simplify the equation, we will make two approximations.
First, we assume that $u(x)\ll 1$, which is expected in the limit $|x|\to\infty$.
This allows us to approximate
\begin{equation}
  (1-u)(-2\rho K+\rho^2L)\approx-2\rho K+\rho^2L
  ,\quad
  L\approx u\ast u\ast S
  .
  \label{simpleq_approx1}
\end{equation}
Second, we will replace $S$ by a Dirac delta function (while preserving its integral):
\begin{equation}
  S(x)\approx\delta(x)\int dx\ S(x)\equiv\frac{2e}\rho\delta(x)
  ,\quad
  e:=\frac\rho2\int dx\ S(x)\equiv\frac\rho2\int dx\ (1-u(x))v(x)
  .
\end{equation}
(The notation $e$ is not innocuous: it is the prediction of the Simple Equation for the ground state energy, see\-~(\ref{E0}).)
The (rough) idea behind this approximation is that, viewed from a large distance, $S$ can be replaced by a $\delta$ function (as long as the behavior is dominated by scattering events, the details of the potential should not matter, just its scattering length).
Using this second approximation, we get
\begin{equation}
  K\approx\frac{2e}\rho u
  ,\quad
  L\approx\frac{2e}\rho u\ast u
  \label{simpleq_approx2}
\end{equation}
which leads us to define the ``{\it Simple Equation}'' as follows.
\bigskip

\theoname{Definition}{Simple Equation of Lieb's Simplified approach}
  \begin{equation}
    -\frac1m\Delta u(x)=(1-u(x))v(x)-4e u(x)+2e\rho u\ast u(x)
    \label{simpleq}
  \end{equation}
  \nopagebreakaftereq
  \begin{equation}
    e:=\frac\rho2\int dx\ (1-u(x))v(x)
    .
    \label{energy}
  \end{equation}
\endtheo
\restorepagebreakaftereq
\bigskip

\indent
Finally, we define an intermediate equation, which is simpler than the Big Equation to compute, but has fewer approximations than the Simple Equation.
As we will see below, this ``{\it Medium Equation}'' agrees quantitatively rather well with the Big Equation and the Bose gas.
To define the Medium Equation, we make the approximation in\-~(\ref{simpleq_approx1}), but not\-~(\ref{simpleq_approx2}).
\bigskip

\theoname{Definition}{Medium Equation of Lieb's Simplified approach}
  \begin{equation}
    -\frac1m\Delta u(x)=(1-u(x))v(x)-2\rho u\ast S(x)+\rho^2u\ast u\ast S(x)
    \label{medeq}
  \end{equation}
  \nopagebreakaftereq
  \begin{equation}
    S(x):=(1-u(x))v(x)
    .
  \end{equation}
\endtheo
\restorepagebreakaftereq

{\bf Remark}:
It is natural to extend the family of equations in Lieb's Simplified approach by interpolating continuously from the Simple equation to the Medium, Big, and Complete.

\section{Existence and uniqueness of solutions of the Simple Equation}\label{sec:existence}
\indent
We have defined four equations in Lieb's Simplified approach.
These have a rather unusual structure.
They are all non-linear, and are also non-local, because of the convolutions.
The first question that arises is whether these equations have solutions.
As of this writing, we only have an existence result for the Simple Equation.
Extending this result to the other equations in Lieb's Simplified approach is an important open problem, see Chapter\-~\ref{sec:open}.
This result holds in all dimensions.
\bigskip

\theoname{Theorem}{{\rm\cite[Theorems 1.1, 1.3]{CJL20}}}\label{theo:existence}
  If $v\in L_1(\mathbb R^d)\cap L_p(\mathbb R^d)$ for $p>\max\{\frac d2,1\}$ and $v\geqslant 0$, then\-~(\ref{simpleq}) has an integrable solution $u$ satisfying $0\leqslant u(x)\leqslant 1$.
\endtheo
\bigskip

(In the rest of this book, we assume that $v$ is not identically zero, otherwise the solution or the Simple equation is $u=0$.)
\bigskip

\indent
The crucial idea of the proof is that the Simple Equation is much easier to solve if we change the point of view: instead of fixing $\rho$ and computing $e$ and $u$, we fix $e$ and compute $\rho$ and $u$.
Doing so turns the term $-4eu$ into a linear term.
Next, we write the Simple Equation as a fixed point equation, by defining
\begin{equation}
  \left(-\frac1m\Delta+v+4e\right)u_n=v+2e\rho_{n-1} u_{n-1}\ast u_{n-1}
  ,\quad
  u_0:=0
  \label{un_def}
\end{equation}
with
\begin{equation}
  \rho_n:=\frac{2e}{\int dx\ (1-u_n(x))v(x)}
  .
\end{equation}
We will then prove the convergence of the iteration using the monotone convergence theorem.
But first, let us prove a simple useful lemma.
\bigskip

\theo{Lemma}\label{lemma:intu}
  If the solution of the simple equation exists in $L_1(\mathbb R^d)$, then
  \nopagebreakaftereq
  \begin{equation}
    \int dx\ u(x)=\frac1\rho
  \end{equation}
\endtheo
\restorepagebreakaftereq
\bigskip

\indent\underline{Proof}:
  We integrate both sides of the Simple equation\-~(\ref{simpleq}):
  \begin{equation}
    0=\frac{2e}\rho-4e\int dx\ u(x)+2e\rho\left(\int dx\ u(x)\right)^2
  \end{equation}
  which we solve:
  \nopagebreakaftereq
  \begin{equation}
    \int dx\ u(x)=\frac1\rho
    .
  \end{equation}
\qed
\restorepagebreakaftereq
\bigskip

\indent
It is convenient to rewrite the iteration\-~(\ref{un_def}): let
\begin{equation}
  K_e:=\left(-\frac1m\Delta+v+4e\right)^{-1}
  \label{Ke}
\end{equation}
which is a well-defined operator when $v\in L_p(\mathbb R^d)$ with $p>\frac d2$, see Appendix\-~\ref{app:Ke}.
As is proved in Lemma\-~\ref{lemma:Ke_pos}, $K_e$ is a bounded operator from $L_p(\mathbb R^d)$ to $W_{2,p}(\mathbb R^d)$ (see Appendix\-~\ref{app:sobolev}).
Actually, we prove in Lemma\-~\ref{lemma:Ke_extend}, that $K_e$ is also well defined on functions in $L_q(\mathbb R^d)$ for any $q\geqslant 1$.
In addition, the operator $K_e$ can be shown to be positivity preserving, see Lemma\-~\ref{lemma:Ke_pos}, that is, if $u\geqslant 0$, then $K_e u\geqslant 0$.
In terms of $K_e$,
\begin{equation}
  u_n=K_e(v+2e\rho_{n-1}u_{n-1}\ast u_{n-1})
  .
\end{equation}
The terms on the right side are well-defined whenever $v$ and $u_{n-1}$ are integrable (it follows from Young's inequality, Theorem\-~\ref{theo:young}, that $u_{n-1}\ast u_{n-1}$ is integrable if $u_{n-1}$ is).
\bigskip

\theo{Lemma}\label{lemma:monotone}
  $\rho_n$ and $u_n$ are pointwise increasing in $n$ and
  \nopagebreakaftereq
  \begin{equation}
    0\leqslant u_n(x)\leqslant 1
    ,\quad
    \int dx\ u_n(x)\leqslant\frac1{2e}\int dx\ (1-u_n(x))v(x)\equiv\frac1{\rho_n}
    .
    \label{inductive_lem}
  \end{equation}
\endtheo
\restorepagebreakaftereq
\bigskip

\indent\underline{Proof}:
  We prove by induction on $n\geqslant 0$ that\-~(\ref{inductive_lem}) holds and that
  \begin{equation}
    u_{n+1}(x)\geqslant u_n(x)
    ,\quad
    \rho_{n+1}\geqslant \rho_n
    .
  \end{equation}
  \bigskip

  \point
  We have
  \begin{equation}
    u_0=0
    ,\quad
    \rho_0=\frac{2e}{\int dx\ v(x)}
  \end{equation}
  so
  \begin{equation}
    0\leqslant u_0\leqslant 1
    ,\quad
    \int dx\ u_0(x)\leqslant\frac1{\rho_n}
    .
  \end{equation}
  We are left with proving that $u_1\geqslant u_0$ and $\rho_1\geqslant \rho_0$.
  We have
  \begin{equation}
    u_1=K_ev
    ,\quad
    \rho_1=\frac{2e}{\int dx\ v(x)-\int dx\ v(x)K_ev(x)}
  \end{equation}
  and since $K_e$ is positivity preserving,
  \begin{equation}
    u_1\geqslant 0\equiv u_0
    ,\quad
    \rho_1\geqslant\frac{2e}{\int dx\ v(x)}\equiv \rho_0
    .
  \end{equation}

  \bigskip

  \point
  Suppose the lemma holds for $n\geqslant 0$.
  Using the fact that $K_e$ is positivity preserving,
  \begin{equation}
    u_{n+1}=K_ev+2e\rho_nK_e u_n\ast u_n\geqslant
    K_ev+2e\rho_{n-1}K_e u_{n-1}\ast u_{n-1}
    \equiv u_n\geqslant 0
    .
  \end{equation}
  In addition,
  \begin{equation}
    \rho_{n+1}=\frac{2e}{\int dx\ v(x)(1-u_{n+1}(x))}
    \geqslant\frac{2e}{\int dx\ v(x)(1-u_{n}(x))}=\rho_n
    .
  \end{equation}
  Finally, by\-~(\ref{un_def}),
  \begin{equation}
    -\frac1m\Delta u_{n+1}+4eu_{n+1}=v(1-u_{n+1})+2e\rho_n u_n\ast u_n
  \end{equation}
  so, integrating both sides,
  \begin{equation}
    \int dx\ u_{n+1}(x)
    =\frac1{4e}\int dx\ v(x)(1-u_{n+1}(x))
    +\frac12\rho_n\left(\int dx\ u_n(x)\right)^2
    \label{intun}
  \end{equation}
  so
  \begin{equation}
    \int dx\ u_{n+1}(x)
    \leqslant\frac1{4e}\int dx\ v(x)(1-u_{n+1}(x))
    +\frac12\rho_n\int dx\ u_n(x)\int dx\ u_{n+1}(x)
  \end{equation}
  and, by the inductive hypothesis,
  \begin{equation}
    \int dx\ u_n(x)
    \leqslant
    \frac1{2e}\int dx\ (1-u_n(x))v(x)
    =\frac1{\rho_n}
  \end{equation}
  so
  \begin{equation}
    \int dx\ u_{n+1}(x)
    \leqslant\frac1{4e}\int dx\ v(x)(1-u_{n+1}(x))
    +\frac12\int dx\ u_{n+1}(x)
  \end{equation}
  which implies
  \begin{equation}
    \int dx\ u_{n+1}(x)
    \leqslant\frac1{2e}\int dx\ v(x)(1-u_{n+1}(x))
    .
  \end{equation}
  \bigskip

  \point
  Finally, we prove that $u_{n+1}\leqslant 1$.
  We have
  \begin{equation}
    \frac1m\Delta u_{n+1}=v(u_{n+1}-1)+4eu_{n+1}-2e\rho_n u_n\ast u_n
    .
  \end{equation}
  By Young's inequality (Theorem\-~\ref{theo:young}),
  \begin{equation}
    |u_n\ast u_n|\leqslant\|u_n\|_1\|u_n\|_\infty
    \leqslant\|u_n\|_1
  \end{equation}
  and, by the inductive hypothesis, $\|u_n\|_1\leqslant\frac1{\rho_n}$ so
  \begin{equation}
    \frac1m\Delta u_{n+1}\geqslant
    v(u_{n+1}-1)+4eu_{n+1}-2e
    .
  \end{equation}
  Let
  \begin{equation}
    A:=\{x:\ u_{n+1}(x)>1\}
    .
  \end{equation}
  Suppose that $A\neq\emptyset$.
  For $x\in A$,
  \begin{equation}
    \frac1m\Delta u_{n+1}
    \geqslant2e>0
  \end{equation}
  Therefore, $u_{n+1}$ is subharmonic on $A$ (see Appendix\-~\ref{app:harmonic}).
  By Theorem\-~\ref{theo:harmonic} $u_{n+1}$ achieves its maximum on the boundary of $A$, but, by definition, $u_{n+1}=1$ on its boundary ($u_{n+1}$ is continuous by\-~(\ref{un_def})).
  Therefore, inside $A$,
  \begin{equation}
    u_{n+1}\leqslant 1
  \end{equation}
  which is a contradiction.
  Therefore $A$ is empty and so
  \nopagebreakaftereq
  \begin{equation}
    u_{n+1}\leqslant 1
    .
  \end{equation}
\qed
\restorepagebreakaftereq
\bigskip

\theo{Lemma}
  Let
  \begin{equation}
    u(x):=\lim_{n\to\infty}u_n(x)
    ,\quad
    \rho(e):=\lim_{n\to\infty}\rho_n
    .
  \end{equation}
  Both limits exist, $u\in W_{2,p}(\mathbb R^d)$ and $u$ solves the Simple equation\-~(\ref{simpleq}).
\endtheo
\bigskip

\indent\underline{Proof}:
  The limit of $u_n$ exist by Lemma\-~\ref{lemma:monotone} and the monotone convergence theorem.
  Furthermore, by Lemma\-~\ref{lemma:monotone},
  \begin{equation}
    \rho_n\leqslant\frac1{\int dx\ u_n(x)}
    \leqslant\frac1{\int dx\ u_1(x)}\equiv
    \frac1{\int dx\ K_ev(x)}<\infty
  \end{equation}
  so, by monotone convergence, $\rho_n$ converges as well.
  In addition, by Lemma\-~\ref{lemma:monotone},
  \begin{equation}
    \int dx\ u(x)\leqslant \frac1{2e}\|v\|_1
  \end{equation}
  so $u\in L_1(\mathbb R^d)$, and so, by dominated convergence, $u_n$ converges in $L_1(\mathbb R^d)$.
  Moreover, since $u\leqslant 1$,
  \begin{equation}
    \|u\|_p^p\leqslant\|u\|_1
  \end{equation}
  and since $u_n-u\leqslant 1$,
  \begin{equation}
    \|u_n-u\|_p^p\leqslant\|u_n-u\|_1
  \end{equation}
  so $u_n$ converges in $L_p$ as well.
  Now, by Young's inequality (Theorem\-~\ref{theo:young}),
  \begin{equation}
    \|u\ast u-u_n\ast u_n\|_p
    =\|(u+u_n)\ast(u-u_n)\|_p
    \leqslant
    (\|u\|_1+\|u_n\|_1)\|u_n-u\|_p
  \end{equation}
  and so
  \begin{equation}
    \|(v+2e\rho_n u_n\ast u_n)-(v+2e\rho u\ast u)\|_p
    \leqslant
    2e\rho\|u_n\ast u_n-u\ast u\|_p
    +2e\|u_n\ast u_n\|_p(\rho-\rho_n)
    .
  \end{equation}
  Thus, $v+2e\rho_n u_n\ast u_n$ converges in $L_p$, so, since $K_e$ is bounded from $L_p$ to $W_{2,p}$ (see above), $K_e(v+2e\rho_n u_n\ast u_n)$ converges in $W_{2,p}$.
  Therefore,
  \begin{equation}
    u=K_e(v+2e\rho u\ast u)
  \end{equation}
  so $u$ satisfies the Simple Equation.
\qed
\bigskip

\indent
We are almost there: we have proved the existence of the solution of the modified problem where $e$ is fixed instead of $\rho$.
To prove Theorem\-~\ref{theo:existence}, we need to prove that $e\mapsto\rho(e)$ can be inverted locally.
\bigskip

\theo{Lemma}\label{lemma:surjective}
  The map $e\mapsto\rho(e)$ is continuous, and $\rho(0)=0$ and $\lim_{e\to\infty}\rho(e)=\infty$.
  Therefore $e\mapsto\rho(r)$ can be inverted locally.
\endtheo
\bigskip

\indent\underline{Proof}:
  The trick of the proof is to consider the sequences
  \begin{equation}
    a_n:=\int dx\ u_n(x)
    ,\quad
    b_n:=\frac1{\rho_n}\equiv\frac1{2e}\int dx\ (1-u_n(x))v(x)
    .
  \end{equation}
  By dominated convergence and Lemma\-~\ref{lemma:intu},
  \begin{equation}
    \lim_{n\to\infty}a_n=\int dx\ u(x)=\frac1\rho
    ,\quad
    \lim_{n\to\infty}b_n=\frac1\rho
  \end{equation}
  and since $u_n$ is increasing,
  \begin{equation}
    a_n\leqslant\frac1\rho\leqslant b_n
    .
  \end{equation}
  Now, by\-~(\ref{intun})
  \begin{equation}
    2a_{n+1}
    =b_{n+1}
    +\rho_na_n^2
  \end{equation}
  and so, since $b_n=1/\rho_n$,
  \begin{equation}
    2(a_{n+1}-a_n)-(b_{n+1}-b_n)=\rho_n(b_n-a_n)^2
  \end{equation}
  which we sum for $n=1\cdots$:
  \begin{equation}
    \frac1\rho+b_1-2a_1=\sum_{n=1}^\infty \rho_n(b_n-a_n)^2
    .
  \end{equation}
  Now, by\-~(\ref{intun}),
  \begin{equation}
    a_1=\frac1{4e}\int dx\ v(x)(1-u_1(x))
    =\frac12b_1
  \end{equation}
  so
  \begin{equation}
    \frac1\rho=\sum_{n=1}^\infty\rho_n(b_n-a_n)^2
  \end{equation}
  and since $\rho_n$ is increasing,
  \begin{equation}
    \sum_{n=1}^\infty(b_n-a_n)^2\leqslant \frac1{\rho_1^2}
    =\left(\frac1{2e}\int dx\ (1-K_ev(x))v(x)\right)^2
    \leqslant\left(\frac1{2e}\int dx\ v(x)\right)^2
  \end{equation}
  which is bounded uniformly in $e$ on any compact interval $e\in[e_1,e_2]$.
  Thus, $b_n-a_n\to0$ uniformly in $e$ on any compact interval and thus $\rho_n$ converges uniformly on $[e_1,e_2]$ so $\rho$ is continuous on $[e_1,e_2]$.
  Since $e_1,e_2$ can be chosen arbitrarily, this holds for all $e$.
  \bigskip

  \indent
  Finally,
  \begin{equation}
    \rho_n(0)=0
  \end{equation}
  so $\rho(0)=0$, and
  \begin{equation}
    \lim_{e\to\infty}\rho_n(e)=\infty
  \end{equation}
  so, since $\rho_n$ converges uniformly on all compacts,
  \nopagebreakaftereq
  \begin{equation}
    \lim_{e\to\infty}\rho(e)=\infty
    .
  \end{equation}
\qed
\restorepagebreakaftereq
\bigskip

\indent
This proves Theorem\-~\ref{theo:existence}, and thus establishes the existence of a solution to the Simple Equation.
But what of the uniqueness?
\bigskip

\theoname{Theorem}{Uniqueness}
  For any fixed $e$, the solution $(u,\rho)$ to the Simple Equation with $u\in L_1(\mathbb R^d)$ and $u(x)\in[0,1]$ is unique.
\endtheo
\bigskip

\indent\underline{Proof}:
  Consider two solutions $(u,\rho)$ and $(u',\rho')$ with $u'(x)\in[0,1]$.
  We first show that $u'\geqslant u_n$ by induction.
  It is trivial at $n=0$, and for $n\geqslant 1$,
  \begin{equation}
    u'-u_n=2eK_e(\rho'u'\ast u'-\mathds 1_{n>1}\rho_{n-1}u_{n-1}\ast u_{n-1})
  \end{equation}
  where $\mathds 1_{n>1}$ is equal to 1 if $n>1$ and $0$ otherwise.
  Since
  \begin{equation}
    \rho'=\frac{2e}{\int dx\ (1-u'(x))v(x)}
  \end{equation}
  we have
  \begin{equation}
    \rho'\geqslant\frac{2e}{\int dx\ (1-u_{n-1}(x))v(x)}\equiv \rho_{n-1}
  \end{equation}
  so
  \begin{equation}
    \rho'u'\ast u'\geqslant\rho_{n-1}u_{n-1}\ast u_{n-1}
  \end{equation}
  and
  \begin{equation}
    u'\geqslant u_n
    .
  \end{equation}
  Thus
  \begin{equation}
    u'\geqslant u
    .
  \end{equation}
  However, by Lemma\-~\ref{lemma:intu},
  \begin{equation}
    \int dx\ u'(x)
    =\int dx\ u(x)=\frac1\rho
  \end{equation}
  so
  \nopagebreakaftereq
  \begin{equation}
    u= u'
    .
  \end{equation}
\qed
\restorepagebreakaftereq
\bigskip

\indent
Thus, taking the point of view in which the energy $e$ is fixed and $\rho$ is computed as a fuction of $e$, the solution is unique.
However, this does not imply that the solution to the problem in which $\rho$ is fixed is unique: the mapping $e\mapsto\rho(e)$ may not be injective (it is surjective by Lemma\-~\ref{lemma:surjective}).
To prove that it is injective, one could prove that $\rho$ is an increasing function of $e$ (physically, it should be: the higher the density is, the higher the energy should be because the potential is repulsive).
This has been proved for small and large values of $e$\-~\cite{CJL21}, but, in general, it is still an open problem, see Chapter\-~\ref{sec:open}.

\section{Predictions of the Simple Equation}\label{sec:predictions}
\indent
Let us now discuss predictions of the Simple equation for the ground state energy and the condensate fraction.
As we will see, the prediction for the ground state energy is asymptotically correct at both low and high density.
These prediction for the condensate fraction coincides at low densities with that from Bogolyubov theory.
In this chapter, we will restrict our attention to the three-dimensional case.
Adapting these results to other dimensions is an open problem, see Chapter\-~\ref{sec:open}.
\bigskip

\subsection{Energy of the Simple Equation}
\indent
Lieb's Simplified approach provides a natural prediction for the ground state energy per-particle (see\-~(\ref{E0})):
\begin{equation}
  e=\frac\rho2\int dx\ (1-u(x))v(x)
  .
\end{equation}
We have already seen numerical evidence for the fact that the Big and Medium equations provide very accurate predictions for the energy at all densities, see Figure\-~\ref{fig:energy}.
The Simple equation may not be so accurate at intermediate, but we can prove that it predicts the correct energy at low and high densities.
\bigskip

\indent
As we discussed in Section\-~\ref{sec:lhy}, the ground state energy of the Bose gas has been computed at low and high densities, see Theorem\-~\ref{theo:lhy_e} and\-~(\ref{e_highrho}).
We have proved that the prediction of the Simple equations agrees with both of these asymptotic expansions.
\bigskip

\theoname{Theorem}{{\rm\cite[Theorem 1.4]{CJL20}}}\label{theo:energy}
  For the Simple Equation, for $d=3$, under the assumptions of Theorem\-~\ref{theo:existence}, as $\rho\to0$,
  \begin{equation}
    e=\frac{2\pi}m\rho a\left(1+\frac{128}{15\sqrt\pi}\sqrt{\rho a^3}+o(\sqrt\rho)\right)
    \label{low_density}
  \end{equation}
  where $a$ is the scattering length of the potential, and as $\rho\to\infty$
  \nopagebreakaftereq
  \begin{equation}
    e\sim_{\rho\to\infty}\frac\rho2\int dx\ v(x)
    .
    \label{high_density}
  \end{equation}
\endtheo
\restorepagebreakaftereq
\bigskip

It is rather striking that the Simple Equation agrees with the Bose gas {\it both} at high and low densities.
Note, however, that whereas\-~(\ref{e_highrho}) holds only for potentials of positive type (that is the Fourier transform of the potential $\hat v$ is non-negative), (\ref{high_density}) holds regardless of the sign of $\hat v$.
The asymptotic agreement at high densities therefore only holds for positive type potentials.
\bigskip

\indent\underline{Proof}:
\medskip

  \point
  Let us first prove\-~(\ref{high_density}), which is easier.
  By\-~(\ref{energy}), it suffices to prove that
  \begin{equation}
    \lim_{\rho\to\infty}\int dx\ u(x)v(x)
    =0
    .
    \label{limintuv}
  \end{equation}
  Let
  \begin{equation}
    \chi_a:=\{x:\ v(x)\geqslant a\}
  \end{equation}
  and decompose
  \begin{equation}
    \int dx\ u(x)v(x)
    =
    \int_{\chi_a}dx\ u(x)v(x)
    +
    \int_{\mathbb R^d\setminus\chi_a}dx\ u(x)v(x)
  \end{equation}
  which, by Lemma\-~\ref{lemma:intu}, is bounded by
  \begin{equation}
    \int dx\ u(x)v(x)
    \leqslant
    \int_{\chi_a}dx\ v(x)
    +
    \frac a\rho
    .
  \end{equation}
  Since $v$ is integrable, $\int_{\chi_a}dx\ v(x)\to0$ as $a\to\infty$.
  Therefore, taking $a=\sqrt{\rho}$,
  \begin{equation}
    \int_{\chi_a}v(x)\ dx
    +
    \frac a\rho
    \mathop{\longrightarrow}_{\rho\to\infty}0
    .
  \end{equation}
  This proves\-~(\ref{high_density}).
  \bigskip

  \point
  Let us now turn to\-~(\ref{low_density}).
  The scheme of the proof is as follows.
  We first approximate the solution $u$ by $w$, which is defined as the decaying solution of
  \begin{equation}
    -\frac1m\Delta w(x)=(1-u(x))v(x)
    .
    \label{u1}
  \end{equation}
  The energy of $w$ is defined to be
  \begin{equation}
    e_w:=\frac\rho2\int(1-w(x))v(x)\ dx
    \label{ew}
  \end{equation}
  and, as we will show, it is {\it close} to $e$, more precisely,
  \begin{equation}
    e-e_w=\frac{16\sqrt 2(me)^{\frac32}}{15\pi^2}\int dx\ v(x)+o(\rho^{\frac32})
    .
    \label{eew}
  \end{equation}
  In addition, (\ref{u1}) is quite similar to the scattering equation\-~(\ref{scateq}).
  In fact we will show that $e_w$ is {\it close} to the energy $2\pi\rho a/m$ of the scattering equation
  \begin{equation}
    e_w-\frac{2\pi}m\rho a
    =
    -\frac{16\sqrt 2(me)^{\frac32}}{15\pi^2}\int dx\ \varphi(x) v(x)+o(\rho^{\frac32})
    \label{ewephi}
  \end{equation}
  where $1-\varphi$ satisfies the scattering equation\-~(\ref{scateq}):
  \begin{equation}
    (-\frac1m\Delta+v)\varphi=v(x)
    .
    \label{scateq_varphi}
  \end{equation}
  Summing\-~(\ref{eew}) and\-~(\ref{ewephi}), we find
  \begin{equation}
    e=\frac{2\pi}m\rho a\left(1+\frac{32\sqrt 2(me)^{\frac32}}{15\pi^2\rho}+o(\sqrt\rho)\right)
    ,
  \end{equation}
  from which\-~(\ref{low_density}) follows.
  We are thus left with proving\-~(\ref{eew}) and\-~(\ref{ewephi}).
  \bigskip

  \subpoint{\bf Proof of\-~(\ref{eew})}.
  By\-~(\ref{energy}) and\-~(\ref{ew}),
  \begin{equation}
    e-e_w=\frac\rho2\int dx\ (w(x)-u(x))v(x)
    .
  \end{equation}
  We will work in Fourier space
  \begin{equation}
    \hat u(k):=\int dx\ e^{ikx}u(x)
    \label{fourieru}
  \end{equation}
  which satisfies, by\-~(\ref{simpleq}),
  \begin{equation}
  \left(\frac1mk^2+4e\right)\hat u(k)
  =\hat S(k)
  +2e\rho \hat u^2(k)
  \end{equation}
  with
  \begin{equation}
    \hat S(k):=\int dx\ e^{ikx}(1-u(x))v(x)
    .
    \label{S}
  \end{equation}
  Therefore,
  \begin{equation}
    \hat u(k)=\frac1\rho\left(\frac{k^2}{4me}+1-\sqrt{\left(\frac{k^2}{4me}+1\right)^2-\frac\rho{2e}\hat S(k)}\right)
    .
    \label{hatu}
  \end{equation}
  Similarly, the Fourier transform of $w$ is
  \begin{equation}
    \hat w(k):=\int dx\ e^{ikx}w(x)=\frac{m\hat S(k)}{k^2}
    .
    \label{u1hat}
  \end{equation}
  Note that, as $|k|\to\infty$, $\hat u\sim \frac{m\hat S(k)}{k^2}$, so, while $\hat u$ is not integrable, $\hat u-\hat w$ is.
  We invert the Fourier transform:
  \begin{equation}
    u(x)-w(x)=\frac1{8\pi^3\rho}\int dk\ e^{-ikx}
    \left(\frac{k^2}{4me}+1-\sqrt{\left(\frac{k^2}{4me}+1\right)^2-\frac\rho{2e}\hat S(k)}-\frac{m\rho\hat S(k)}{k^2}\right)
    .
  \end{equation}
  We change variables to $\tilde k:=\frac k{2\sqrt{me}}$:
  \begin{equation}
    u(x)-w(x)
    =
    \frac{(me)^{\frac32}}{\rho\pi^3}\int d\tilde k\ e^{-i2\sqrt{me}\tilde kx}
    \left(\tilde k^2+1-\sqrt{(\tilde k^2+1)^2-\frac\rho{2e}\hat S(2\tilde k\sqrt{me})}-\frac{\frac\rho{2e}\hat S(2\tilde k\sqrt{me})}{2\tilde k^2}\right)
    .
  \end{equation}
  Furthermore,
  \begin{equation}
    s\mapsto\left|\tilde k^2+1-\sqrt{(\tilde k^2+1)^2-s}-\frac{s}{2\tilde k^2}\right|
  \end{equation}
  is monotone increasing.
  In addition, by~\-(\ref{S}) and~\-(\ref{simpleq}), and using the fact that $u(x)\leqslant 1$ and $v(x)\geqslant 0$,
  \begin{equation}
    |\hat S(k)|\leqslant\int dx\ (1-u(x))v(x)=\frac{2e}\rho
    .
  \end{equation}
  Therefore
  \begin{equation}
    \left|
      \tilde k^2+1-\sqrt{(\tilde k^2+1)^2-\frac\rho{2e}\hat S(2\tilde k\sqrt{me})}-\frac{\frac\rho{2e}\hat S(2\tilde k\sqrt{me})}{2\tilde k^2}
    \right|
    \leqslant
    \left|
      \tilde k^2+1-\sqrt{(\tilde k^2+1)^2-1}-\frac{1}{2\tilde k^2}
    \right|
    .
  \end{equation}
  Therefore,
  \begin{equation}
    |u(x)-w(x)|\leqslant
    \frac{(me)^{\frac32}}{\rho\pi^3}\int d\tilde k\ \left|\tilde k^2+1-\sqrt{(\tilde k^2+1)^2-1}-\frac{1}{2\tilde k^2}\right|
    =\frac{32\sqrt 2(me)^{\frac32}}{15\pi^2\rho}
    .
    \label{bound_absuw}
  \end{equation}
  By dominated convergence, and using the fact that $\hat S(0)=\frac{2e}\rho$,
  \begin{equation}
    \begin{largearray}
      \lim_{e\to 0}\frac1{(me)^{\frac32}}(e-e_w)
      =-\lim_{e\to 0}\frac\rho{2(me)^{\frac32}}\int dx\ (u(x)-w(x))v(x)
      =\\[0.5cm]\hfill
      =
      -\frac12\int dx\ v(x)\left(
	\frac1{\pi^3}\int\left(\tilde k^2+1-\sqrt{(\tilde k^2+1)^2-1}-\frac1{2\tilde k^2}\right)d \tilde k
      \right)
      =\frac{16\sqrt 2}{15\pi^2}\int dx\ v(x)
      .
    \end{largearray}
  \end{equation}
  Finally, since $u(x)\geqslant 0$,
  \begin{equation}
    e\leqslant \frac\rho 2\int dx\ v(x)
  \end{equation}
  so $e=O(\rho)$, and $e^{\frac32}=O(\rho^{\frac32})$.
  This proves\-~(\ref{eew}).
  Incidentally, again by dominated convergence,
  \begin{equation}
    \begin{largearray}
      u(x)-w(x)=
      \frac{(me)^{\frac32}}{\rho\pi^3}\int\left(\tilde k^2+1-\sqrt{(\tilde k^2+1)^2-1}-\frac{1}{2\tilde k^2}\right)\ d\tilde k
      +\sqrt\rho f_\rho(x)
      =\\[0.5cm]\hfill
      =-\frac{32\sqrt 2(me)^{\frac32}}{15\pi^2\rho}+\sqrt\rho f_\rho(x)
      \label{uapproxw}
    \end{largearray}
  \end{equation}
  with
  \begin{equation}
    f_\rho(x)\mathop{\longrightarrow}_{\rho\to0}0
  \end{equation}
  and by\-~(\ref{bound_absuw}),
  \begin{equation}
    0\leqslant f_\rho(x)\leqslant\frac{64\sqrt2(me)^{\frac32}}{15\pi^2\rho^{\frac32}}=O(1)
  \end{equation}
  pointwise in $x$.

  \bigskip

  \subpoint{\bf Proof of\-~(\ref{ewephi})}.
  Let
  \begin{equation}
    \xi(r):=w(r)-\varphi(r).
    \label{xi}
  \end{equation}
  By\-~(\ref{u1}), (\ref{scateq_varphi}) and\-~(\ref{simpleq}),
  \begin{equation}
    (-\Delta+v(x))\xi(x)=-(u(x)-w(x))v(x)
    .
  \end{equation}
  Therefore, by Lemma\-~\ref{lemma:scattering},
  \begin{equation}
    e_w-\frac{2\pi}m\rho a=-\frac\rho2\int dx\ \xi(x)v(x)
  \end{equation}
  and so, by\-~(\ref{scateq_varphi}),
  \begin{equation}
    e_w-\frac{2\pi}m\rho a
    =-\frac\rho2\int dx\ \xi(x)(-\Delta+v)\varphi(x)
  \end{equation}
  and so, since $-\Delta +v$ is symmetric,
  \begin{equation}
    e_w-\frac{2\pi}m\rho a
    =-\frac\rho2\int dx\ \varphi(x)(u(x)-w(x))v(x)
    .
  \end{equation}
  By\-~(\ref{uapproxw}),
  \begin{equation}
    e_w-\frac{2\pi}m\rho a
    =\frac{16\sqrt 2(me)^{\frac32}}{15\pi^2}\int dx\ \varphi(x)v(x)
    -\frac{\rho^{\frac32}}2\int dx\ \varphi(x)f_\rho(x)v(x)
    .
  \end{equation}
  Since $x\mapsto f_\rho(x)$ is bounded, we can use dominated convergence to show\-~(\ref{ewephi}).
\qed

\subsection{Condensate fraction}\label{sec:condensate_fraction}
\indent
The computation of the condensate fraction is not a straightforward as the energy.
In fact, Lieb's Simplified approach really only gives a clear prediction for the energy, through\-~(\ref{E0}) which translates into\-~(\ref{energy}).
How do we compute the condensate fraction without knowing the expression of the wavefunction?
The idea is to rewrite the condensate fraction in terms of the ground-state energy of an effective Hamiltonian:
\begin{equation}
  \tilde H_N(\epsilon):=
  H_N+\epsilon\sum_{i=1}^NP_i
\end{equation}
where $P_i$ is the projector onto the constant state\-~(\ref{Pi}) (recall\-~(\ref{eta_Pi})).
Let $\tilde E_0(\epsilon)$ denote the ground state energy of $\tilde H_N(\epsilon)$, and $\tilde\psi_0(\epsilon)$ denote the ground state of $\tilde H_N(\epsilon)$ with $\|\tilde\psi_0(\epsilon)\|_2=1$.
We have
\begin{equation}
  \tilde E_0(\epsilon)
  =\left<\tilde\psi_0(\epsilon)\right|\tilde H_N(\epsilon)\left|\tilde\psi_0(\epsilon)\right>
\end{equation}
so
\begin{equation}
  \partial_\epsilon\tilde E_0|_{\epsilon=0}
  =
  2\left<\partial_\epsilon\tilde\psi_0|_{\epsilon=0}\right|H_N\left|\psi_0\right>
  +
  \left<\psi_0\right|\partial_\epsilon\tilde H_N|_{\epsilon=0}\left|\psi_0\right>
\end{equation}
that is
\begin{equation}
  \partial_\epsilon\tilde E_0|_{\epsilon=0}
  =
  2E_0\left<\partial_\epsilon\tilde\psi_0|_{\epsilon=0}|\psi_0\right>
  +
  \left<\psi_0\right|\sum_{i=1}^NP_i\left|\psi_0\right>
  .
\end{equation}
In addition,
\begin{equation}
  2\left<\partial_\epsilon\tilde\psi_0|_{\epsilon=0}|\psi_0\right>
  =
  \partial_\epsilon\left<\tilde\psi_0(\epsilon)|\tilde\psi_0(\epsilon)\right>|_{\epsilon=0}
  =
  \partial_\epsilon1|_{\epsilon=0}
  =0
  .
\end{equation}
Thus,
\begin{equation}
  \frac1N
  \partial_\epsilon\tilde E_0|_{\epsilon=0}
  =
  \frac1N\sum_{i=1}^N\left<\psi_0\right|P_i\left|\psi_0\right>
  .
\end{equation}
Therefore,
\begin{equation}
  \eta_0=
  \lim_{\displaystyle\mathop{\scriptstyle N,V\to\infty}_{\frac NV=\rho}}
  \frac1N
  \partial_\epsilon\tilde E_0|_{\epsilon=0}
  .
  \label{condensate_feynman_hellman}
\end{equation}
\bigskip

\indent
This construction gives us a natural prediction of the condensate fraction for the Simple equation.
To define it, we repeat the arguments of Sections\-~\ref{sec:simplified_construction}-\ref{sec:equations} and Appendix\-~\ref{app:proof_factorization} to the modified Hamiltonian.
The corresponding modified Simple equation is\-~\cite{CJL21}:
\begin{equation}
  (-\Delta+2\epsilon+4e)u_\epsilon=(1-u_\epsilon)v+2\rho e_\epsilon u_\epsilon\ast u_\epsilon
  ,\quad
  e_\epsilon=\frac\rho2\int dx\ (1-u_\epsilon(x))
  \label{simpleq_eta}
\end{equation}
from which we define the prediction of the condensate fraction by analogy with\-~(\ref{condensate_feynman_hellman}):
\begin{equation}
  \eta:=\partial_\epsilon e_\epsilon|_{\epsilon=0}
  .
\end{equation}
\bigskip

\indent
As was shown in Figure\-~\ref{fig:condensate}, the prediction of the Big and Medium equations are rather accurate for all densities, as the energy was.
For the Simple Equation, we can prove an asymptotic expansion for low densities, which agrees with the prediction of Bogolyubov theory in Conjecture\-~\ref{conjecture:lhy_h}.
\bigskip

\theoname{Theorem}{{\rm\cite[Theorem 1.6]{CJL21}}}
  For the Simple Equation, in $d=3$, if $(1+|x|^4)v(x)\in L_1(\mathbb R^3)\cap L_2(\mathbb R^3)$ and $v\geqslant 0$, as $\rho\to0$,
  \begin{equation}
    1-\eta\sim\frac{8\sqrt{\rho a^3}}{3\sqrt\pi}
  \end{equation}
  where $a$ is the scattering length of the potential.
\endtheo
\bigskip

\indent
This statement is proven in\-~\cite{CJL21}.
Here, let us present a different proof: the proof in\-~\cite{CJL21} is a bit more general, but uses a more sophisticated discussion.
Instead, here we will present a more elementary proof that has slightly different requirements on $v$.
\bigskip

\theo{Theorem}
  For the Simple Equation, in $d=3$, suppose $v\in L_1(\mathbb R^3)\cap L_2(\mathbb R^3)$, $v\geqslant 0$, and $\hat v$ is H\"older continuous: there exist $C,\alpha>0$ such that, for $k,q\in\mathbb R^3$, $|\hat v(k+q)-\hat v(k)|\leqslant C|q|^\alpha$.
  As $\rho\to0$,
  \nopagebreakaftereq
  \begin{equation}
    1-\eta\sim\frac{8\sqrt{\rho a^3}}{3\sqrt\pi}
    .
  \end{equation}
\endtheo
\restorepagebreakaftereq
\bigskip

\indent\underline{Proof}:
  We write\-~(\ref{simpleq_eta}) in Fourier space:
  \begin{equation}
    \rho\hat u_\epsilon(k)=\left(\frac{k^2+2\epsilon}{4e_\epsilon}+1\right)
    \left(1-\sqrt{1-\frac{\frac\rho{2e_\epsilon}\hat S_\epsilon(k)}{(\frac{k^2+2\epsilon}{4e_\epsilon}+1)^2}}\right)
    ,\quad
    \hat S_\epsilon(k):=\int dx\ e^{ikx}(1-u_\epsilon(x))v(x)
    .
    \label{fmu}
  \end{equation}
  Thus
  \begin{equation}
    \rho\partial_\epsilon\hat u_\epsilon(k)|_{\epsilon=0}
    =
    \frac1{2e_0}\left(
      1-\frac{k^2}{2e_0}\eta
      -
      \frac{(\frac{k^2}{4e_0}+1)(1-\frac{k^2}{2e_0}\eta)+\frac\rho{2e_0}\eta\hat S_0(k)-\frac\rho2\partial_\epsilon\hat S_\epsilon(k)|_{\epsilon=0}}{\sqrt{(\frac{k^2}{4e_0}+1)^2-\frac \rho{2e_0}\hat S_0(k)}}
    \right)
    .
    \label{du}
  \end{equation}
  Furthermore,
  \begin{equation}
    \eta=-\frac12\int\frac{dk}{8\pi^3}\ \hat v(k)\rho\partial_\epsilon\hat u_\epsilon(k)|_{\epsilon=0}
  \end{equation}
  which we split into three terms:
  \begin{equation}
    \eta=H_1+H_2+H_3
    \label{etasplit}
  \end{equation}
  with
  \begin{equation}
    H_1:=\frac1{32e_0\pi^3}\int dk\ \hat v(k)
    \left(
      \frac{\frac{k^2}{4e_0}+1}{\sqrt{(\frac{k^2}{4e_0}+1)^2-\frac \rho{2e_0}\hat S_0(k)}}-1
    \right)
  \end{equation}
  \begin{equation}
    H_2:=-\eta\frac1{64e_0^2\pi^3}\int dk\ \hat v(k)
    k^2\left(
      \frac{\frac{k^2}{4e_0}+1-\frac{\rho}{k^2}\hat S_0(k)}{\sqrt{(\frac{k^2}{4e_0}+1)^2-\frac \rho{2e_0}\hat S_0(k)}}
      -1
    \right)
  \end{equation}
  \begin{equation}
    H_3:=-\frac1{32e_0\pi^3}\int dk\ \hat v(k)
    \frac{\frac\rho2\partial_\epsilon\hat S_\epsilon(k)|_{\epsilon=0}}{\sqrt{(\frac{k^2}{4e_0}+1)^2-\frac \rho{2e_0}\hat S_0(k)}}
    .
  \end{equation}
  \bigskip

  \point We first compute $H_1$.
  We rescale the integral by $2\sqrt{e_0}$:
  \begin{equation}
    H_1=\frac{\sqrt{e_0}}{4\pi^3}\int dk\ \hat v(2\sqrt{e_0}k)
    \left(
      \frac{k^2+1}{\sqrt{(k^2+1)^2-\frac \rho{2e_0}\hat S_0(2\sqrt{e_0}k)}}-1
    \right)
    .
  \end{equation}
  Since $S_0(x)\geqslant 0$, $|\hat S_0(k)|\leqslant|\hat S_0(0)|=\frac{2e_0}\rho$, so
  \begin{equation}
    \left|\frac{k^2+1}{\sqrt{(k^2+1)^2-\frac \rho{2e_0}\hat S_0(2\sqrt{e_0}k)}}-1\right|
    \leqslant
    \frac{k^2+1}{\sqrt{(k^2+1)^2-1}}-1
    \label{ineqH1}
  \end{equation}
  which is integrable, and since $v(x)\geqslant 0$, $|\hat v(k)|\leqslant|\hat v(0)|$, so, by dominated convergence,
  \begin{equation}
    H_1\sim\frac{\sqrt{e_0}}{4\pi^3}\hat v(0)\int dk
    \left(\frac{k^2+1}{\sqrt{(k^2+1)^2-1}}-1\right)
    =
    \frac{\sqrt{2e_0}}{3\pi^2}\hat v(0)
    .
    \label{H1}
  \end{equation}
  ($\hat v$ is continuous since $v\in L_1(\mathbb R^3)$, see e.g. \cite[Section\-~5.1]{LL01}.)
  \bigskip

  \point We now turn to $H_2$.
  We rescale the integral by $2\sqrt{e_0}$:
  \begin{equation}
    H_2=
    -\eta\frac{\sqrt{e_0}}{2\pi^3}\int dk\ \hat v(2\sqrt{e_0}k)
    k^2\left(
    \frac{k^2+1-\frac\rho{4e_0k^2}\hat S_0(2\sqrt e_0k)}{\sqrt{(k^2+1)^2-\frac \rho{2e_0}\hat S_0(2\sqrt{e_0}k)}}-1
    \right)
    .
  \end{equation}
  Furthermore, consider the function
  \begin{equation}
    \mathfrak G_k(\sigma):=
    \frac{k^2+1-\frac\sigma{2k^2}}{\sqrt{(k^2+1)^2-\sigma}}-1
  \end{equation}
  its critical points $\partial_\sigma\mathfrak G_k(\sigma)=0$ satisfy
  \begin{equation}
    \partial_\sigma\mathfrak G_k(\sigma)=\frac{k^2+1-\frac\sigma{2k^2}}{2((k^2+1)^2-\sigma)^{\frac32}}-\frac1{2k^2\sqrt{(k^2+1)^2-\sigma}}=0
  \end{equation}
  that is
  \begin{equation}
    \sigma=2(1+k^2)>1
    .
  \end{equation}
  In addition,
  \begin{equation}
    \partial_\sigma\mathfrak G_k(0)
    =
    -\frac1{2k^2(k^2+1)^2}<0
  \end{equation}
  so $\mathfrak G_k(\sigma)$ is decreasing and negative for $|\sigma|\leqslant 1$.
  Therefore,
  \begin{equation}
    |\mathfrak G_k(\sigma)|\leqslant|\mathfrak G_k(1)|
  \end{equation}
  and, since $\hat S_0\leqslant\frac{2e_0}\rho$,
  \begin{equation}
    k^2\left|
    \frac{k^2+1-\frac\rho{4e_0k^2}\hat S_0(2\sqrt e_0k)}{\sqrt{(k^2+1)^2-\frac \rho{2e_0}\hat S_0(2\sqrt{e_0}k)}}-1
    \right|
    \leqslant
    k^2\left|
    \frac{k^2+1-\frac1{2k^2}}{\sqrt{(k^2+1)^2-1}}-1
    \right|
    \label{ineqH2}
  \end{equation}
  which is integrable.
  Therefore, by dominated convergence,
  \begin{equation}
    H_2\sim
    -\eta\frac{\sqrt{e_0}}{2\pi^3}\hat v(0)\int dk\ 
    k^2\left(
    \frac{k^2+1-\frac1{2k^2}}{\sqrt{(k^2+1)^2-1}}-1
    \right)
    =O(\eta\sqrt{e_0})
    .
    \label{H2}
  \end{equation}
  \bigskip

  \point We now turn to $H_3$.
  We expand
  \begin{equation}
    \frac1{\sqrt{(\frac{k^2}{4e_0}+1)^2-\frac \rho{2e_0}\hat S_0(k)}}
    =
    \frac1{\frac{k^2}{4e_0}+1}(1+O(e_0^2k^{-4}))
    =
    \frac{4e_0}{k^2}(1+O(e_0k^{-2}))
    .
  \end{equation}
  We extract the non-integrable term by defining
  \begin{equation}
    H_3=h_3-
    \frac1{8\pi^3}\int dk\ \hat v(k)
    \frac{\frac\rho{2}\partial_\epsilon\hat S_\epsilon(k)|_{\epsilon=0}}{k^2}
    \label{H3split}
  \end{equation}
  with
  \begin{equation}
    h_3:=\frac1{32e_0\pi^3}\int dk\ \hat v(k)
    \frac\rho2\partial_\epsilon\hat S_\epsilon(k)|_{\epsilon=0}\mathfrak R_3(k)
    ,\quad
    \mathfrak R_3(k):=\frac{4e_0}{k^2}-\frac1{\sqrt{(\frac{k^2}{4e_0}+1)^2-\frac \rho{2e_0}\hat S_0(k)}}
    .
  \end{equation}
  \bigskip

  \subpoint To compute $h_3$, we rescale the integral by $2\sqrt{e_0}$:
  \begin{equation}
    h_3=\frac{\sqrt{e_0}}{4\pi^3}\int dk\ \hat v(2\sqrt{e_0}k)
    \frac\rho2\partial_\epsilon\hat S_\epsilon(2\sqrt{e_0}k)|_{\epsilon=0}\mathfrak R_3(2\sqrt{e_0}k)
  \end{equation}
  and
  \begin{equation}
    0\leqslant \mathfrak R_3(2\sqrt{e_0}k)=
    \frac1{k^2}-\frac1{\sqrt{(k^2+1)^2-\frac \rho{2e_0}\hat S_0(2\sqrt{e_0}k)}}
    \leqslant
    \frac1{k^2}-
    \frac1{\sqrt{(k^2+1)^2+1}}
    \label{ineqH3}
  \end{equation}
  which is integrable, so, using the fact that
  \begin{equation}
    \eta=\frac\rho2\partial_\epsilon\hat S_\epsilon(0)|_{\epsilon=0}
  \end{equation}
  by dominated convergence,
  \begin{equation}
    h_3\sim
    \frac{\eta\sqrt{e_0}}{4\pi^3}\hat v(0)\int dk\ \left(\frac1{k^2}-\frac1{\sqrt{(k^2+1)^2-1}}\right)
    =O(\eta\sqrt{e_0})
    .
    \label{h3}
  \end{equation}
  \bigskip

  \subpoint Furthermore,
  \begin{equation}
    \frac\rho2\partial_\epsilon\hat S_\epsilon(q)|_{\epsilon=0}=-\frac\rho2\int\frac{dk}{8\pi^3}\ \hat v(k-q)\partial_\epsilon\hat u_\epsilon(k)|_{\epsilon=0}
  \end{equation}
  so, by\-~(\ref{du}),
  \begin{equation}
    \frac\rho2\partial_\epsilon\hat S_\epsilon(q)|_{\epsilon=0}=F_1(q)+F_2(q)+F_3(q)
    \label{dSsplit}
  \end{equation}
  with
  \begin{equation}
    F_1(q):=\frac1{32e_0\pi^3}\int dk\ \hat v(k-q)
    \left(
      \frac{\frac{k^2}{4e_0}+1}{\sqrt{(\frac{k^2}{4e_0}+1)^2-\frac \rho{2e_0}\hat S_0(k)}}-1
    \right)
    \label{F1}
  \end{equation}
  \begin{equation}
    F_2(q):=-\eta\frac1{64e_0^2\pi^3}\int dk\ \hat v(k-q)
    k^2\left(
      \frac{\frac{k^2}{4e_0}+1-\frac{\rho}{k^2}\hat S_0(k)}{\sqrt{(\frac{k^2}{4e_0}+1)^2-\frac \rho{2e_0}\hat S_0(k)}}
      -1
    \right)
    \label{F2}
  \end{equation}
  \begin{equation}
    F_3(q):=-\frac1{32e_0\pi^3}\int dk\ \hat v(k-q)
    \frac{\frac\rho2\partial_\epsilon\hat S_\epsilon(k)|_{\epsilon=0}}{\sqrt{(\frac{k^2}{4e_0}+1)^2-\frac \rho{2e_0}\hat S_0(k)}}
    .
    \label{F3}
  \end{equation}
  Now, let
  \begin{equation}
    r_1(q):=F_1(q)-\frac{\sqrt{2e_0}}{3\pi^2}\hat v(q)
    ,\quad
    r_2(q):=F_2(q)
  \end{equation}
  \begin{equation}
    r_3(q):=F_3(q)
    +\frac1{8\pi^3}\int dk\ \hat v(k-q)\frac{\frac\rho2\partial_\epsilon\hat S_\epsilon(k)|_{\epsilon=0}}{k^2}
    .
  \end{equation}
  Furthermore, by\-~(\ref{ineqH2}), and using the fact that $|\hat v(k)|\leqslant\hat v(0)$,
  \begin{equation}
    |r_2(q)|
    \leqslant
    |\eta|\frac{\sqrt{e_0}}{2\pi^3}\hat v(0)\int dk\ 
    k^2\left|
    \frac{k^2+1-\frac1{2k^2}}{\sqrt{(k^2+1)^2-1}}-1
    \right|
    =O(\eta\sqrt{e_0})
    \label{r2}
  \end{equation}
  uniformly in $q$.
  Similarly, by\-~(\ref{ineqH3}),
  \begin{equation}
    |r_3(q)|\leqslant
    \frac{\sqrt{e_0}}{4\pi^3}\hat v(0)\int dk\ \frac\rho2\partial_\epsilon\hat S_\epsilon(2\sqrt{e_0}k)|_{\epsilon=0}\left|\frac1{\sqrt{(k^2+1)^2-1}}-\frac1{k^2}\right|
    =O(\eta\sqrt{e_0})
    \label{r3}
  \end{equation}
  uniformly in $q$.
  Bounding $r_1$ is a bit more of a challenge.
  By\-~(\ref{H1}),
  \begin{equation}
    \begin{largearray}
      r_1(q)=
      \frac{\sqrt{e_0}}{4\pi^3}\int dk\ (\hat v(q-2\sqrt{e_0}k)-\hat v(q))
      \left(
	\frac{k^2+1}{\sqrt{(k^2+1)^2-\frac \rho{2e_0}\hat S_0(2\sqrt{e_0}k)}}-1
      \right)
      \\[0.5cm]\hfill
      +
      \hat v(q)\frac{\sqrt{e_0}}{4\pi^3}\int dk\ 
      \left(
	\frac{k^2+1}{\sqrt{(k^2+1)^2-\frac \rho{2e_0}\hat S_0(2\sqrt{e_0}k)}}
	-\frac{k^2+1}{\sqrt{(k^2+1)^2-1}}
      \right)
      .
    \end{largearray}
  \end{equation}
  Furthermore, by\-~(\ref{ineqH1}), and using the H\"older continuity of $\hat v$,
  \begin{equation}
    \begin{largearray}
      \frac{\sqrt{e_0}}{4\pi^3}\int dk\ |\hat v(q-2\sqrt{e_0}k)-\hat v(q)|
      \left|
	\frac{k^2+1}{\sqrt{(k^2+1)^2-\frac \rho{2e_0}\hat S_0(2\sqrt{e_0}k)}}-1
      \right|
      \\[0.5cm]\hfill
      \leqslant
      C(2\sqrt{e_0})^{\alpha}\frac{\sqrt{e_0}}{4\pi^3}\int dk\ 
      k^\alpha\left|
	\frac{k^2+1}{\sqrt{(k^2+1)^2-1}}-1
      \right|
      =o(\sqrt{e_0})
      .
    \end{largearray}
  \end{equation}
  uniformly in $q$.
  Furthermore,
  \begin{equation}
    \left|
      \frac{k^2+1}{\sqrt{(k^2+1)^2-\frac \rho{2e_0}\hat S_0(2\sqrt{e_0}k)}}
      -\frac{k^2+1}{\sqrt{(k^2+1)^2-1}}
    \right|
    \leqslant
    \frac{k^2+1}{((k^2+1)^2-1)^{\frac32}}\frac{1-\frac\rho{2e_0}\hat S_0(2\sqrt{e_0}k)}2
    .
  \end{equation}
  Now,
  \begin{equation}
    \hat S_0(k)=\hat v(k)-\hat u_0\ast\hat v(k)
  \end{equation}
  so
  \begin{equation}
    |\hat S_0(k+\epsilon)-\hat S_0(k)|
    \leqslant
    |\hat v(k+\epsilon)-\hat v(k)|
    +\int\frac{dq}{8\pi^3}|\hat v(k+\epsilon-q)-\hat v(k-q)||\hat u_0(q)|
  \end{equation}
  and
  \begin{equation}
    |\hat S_0(k+\epsilon)-\hat S_0(k)|
    \leqslant
    C\epsilon^\alpha\left(1+\int\frac{dq}{8\pi^3}|\hat u_0(q)|\right)
  \end{equation}
  and, since $u_0(x)\leqslant 1$, $\|\hat u_0\|_1\leqslant 8\pi^3$, so $\hat S_0$ is H\"older continuous.
  Therefore
  \begin{equation}
    \begin{largearray}
      \frac{\sqrt{e_0}}{4\pi^3}|\hat v(q)|\int dk\ 
      \left|
	\frac{k^2+1}{\sqrt{(k^2+1)^2-\frac \rho{2e_0}\hat S_0(2\sqrt{e_0}k)}}
	-\frac{k^2+1}{\sqrt{(k^2+1)^2-1}}
      \right|
      \\[0.5cm]\hfill
      \leqslant
      C\sqrt{2e_0}^{\alpha}\frac{\rho}{2e_0}
      \frac{\sqrt{e_0}}{4\pi^3}|\hat v(q)|\int dk\ 
      k^\alpha\frac{k^2+1}{((k^2+1)^2-1)^{\frac32}}
      =o(\sqrt{e_0})
    \end{largearray}
  \end{equation}
  uniformly in $q$.
  Thus,
  \begin{equation}
    r_1(q)=o(\sqrt{e_0})
    \label{r1}
  \end{equation}
  uniformly in $q$.
  \bigskip

  \subpoint By inserting\-~(\ref{r1}), (\ref{r2}) and\-~(\ref{r3}) into\-~(\ref{dSsplit}), we find
  \begin{equation}
    \frac\rho2\partial_\epsilon\hat S_\epsilon(q)|_{\epsilon=0}
    +
    \frac1{8\pi^3}\int dk\ \hat v(k-q)\frac{\frac\rho2\partial_\epsilon\hat S_\epsilon(k)|_{\epsilon=0}}{k^2}
    =
    \frac{\sqrt{2e_0}}{3\pi^2}\hat v(q)
    +o(\sqrt{e_0})
    +O(\eta\sqrt{e_0})
  \end{equation}
  where $o(\sqrt{e_0})$ and $O(\eta\sqrt{e_0})$ are uniform in $q$.
  In other words, if we define
  \begin{equation}
    \hat f(k):=
    \frac{3\pi^2}{\sqrt{2e_0}}\frac{\frac\rho2\partial_\epsilon\hat S_\epsilon(k)|_{\epsilon=0}}{k^2}
  \end{equation}
  we have
  \begin{equation}
    (k^2+\hat v\ast)\hat f(k)=\hat v(k)+o(1)+O(\eta)
    .
  \end{equation}
  Now, denoting the solution of the scattering equation\-~(\ref{scateq_varphi}) by $\varphi$:
  \begin{equation}
    (-\Delta+v)\varphi(x)=v(x)
  \end{equation}
  and its Fourier transform by $\hat\varphi$, we have
  \begin{equation}
    -\frac1{8\pi^3}\int dk\ \hat v(k)
    \frac{\frac\rho{2}\partial_\epsilon\hat S_\epsilon(k)|_{\epsilon=0}}{k^2}
    =
    -\frac{\sqrt{2e_0}}{3\pi^2}\int \frac{dk}{8\pi^3}\ \hat v(k)\hat f(k)
    =
    -\frac{\sqrt{2e_0}}{3\pi^2}\int \frac{dk}{8\pi^3}\ (k^2+\hat v\ast)\hat\varphi(k)\hat f(k)
  \end{equation}
  and since $(k^2+\hat v\ast)$ is symmetric,
  \begin{equation}
    -\frac1{8\pi^3}\int dk\ \hat v(k)
    \frac{\frac\rho{2}\partial_\epsilon\hat S_\epsilon(k)|_{\epsilon=0}}{k^2}
    =
    -\frac{\sqrt{2e_0}}{3\pi^2}\int \frac{dk}{8\pi^3}\ \hat v(k)\hat\varphi(k)
    -\int\frac{dk}{8\pi^3}\ \hat\varphi(k)(o(\sqrt{e_0})+O(\sqrt{e_0}\eta))
    .
  \end{equation}
  and so
  \begin{equation}
    -\frac1{8\pi^3}\int dk\ \hat v(k)
    \frac{\frac\rho{2}\partial_\epsilon\hat S_\epsilon(k)|_{\epsilon=0}}{k^2}
    =
    -\frac{\sqrt{2e_0}}{3\pi^2}\int \frac{dk}{8\pi^3}\ \hat v(k)\hat\varphi(k)
    +(o(\sqrt{e_0})+O(\sqrt{e_0}\eta))
    .
  \end{equation}
  Finally, by Lemma\-~\ref{lemma:scattering}
  \begin{equation}
    \int \frac{dk}{8\pi^3}\ \hat v(k)\hat\varphi(k)
    =
    \int dx\ v(k)\varphi(k)
    =
    \hat v(0)-4\pi a
    .
  \end{equation}
  Therefore, by\-~(\ref{H3split}) and\-~(\ref{h3}),
  \begin{equation}
    H_3=
    \frac{\sqrt{2e_0}}{3\pi^2}(4\pi a-\hat v(0))
    +o(\sqrt{e_0})+O(\sqrt{e_0}\eta)
    .
    \label{H3}
  \end{equation}
  Inserting\-~(\ref{H1}), (\ref{H2}) and\-~(\ref{H3}) into\-~(\ref{etasplit}), we find
  \begin{equation}
    \eta\sim\frac{4\sqrt{2e_0}}{3\pi}a
    .
  \end{equation}
  We conclude the proof using Theorem\-~\ref{theo:energy}.
\qed

\subsection{Other observables}
\indent
As we have seen, the Simple equation makes predictions for the ground state energy that agree with that of the many-body interacting Boson system for high and low densities, and its prediction for the condensate fraction agrees with Bogolyubov theory at low density.
We have studied the predictions of the Simple equation for other observables as well.
We will state these in this section, but will not discuss the proofs.
Instead, we will refer the readers to the appropriate publications for details.
\bigskip

\subsubsection{Two-point correlation function}
\indent
The two-point correlation function is the joint probability (in the usual quantum mechanical probability distribution $|\psi|^2$, not the probability distribution $\psi/\int\psi$ that we considered in Chapter\-~\ref{sec:simplified_def}) of finding a particle at $y$ and another at $y'$:
\begin{equation}
  C_2^{(0)}(y-y'):=
  \lim_{\displaystyle\mathop{\scriptstyle N,V\to\infty}_{\frac NV=\rho}}
  \sum_{i\neq j}\left<\psi_0\right|\delta(x_i-y)\delta(x_j-y')\left|\psi_0\right>
  .
  \label{C2}
\end{equation}
Lee, Huang and Yang made the following prediction, using Bogolyubov theory.
\bigskip

\theoname{Conjecture}{{\rm\cite[(48)]{LHY57}}}\label{conjecture:lhy_C2}
  For any potential $v\in L_1(\mathbb R^3)$ with $v\geqslant 0$ (actually, the potential may include a hard-core component), as $\sqrt{\rho a}|x|\to \infty$,
  \begin{equation}
    \frac1{\rho^2}C_2^{(0)}(x)-1\sim\frac{16\rho a^3}{\pi^3(\sqrt{\rho a}|x|)^4}
  \end{equation}
  where $a$ is the scattering length of the potential.
\endtheo
\bigskip

\indent
As was the case for the condensate fraction, see Section\-~\ref{sec:condensate_fraction}, we must define a natural prediction of the Simple equation for the two-point correlation function.
As we mentioned above, the Simple equation does not predict the ground state wavefunction $\psi_0$, so we must relate the correlation function to the ground state energy.
To do so, we simplify\-~(\ref{C2}) by taking advantage of the translation invariance: if $z=y-y'$, then
\begin{equation}
  \sum_{i\neq j}\left<\psi_0\right|\delta(x_i-y)\delta(x_j-y')\left|\psi_0\right>
  =
  \sum_{i\neq j}\left<\psi_0\right|\delta(x_i-z-y')\delta(x_j-y')\left|\psi_0\right>
\end{equation}
but, by translation invariance, this quantity is independent of $y'$, so we can take an average over $y'$:
\begin{equation}
  \begin{largearray}
    \sum_{i\neq j}\left<\psi_0\right|\delta(x_i-y)\delta(x_j-y')\left|\psi_0\right>
    =
    \frac1V\int dy'
    \sum_{i\neq j}\left<\psi_0\right|\delta(x_i-z-y')\delta(x_j-y')\left|\psi_0\right>
    =\\\hfill=
    \frac1V\sum_{i\neq j}\left<\psi_0\right|\delta(x_i-x_j-z)\left|\psi_0\right>
  \end{largearray}
\end{equation}
that is,
\begin{equation}
  \sum_{i\neq j}\left<\psi_0\right|\delta(x_i-y)\delta(x_j-y')\left|\psi_0\right>
  =
  \frac2V\sum_{i<j}\left<\psi_0\right|\delta(x_i-x_j-z)\left|\psi_0\right>
\end{equation}
which is
\begin{equation}
  \sum_{i\neq j}\left<\psi_0\right|\delta(x_i-y)\delta(x_j-y')\left|\psi_0\right>
  =
  \frac2V\frac{\delta}{\delta v(z)}\left<\psi_0\right|H\left|\psi_0\right>
\end{equation}
and so
\begin{equation}
  C_2^{(0)}(z)=2\rho\frac{\delta e_0}{\delta v(z)}
  .
\end{equation}
\bigskip

\indent
Therefore, the prediction of the Simple equation for the two-point correlation function is defined to be
\begin{equation}
  C_2(z):=2\rho\frac{\delta e}{\delta v(z)}
  .
\end{equation}
This quantity can be proved to decay as $|x|^{-4}$ for any value of the density, which agrees with the Bogolyubov prediction in Conjecture\-~\ref{conjecture:lhy_C2} (though the result for the Simple equation goes beyond, as it is not restricted to small densities).
\bigskip

\theoname{Theorem}{{\rm\cite[Theorem 4.5]{Ja22}}}\label{theo:C2}
  Under the assumptions of Theorem\-~\ref{theo:existence}, if $(1+|x|^6)v(x)\in L_1(\mathbb R^d)$, then
  \begin{equation}
    \lim_{|x|\to\infty}|x|^4\left|\frac{C_2}{\rho^2}-1-r(x)\right|<\infty
  \end{equation}
  where $|x|^4r\in L_2(\mathbb R^3)\cap L_\infty(\mathbb R^3)$.
\endtheo
\bigskip

\indent
Strictly speaking, this does not necessarily mean that $C_2\sim|x|^{-4}$: the function $|x|^4r$ is not guaranteed to decay.
However it is square integrable and bounded, so, at worst, there could be small intervals where $|x|^4r$ is not small (though also not large), and these intervals would get more and more spaced out as $|x|\to\infty$.
This is not far from saying that $r$ decays faster than $|x|^{-4}$.
\bigskip

\indent
The proof uses some tools that we have not introduced here.
Namely, it relies on the analysis of the operator
\begin{equation}
  \mathfrak K_e:=(-\Delta+v+4e(1-\rho u\ast))^{-1}
  .
\end{equation}
This operator is studied in depth in\-~\cite{CJL21}.
The full proof of Theorem\-~\ref{theo:C2} can be found in\-~\cite[Appendix B]{Ja22}.

\subsubsection{Momentum distribution}
\indent
The condensate fraction is the proportion of particles in the condensate state, which is the constant state.
The constant state can also be seen as the zero-momentum state: $e^{ikx}|_{k=0}$.
A natural extension is to compute the probability of finding a particle in a state $e^{ikx}$ with momentum $k$.
This quantity is called the {\it momentum distribution}, and is defined as
\begin{equation}
  \mathcal M_0(k):=
  \lim_{\displaystyle\mathop{\scriptstyle N,V\to\infty}_{\frac NV=\rho}}
  \frac1N\sum_{i=1}^N\left<\psi_0\right|K_i(k)\left|\psi_0\right>
  \label{momentum_distribution_def}
\end{equation}
where $K_i(k)$ is the projector on the subspace where particle $i$ has momentum $k$:
\begin{equation}
  K_i(k)\psi(x_1,\cdots,x_N):=e^{ikx_i}\int dy_i\ e^{-iky_i}\psi(x_1,\cdots,x_{i-1},y_i,x_{i+1},\cdots,x_N)
  .
\end{equation}
The prediction from Bogolyubov theory can be recovered from\-~\cite[Appendix\-~A]{LSe05}:
\bigskip

\theo{Conjecture}\label{conjecture:Nk}
  For any potential $v\in L_1(\mathbb R^3)$ with $v\geqslant 0$ (actually, the potential may include a hard-core component), as $\rho\to0$ and
  \begin{equation}
    \frac{|k|}{\sqrt{8m\pi\rho a}}=:\kappa
    \label{kappak_conj}
  \end{equation}
  is fixed,
  \begin{equation}
    \mathcal M_0(k)\sim\frac1{2\rho}\left(\frac{\kappa^2+1}{\sqrt{(\kappa^2+1)^2-1}}-1\right)
    \label{Nk_conj}
  \end{equation}
  where $a$ is the scattering length of the potential.
\endtheo
\bigskip

See Appendix\-~\ref{app:bog_Nk} for a derivation of this expression from\-~\cite[Appendix\-~A]{LSe05}.
\bigskip

\indent
To obtain a prediction for the momentum distribution using the Simple equation, we proceed in a similar way as for the condensate fraction: we add a term to the Hamiltonian
\begin{equation}
  \tilde H_N(\epsilon):=H_N+\epsilon\sum_{i=1}^NK_i(k)
\end{equation}
after which
\begin{equation}
  \mathcal M_0(k)=
  \lim_{\displaystyle\mathop{\scriptstyle N,V\to\infty}_{\frac NV=\rho}}
  \frac1N\partial_\epsilon \tilde E_0|_{\epsilon=0}
  .
\end{equation}
We then repeat the arguments from Sections\-~\ref{sec:simplified_construction}-\ref{sec:equations} and Appendix\-~\ref{app:proof_factorization} to the modified Hamiltonian.
However, there is one added difficulty: the modified Hamiltonian no longer necessarily has a unique ground state.
In particular, it is not guaranteed that its ground states are translation invariant.
However, translation invariance is used repeatedly in the derivation of Lieb's Simplified approach.
In order to carry this out carefully, this derivation needs to be adapted to cases without translation invariance.
This was done in\-~\cite{Ja24}.
The modified Simple equation is\-~\cite[(60)]{Ja24}:
\begin{equation}
  -\Delta u_\epsilon=(1-u_\epsilon)v-4e_\epsilon u_\epsilon+2\rho e_\epsilon u_\epsilon\ast u_\epsilon+\epsilon F(x)
  ,\quad
  e_\epsilon:=\frac\rho2\int dx\ (1-u_\epsilon(x))v(x)
\end{equation}
with
\begin{equation}
  F(x):=-2\hat u_\epsilon(k)\cos(kx)
\end{equation}
in terms of which
\begin{equation}
  \mathcal M(k):=\partial_\epsilon e_\epsilon|_{\epsilon=0}
  .
\end{equation}
\bigskip

\theoname{Theorem}{{\rm\cite[Theorem 3]{Ja24}}}\label{theo:Nk}
  Under the assumptions of Theorem\-~\ref{theo:existence}, for $d=3$, if
  \begin{equation}
    \kappa:=\frac{|k|}{2\sqrt{e}}
  \end{equation}
  then, for all $\kappa\in\mathbb R^3$,
  \begin{equation}
    \lim_{\rho\to0}\rho\mathcal M(2\sqrt{e}\kappa)=\frac12\left(\frac{\kappa^2+1}{\sqrt{(\kappa^2+1)^2-1}}-1\right)
    .
    \label{Nk}
  \end{equation}
\endtheo
\bigskip

The proof of this Theorem can be found in\-~\cite[Section\-~4.1]{Ja24}.
\bigskip

\indent
The scaling $|k|\sim\sqrt{8m\pi\rho a}$ in Conjecture\-~\ref{conjecture:Nk} comes from Theorem\-~\ref{theo:Nk} (using the leading order in\-~(\ref{lhy})), which suggests that the asymptotic behavior in\-~(\ref{Nk_conj}) holds for values of $\sqrt{\rho a}\lesssim |k|\ll 1$.
Now, if we were to take the limit $\kappa\to\infty$ {\it after} the limit $\rho\to0$ in\-~(\ref{Nk}), then we find
\begin{equation}
  \lim_{\rho\to0}\rho\mathcal M(2\sqrt e\kappa)\sim_{\kappa\to\infty}\frac1{4\kappa^4}\sim\frac{4e^2}{|k|^4}\sim\frac{16\pi^2m^2\rho^2 a^2}{|k|^4}
  .
\end{equation}
The asymptotics $\rho\mathcal M_0(k)\sim\frac C{|k|^4}$ is called the {\it universal Tan relation}, which was first derived for Fermions\-~\cite{Ta08,Ta08b,Ta08c}, and extended to Bosonic systems\-~\cite{CAL09}.
In fact, the value $C=16\pi^2m^2\rho^2a^2$ was predicted by\-~\cite[6.2.1.2]{NE17} from\-~\cite{CAL09}.
Thus the Simple equation reproduces the universal Tan relation, but it also shows that the order of the limits matters when considering low density and large $|k|$.
Indeed, it predicts that the Tan relation holds if $\rho\to0$ with $\frac{|k|}{\sqrt{8m\pi\rho a}}$ fixed (which means that $|k|$ goes to $0$ with $\rho$), and {\it then}, $\kappa$ is taken to infinity.
In other words, the Simple equation predicts that the Tan relation holds in the regime $\sqrt{\rho a}\ll|k|\ll 1$.
Now, this is only relevant for low-densities: if $\sqrt{\rho a}$ is not small, then the range $\sqrt{\rho a}\ll|k|\ll 1$ is empty.
In fact, it was verified numerically, using the Medium and Big equations (which are more accurate for intermediate densities), that the Tan relation does {\it not} hold in the intermediate density regime\-~\cite[Fig.\-~6]{CHe21}.

\subsubsection{Decay of the solution $u$}
\indent
A useful lemma that was used to compute the decay of two-point correlation function, as well as several other properties of the solution of the Simple equation is an estimate of the decay rate of the solution $u$.
\bigskip

\theoname{Lemma}{{\rm\cite[Theorem 1.2]{CJL21}}}
  In $d=3$, under the assumptions of Theorem\-~\ref{theo:existence}, if $(1+|x|^4)v(x)\in L_1(\mathbb R^3)\cap L_2(\mathbb R^3)$, then
  \begin{equation}
    \rho u(x)=\frac{\sqrt{2+\beta}}{2\pi^2\sqrt e}\frac1{|x|^4}+R(x)
  \end{equation}
  where
  \begin{equation}
    \beta=\rho\int dx\ |x|^2v(x)(1-u(x))dx\leqslant\rho\||x|^2v\|_1
  \end{equation}
  and $|x|^4R(x)\in L_2(\mathbb R^3)\cap L_\infty(\mathbb R^3)$, uniformly in $e$ on all compact sets.
  In addition, $\forall\rho_0>0$, there exists a constant $C$ that only depends on $\rho$ such that, for all $\rho<\rho_0$ and all $x\in\mathbb R^3$,
  \nopagebreakaftereq
  \begin{equation}
    u(x)\leqslant\min\left\{1,\frac C{\rho\sqrt e|x|^4}\right\}
    .
  \end{equation}
\endtheo
\restorepagebreakaftereq
\bigskip

The proof of this Theorem can be found in\-~\cite[Section\-~2]{CJL21}.
\bigskip

\indent
The decay of $u$ translates to the decay of the two-point correlation function in Theorem\-~\ref{theo:C2}, see\-~\cite{Ja22}, and is used to prove various results in\-~\cite{CJL21}, as it is used to prove bounds on the operator $\mathfrak K_e$, see\-~\cite[Section\-~1.2]{CJL21}.

\section{Numerical computation of the solution to the Big and Medium equations}\label{sec:numerics}
\indent
We have seen that the predictions of the Simple equation can be studied analytically.
However, as one can see from Figures\-~\ref{fig:energy}-\ref{fig:condensate}, the Simple equation is not so accurate in the intermediate density regime.
The predictions of the Big and Medium equations are much more accurate.
However, no analytical results about those equations have been proved so far: even the existence of a solution remains an outstanding open problem, see Chapter\-~\ref{sec:open}.
On the other hand, these equations are PDEs in three dimensions, and solutions can be computed numerically with modest hardware.
\bigskip

\indent
This numerical computation has been implemented in a software package called {\tt simplesolv}\-~\cite{ss} that is available for use under the free Apache License.
The program is written in the Julia language\-~\cite{Julia}.
Documentation is bundled with the software package that explains the details of the numerical algorithms used, and their accuracy.
\bigskip

\indent
The numerical computation is carried out in Fourier space, as the equations of Lieb's Simplified approach have fewer convolutions in Fourier space, and thus fewer numerical integrals need to be evaluated.
The equations are then written in the form
\begin{equation}
  \hat u(k)=F(\hat u)(k)
  \label{newton_setup}
\end{equation}
where $F$ is a functional that is specific to each equation of the approach.
Note that we are searching for spherically symmetric solutions of the equations, because $v$ is spherically symmetric, so the unknown function $\hat u$ is a function of $|k|$ alone.
Next, we discretize the space of functions $\hat u(|k|)$ by truncating its Chebyshev polynomial expansion.
In doing so, we represent the function $\hat u$ as a finite vector, from which we can recover an approximation for $\hat u$ in terms of Chebyshev polynomials.
The approximation is exponentially accurate if $\hat u$ is analytic (see the documentation for\-~\cite{ss}).
(Note that $\hat u(k)$ is not analytic, as $u(x)\sim|x|^{-4}$, but $|k|\mapsto \hat u(|k|)$ presumably is.)
Some terms in $F$ are integrals of $\hat u$.
To evaluate them, we approximate the integrals using Gauss quadratures.
This approximation is also exponentially accurate if $\hat u$ is analytic\-~\cite{ss}.
Finally, to solve\-~(\ref{newton_setup}) (which, having discretized space as we did, is now a vector valued equation), we use the Newton algorithm, which converges super-exponentially fast.
\bigskip

\indent
In this way, we can compute numerical values for various observables for any of the equations of Lieb's Simplified approach.
Note that it is significantly faster to run the computation for the Medium equation than the Big equation (see the documentation of\-~\cite{ss} for an explanation), so it provides a good middle ground between computational complexity and accuracy.
\bigskip

\indent
The plots for the energy and condensate fraction were already discussed above, see Figures\-~\ref{fig:energy}-\ref{fig:condensate}.
In both cases we see that the simple equation is accurate at low and high density, but rather far off the mark in the intermediate density regime.
On the other hand, the Big equation is remarkably accurate (especially so for the energy) at {\it all} densities.
The Medium equation lies in the middle of the Simple and Big.
(Note that we have found that larger potentials lead to worse accuracy\-~\cite{CHe21}.)
The plot of the condensate fraction further reveals an interesting fact: the condensate fraction reaches a minimum at intermediate densities.
\bigskip

\begin{figure}
  \hfil\includegraphics[width=7.5cm]{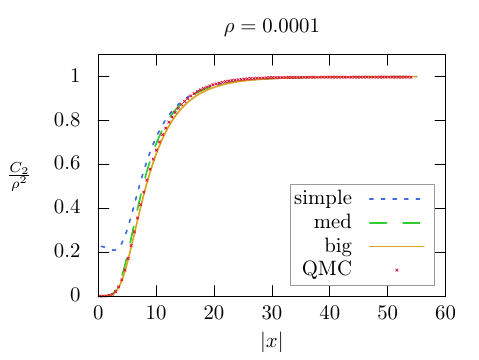}
  \hfil\includegraphics[width=7.5cm]{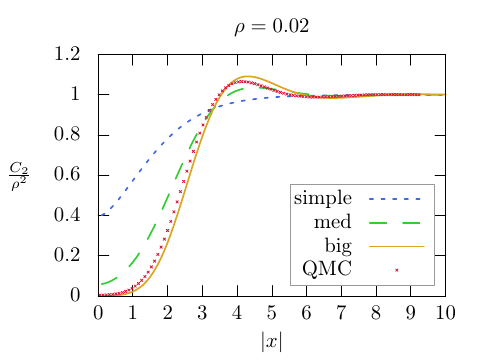}\par
  \caption{The predictions of the two-point correlation function for the \eqformat{Simple Equation}, \eqformat{Medium Equation}, and \eqformat{Big Equation}, compared to a \eqformat{Quantum Monte Carlo} (QMC) simulation (computed by M.\-~Holzmann), for $\rho=0.0001$ and $\rho=0.02$ and $v=16e^{-|x|}$.}
  \label{fig:2pt}
\end{figure}

\indent
A computation of the two-point correlation function is shown in Figure\-~\ref{fig:2pt}.
As was the case for the energy and condensate fraction, the prediction of the Simple equation is only accurate at small densities (and values of $|x|$ that are not too small).
On the other hand, the Big equation reproduces the results of the Quantum Monte Carlo computation rather accurately at all densities.
In particular, we see that the correlation function exhibits a local maximum above $\rho^2$ at intermediate densities.
This is a rather interesting fact: such a local maximum implies the existence of a preferred length scale, at which it is more probable to find pairs of particles.
This behavior is only observed at intermediate densities, and points to non-trivial behavior, which has yet to be fully studied.
A more systematic study of the two-point correlation function in the intermediate density regime is in progress, see\-~\cite{Ja23}.

\section{Open problems}\label{sec:open}
\indent
Lieb's Simplified approach is thus a powerful tool to study the ground state of systems of interacting Bosons.
It reproduces the predictions of Bogolyubov theory, but goes far beyond: it is asymptotically correct at high densities in addition to low ones, and is very accurate at intermediate densities as well.
We have discussed a number of results on the approach, but there still are many problems that remain open.
\bigskip

\subsection{Connecting Lieb's Simplified approach to the many-body Bose gas}
\indent
For one, there is still no rigorous connection between Lieb's Simplified approach and the many-body Bose gas, other than the fact that the energy asymptotics for the Simple equation agree at low and high densities.
Even in that case, the computation of the energy for the Simple equation is independent from that for the many-body problem, and our understanding of the connection between the two models is still virtually non-existent.
\bigskip

\indent
One way to approach this question would be to understand the approximation made in the factorization assumption.
This would require to bound the difference between the correlation functions $g_3,g_4$ and their factorized counterparts.
This may be doable for small enough densities, where inter-particle distances are large, and the interaction between particles mostly affects isolated clusters of particles.
One of the major difficulties of proceeding in this way is that it is hard to compute much about the ground state wavefunction, so establishing this clustering picture could be a challenge.
\bigskip

\indent
Another idea (or, perhaps a first step) could be to use Lieb's Simplified approach to find an approximation for the ground state wavefunction itself.
As we have seen, the approach makes a prediction for the ground state energy, but not the wavefunction.
A natural candidate may be to construct such a wavefunction in Bijl-Dingle-Jastrow (BDJ) form:
\begin{equation}
  \psi(x_1,\cdots,x_N)=\prod_{i<j}e^{-u(x_i-x_j)}
  \label{bdj}
\end{equation}
where $u$ is the solution to the Simple equation.
Taking $\psi$ to be in BDJ form is nothing new: the upper bounds for the energy of the Bose gas are obtained by using modified BDJ wavefunctions\-~\cite{Dy57,YY09,BCS21}.
However, the BDJ functions used so far are mostly based on the solution to the scattering equation.
Because the Simple equation seems to capture the properties of the system much more accurately (at low densities), it would seem reasonable to use it instead.
One distinct advantage of the BDJ form is that $\psi/\int\psi$ is formally equal to the Gibbs measure for a classical statistical mechanical model of particles interacting via the potential $u$.
So one might imagine using the many tools developed for the study of equilibrium statistical mechanics to study such a wavefunction.
However, there is one important difficulty, which is that $u$ depends on the density of particles, which is not the usual situation in statistical mechanics.
In particular, $\rho\int u=1$ so, even when $\rho$ is small, the effective classical statistical mechanical model associated to this BDJ function is {\it not} a low-density model.
If this could be overcome, then it should be possible to use an expression of the form\-~(\ref{bdj}) to prove an upper bound on the ground state energy.
It is encouraging to note that the numerics indicate that the energy predicted by the Simple equation seems to be an upper bound on the ground state energy.
Proving a corresponding lower bound would be more of a challenge, as would computing observables other than the energy.
But having a good, and simple candidate for the ground state wavefunction would be quite helpful to develop an intuition on how to tackle this problem.
\bigskip

\indent
Proving an upper bound on the ground state energy on its own could already be interesting.
In fact, in the current state of the art, the Lee-Huang-Yang formula\-~(\ref{lhy}) has still not been proved for the hard-core potential.
Unusually, the lower bound has been proved\-~\cite{FS23}, but the upper bound is still missing.
It may be possible to study the Simple equation with a hard-core potential in ways that are rather similar to the ones described in this book.
If one could show that the Simple equation provides an upper bound for the ground state energy of the many-body problem, then this could be an avenue to proving the Lee-Huang-Yang formula for the hard-core potential.

\subsection{Proving more properties of the solutions of Lieb's Simplified approach}

\indent
Almost every question on connecting Lieb's Simplified approach to the many-body Bose gas is open.
In addition, there are many open questions about the study of the approach itself.
\bigskip

\indent
Next to nothing is known analytically about the Big, or even the Medium equation.
Even the existence of solutions has remained an open problem.
This is somewhat surprising, as the numerical computations converge very quickly and reliably in practice\-~\cite{ss}.
Perhaps one could prove that the algorithm we use in our numerics converges to a solution.
A natural first step would be to attempt this for the Medium equation, which is much simpler than the Big equation.
\bigskip

\indent
There also are outstanding open questions on the analysis of the Simple equation.
One is to study the solution for a hard-core potential, as we already discussed.
Another big gap in our knowledge is the absence of a proof of the uniqueness of solutions.
In Chapter\-~\ref{sec:existence}, we proved that the modified problem where we fix $e$ and compute $\rho$ and $u$ has a unique solution, but to conclude the uniqueness of the solution of the original problem, where $\rho$ is fixed, we must prove that $e\mapsto\rho(e)$ is (globally) invertible.
Because $\rho(e)$ is continuous (see Lemma\-~\ref{lemma:surjective}), this is equivalent to the fact that $e\mapsto\rho(e)$ is strictly monotone, or, equivalently, that $\rho\mapsto e(\rho)$ is strictly monotone.
For the many-body system\-~(\ref{H}), this is certainly true: $e_0(\rho)$ is monotone increasing: consider the dependence of the ground state energy $E_0$ on the volume $V$; if the volume expands, then the Hilbert space becomes larger, so the ground state energy cannot decrease (it can also not stay constant by translation invariance).
However, we do not have a formulation of the Simple equation in terms of a variational problem, so we cannot use this argument for the Simple equation.
We have proved\-~\cite[Theorem\-~1.3]{CJL21} that $e\mapsto\rho(e)$ is monotone increasing for small and for large values of $\rho$, but not for intermediate values.
\bigskip

\indent
A related open problem is the convexity of $\rho e(\rho)$.
For the many-body system\-~(\ref{H}), $\rho e_0(\rho)$ is the energy per unit volume.
This quantity must be convex, otherwise the compressibility would be negative, and the system would be unstable.
(A more direct argument is that the density cannot be uniform in this case: if $\rho e_0(\rho)$ is not convex, then for every $\alpha\in(0,1)$, there exists $\rho_1<\rho<\rho_2$ such that $\alpha\rho_1+(1-\alpha)\rho_2=\rho$ and $\alpha\rho_1 e_0(\rho_1)+(1-\alpha)\rho_2 e_0(\rho_2)<\rho e(\rho)$; then if we split the volume into two parts of volumes $V_1:=\alpha V$ and $V_2:=(1-\alpha)V$, and consider a configuration in which the density in $V_1$ is $\rho_1$ and the density in $V_2$ is $\rho_2$, then the total energy would be $V_1\rho_1e_0(\rho_1)+V_2\rho_2e_0(\rho_2)<V\rho e(\rho)$; thus the system would spontaneously split into one region with density $\rho_1$ and one with density $\rho_2$, so the density would not be uniform.)
It would therefore be desirable to prove that $\rho e(\rho)$ is convex.
We have done this\-~\cite[Theorem\-~1.4]{CJL21} for small enough values of $e$, but the problem is still open for intermediate and large densities.

\subsection{Extending Lieb's Simplified approach to other settings}
\indent
As we have seen, Lieb's Simplified approach is an effective tool to study the ground state of interacting Bosons.
A natural question is whether it can be extended to other settings.
For one, the derivation of the approach in Chapter\-~\ref{sec:simplified_def} is tailored to the ground state of the system.
Can a simplified approach be derived for positive temperature states?
This would involve studying the heat kernel $e^{-\beta H_N}$ instead of the ground state wavefunction, which would entail studying the equation $-\partial_\beta e^{-\beta H_N}=H_N e^{-\beta H_N}$ instead of\-~(\ref{eigenvalue}).
\bigskip

\indent
In addition, the derivation of Lieb's Simplified approach in Chapter\-~\ref{sec:simplified_def} relies on the translation invariance of the system.
The derivation has been extended to systems that break the translation invariance in\-~\cite{Ja24}, but what still remains open is to study interacting Bosons in a trap.
That is, consider the Hamiltonian\-~(\ref{H}) to which we add a trapping term $\sum_i v_0(\epsilon|x_i|)$, which grows at infinity, so that the particles are trapped inside a volume of order $\epsilon^{-3}$.
Now, if $\epsilon$ is small, one would expect to find that, locally, the system behaves like the translation invariant one, but the density should be modulated over the trap: it should be largest in the center, and smaller towards the edges.
However, the Simplified approach in the presence of the trap has been found\-~\cite{Ja24} to yield a slightly different picture: the equation in the bulk includes extra terms that are not expected.
A deeper analysis of these extra terms is required to understand this.
\bigskip

\indent
We have studied Lieb's Simplified approach for potentials that are non-negative and integrable.
It could be quite interesting to extend this analysis to more general potentials.
We have already discussed the interest in the hard-core potential.
Another candidate is the Coulomb potential.
A regularized Coulomb potential was studied in\-~\cite{LS64}, but an investigation of the unadorned Coulomb potential is still open.
Relaxing the non-negativity of the potential is another challenge.
One would expect that adding a small negative part, small enough for the Hamiltonian to not have bound states, may not change the picture so much.
For instance, one could consider a shallow Lennard-Jones potential.
On the other hand, if the potential generates bound states, the physical picture, even at low densities, would be quite different.
\bigskip

\indent
Most of the results discussed here (other than the derivation of Lieb's Simplified approach and the existence of solutions to the Simple equation) have focused on three dimensions.
The one-dimensional Bose gas was studied using the Simplified approach in\-~\cite{LL64}, and compared to the exactly solvable one-dimensional interacting Boson model\-~\cite{LL63}.
It was found that the Simplified approach agrees with the many-body model in the high-density limit, but not at low densities.
The natural next question is: what about two dimensions?
\bigskip

\indent
Finally, Lieb's Simplified approach concerns interacting Bosons.
Could it be extended to a Fermionic system?
The factorization assumption is (loosely) justified using the non-negativity of the wavefunction $\psi_0$.
The non-negativity of the wavefunction can only hold for one of the eigenstates of the system (regardless of whether the state is Bosonic or Fermionic), as all others must be orthogonal to $\psi_0$.
Thus, the Fermionic ground state would not be non-negative, and the factorization assumption would need to be changed accordingly.
Understanding how to do this change is an open problem.

\vfill
\eject

\appendix

\section{Elements of functional analysis}\label{app:functional_analysis}
\indent
In this appendix we gather a few useful definitions and results from functional analysis.
\bigskip

\subsection{Compact and trace-class operators}\label{app:compact}
\theoname{Definition}{Trace}
  Given a separable Hilbert space with an orthonormal basis $\{\varphi_i\}_{i\in\mathbb N}$, the trace of an operator $A$ is formally defined as
  \begin{equation}
    \mathrm{Tr}(A):=\sum_{i=0}^\infty \left<\varphi_i\right|A\left|\varphi_i\right>
    .
  \end{equation}
  If this expression is finite, then $A$ is said to be trace-class.
\endtheo
\bigskip

\theoname{Definition}{Compact operators}
  The set of compact operators on a Hilbert space is the closure of the set of finite-rank operators.
\endtheo
\bigskip

\theoname{Theorem}{{\rm\cite[Theorem 6.6]{Te14}}}
  If $A$ is a compact self-adjoint operator on a Hilbert space, then its spectrum consists of an at most countable set of eigenvalues.
\endtheo
\bigskip

\theoname{Theorem}{{\rm\cite[Section 6.3]{Te14}}}\label{theo:trace_compact}
  Trace class operators are compact.
\endtheo
\bigskip

\theoname{Theorem}{{\rm\cite[Lemma 5.5]{Te14}}}\label{theo:compact_ideal}
  If $A$ is compact and $B$ is bounded, then $KA$ and $AK$ are compact.
\endtheo
\bigskip

\theo{Theorem}\label{theo:compact_schrodinger}
  Let $\Delta$ be the Laplacian on $L_2(\mathbb R^d/(L\mathbb Z)^d)$ (the same result holds for Dirichlet and Neumann boundary conditions as well).
  For any function $v$ on $[0,L]^d$ that is bounded below, $e^{-t(-\Delta+v)}$ is compact for any $t\geqslant 0$.
\endtheo
\bigskip

\indent\underline{Proof}:
  First of all, $e^{t\Delta}$ is trace class: using the basis $e^{ikx}$ with $k\in(\frac{2\pi}L\mathbb Z)^d$
  \begin{equation}
    \mathrm{Tr}(e^{t\Delta})
    =\sum_{k\in(\frac{2\pi}L\mathbb Z)^d}e^{-tk^2}<\infty
    .
  \end{equation}
  By Theorem\-~\ref{theo:trace_compact}, $e^{t\Delta}$ is compact.
  By the Trotter product formula, Theorem\-~\ref{theo:trotter_hilbert},
  \begin{equation}
    e^{t\Delta-tv}
    =\lim_{N\to\infty}(e^{\frac tN\Delta}e^{-\frac tNv})^N
  \end{equation}
  and by Theorem\-~\ref{theo:compact_ideal}, $(e^{\frac tN\Delta}e^{-\frac tNv})^N$ is compact.
  Since the set of compact operators is closed, so is $e^{t(\Delta-v)}$.
\qed

\subsection{$L_p$ and Sobolev spaces}\label{app:sobolev}

\theoname{Definition}{$L_p$ and Sobolev spaces}
  For $p\geqslant 1$ and $\Omega$ a metric space, we define the $L_p$ space:
  \begin{equation}
    L_p(\Omega):=\left\{f:\Omega\to\mathbb R,\ \left(\int dx\ |f(x)|^p\right)^{\frac1p}<\infty\right\}
    ,\quad
    L_\infty(\Omega):=\sup_{x\in\Omega} |f(x)|
  \end{equation}
  and the Sobolev space
  \begin{equation}
    W_{k,p}(\Omega):=
    \left\{f:\Omega\to\mathbb R\ \forall l\in\{0,\cdots,k\},\ \partial^lf\in L_p(\Omega)\right\}
  \end{equation}
  in which $\partial^l f$ is the $l$-th `weak' derivative (for our purposes, you can think of these as being regular derivatives, see\-~\cite[p.271]{Br11} for a more precise definition).
  These are Banach spaces (complete metric spaces), and $W_{k,2}$ is a Hilbert space (it has a scalar product).
\endtheo
\bigskip

\theoname{Theorem}{H\"older's inequality {\rm\cite[Problem 0.26]{Te14}}}\label{theo:holder}
  If $f\in L_p(\Omega)$ and $g\in L_q(\Omega)$, then $fg\in L_r(\Omega)$ with
  \begin{equation}
    \frac1r=\frac1p+\frac 1q
  \end{equation}
  and
  \nopagebreakaftereq
  \begin{equation}
    \|fg\|_r\leqslant \|f\|_p\|g\|_q
    .
  \end{equation}
\endtheo
\restorepagebreakaftereq
\bigskip

\theoname{Theorem}{Sobolev embedding theorem {\rm\cite[Theorem 8.8]{LL01}}} 
  \label{theo:sobolev}
  Given $q\geqslant p\geqslant 1$, $m\geqslant 1$, $\ell\in[0,m]$, and $d\geqslant 1$.
  If $(m-\ell)p\leqslant d$ and
  \begin{equation}
    \frac1p-\frac md\leqslant \frac 1q-\frac\ell d
  \end{equation}
  we have
  \begin{equation}
    W_{m,p}(\mathbb R^d)\subset W_{\ell,q}(\mathbb R^d)
  \end{equation}
  in particular, setting $\ell=0$,
  \begin{equation}
    W_{m,p}(\mathbb R^d)\subset L_q(\mathbb R^d)
    .
  \end{equation}
  In addition, if $(m-\ell)p>d$, then
  \begin{equation}
    |\partial^\alpha f|\in L_\infty(\mathbb R^d)
  \end{equation}
  for $|\alpha|\leqslant \ell$, in particular, if $p>\frac dm$, then
  \nopagebreakaftereq
  \begin{equation}
    W_{m,p}(\mathbb R^d)\subset L_\infty(\mathbb R^d)
    .
  \end{equation}
\endtheo
\restorepagebreakaftereq
\bigskip

\theo{Lemma}\label{lemma:Lq_from_Wmp}
  If $f\in W_{m,p}(\mathbb R^d)$ with $mp\geqslant d$, then $f\in L_q(\mathbb R^d)$ for all $q\geqslant p$.
\endtheo
\bigskip

\indent\underline{Proof}:
  We apply the Sobolev Embedding Theorem\-~\ref{theo:sobolev} with $\ell=m-\frac dp$, so
  \begin{equation}
    (m-\ell)p=d
    ,\quad
    \frac1p-\frac md+\frac\ell d=0
  \end{equation}
  and find that $f\in W_{\ell,q}(\mathbb R^d)\subset L_q(\mathbb R^d)$ for any $q\geqslant p$.
\qed
\bigskip

\theoname{Theorem}{Young's inequality {\rm\cite[Section 4.2, Remark 2]{LL01}}}\label{theo:young}
  If $f\in L_p(\Omega)$ and $g\in L_q(\Omega)$, then $f\ast g\in L_r(\Omega)$ with
  \begin{equation}
    1+\frac1r=\frac1p+\frac 1q
  \end{equation}
  and
  \nopagebreakaftereq
  \begin{equation}
    \|f\ast g\|_r\leqslant \|f\|_p\|g\|_q
    .
  \end{equation}
\endtheo
\restorepagebreakaftereq
\bigskip

\subsection{Operators on Banach spaces}
\indent
To prove properties of the Simple Equation, we study the operator $K_e$, see\-~(\ref{Ke}), which acts on $W_{2,p}(\mathbb R^d)$.
Because $W_{2,p}(\mathbb R^d)$ is a Banach space (and only a Hilbert space for $p=2$), we will need some theorems on operators acting on Banach spaces as opposed to Hilbert spaces.
The tool we will use the most are ``generators of contraction semigroups'', which generalize the notion of positive definite self-adjoint operators on Hilbert spaces.
\bigskip

\theo{Definition}
  A contraction semigroup on a Banach space $X$ is a family $\{T(t),\ t\in[0,\infty)\}$ of bounded operators satisfying
  \begin{equation}
    T(0)=\mathds 1
    ,\quad
    T(s)T(t)=T(s+t)
    ,\quad
    t\mapsto T(t)\varphi\mathrm{\ is\ continuous}
    ,\quad
    \|T(t)\|\leqslant 1
  \end{equation}
  for any $s,t\in[0,\infty)$, $\varphi\in X$.
\endtheo
\bigskip

\theo{Definition}\label{def:generate_semigroup}
  An operator $A$ on a Banach space $X$ is said to generate a contraction semigroup if $A$ is closed, for any $\lambda>0$, $A+\lambda$ is invertible, and
  \begin{equation}
    \|(A+\lambda)^{-1}\|\leqslant\frac1\lambda
    .
  \end{equation}
\endtheo
\bigskip

This is not the usual definition of the generator of a contraction semigroup, but it is equivalent to it, by the Hille-Yosida Theorem\-~\cite[Theorem X.47a]{RS75b}.
The Hille-Yosida Theorem also shows the following.
\bigskip

\theoname{Theorem}{Hille-Yosida Theorem {\rm\cite[Theorem X.47a]{RS75b}}}\label{theo:hille_yosida}
  If $A$ generates a contraction semigroup, then it is associated a unique contraction semigroup, denoted by $e^{-tA}$, and, for $\lambda>0$,
  \nopagebreakaftereq
  \begin{equation}
    (A+\lambda)^{-1}=-\int_0^\infty dt\ e^{-\lambda t}e^{-tA}
    .
  \end{equation}
\endtheo
\restorepagebreakaftereq
\bigskip

\theoname{Theorem}{Kato-Rellich Theorem on Banach space {\rm\cite[Theorem X.51]{RS75b}}}\label{theo:kato_rellich}
  Let $A$ and $B$ generate contraction semigroups on a Banach space, such that the domain $\mathcal D(A)$ of $A$ is a subset of the domain $\mathcal D(B)$ of $B$.
  If, for $\varphi\in\mathcal D(A)$,
  \begin{equation}
    \|B\varphi\|\leqslant a\|A\varphi\|+b\|\varphi\|
  \end{equation}
  for some $b\geqslant 0$ and $a\in[0,1)$, then $A+B$ generates a contraction semigroup.
\endtheo
\bigskip

Note that this theorem holds under the more general condition that $B$ is ``accretive'', see\-~\cite[p.\-~240]{RS75b}, but this less general formulation is sufficient for our purposes.
\bigskip

\theoname{Theorem}{Trotter product formula {\rm\cite[Theorem X.51]{RS75b}}}\label{theo:trotter}
  Given two operators $A$ and $B$ on a Banach space such that $A$, $B$ and $A+B$ generate contraction semigroups, then for $t\geqslant 0$,
  \nopagebreakaftereq
  \begin{equation}
    e^{-t(A+B)}
    =
    \lim_{N\to\infty}\left(e^{-t\frac 1N A}e^{-t\frac1N B}\right)^N
    .
  \end{equation}
\endtheo
\restorepagebreakaftereq
\bigskip

\indent
We will also use the Hilbert space version of this theorem, stated below.
\bigskip

\theoname{Theorem}{Trotter product formula on a Hilbert space {\rm\cite[Theorem 5.12]{Te14}}}\label{theo:trotter_hilbert}
  Given two operators $A$ and $B$ on a Hilbert space such that $A$, $B$ and $A+B$ are self-adjoint, and bounded from below, then for $t\geqslant 0$,
  \nopagebreakaftereq
  \begin{equation}
    e^{-t(A+B)}
    =
    \lim_{N\to\infty}\left(e^{-t\frac 1N A}e^{-t\frac1N B}\right)^N
    .
  \end{equation}
\endtheo
\restorepagebreakaftereq
\bigskip

The connection to Theorem\-~\ref{theo:trotter} is that a self-adjoint operator on a Hilbert space generates a contraction semigroup if and only if its spectrum is $>\epsilon>0$ (this follows from the spectral theorem).
One can then extend the result to operators that are bounded below by adding a multiple of the identity to raise the spectrum above $\epsilon$.
\bigskip

\theoname{Theorem}{Heat kernel}\label{theo:heat_contractive}
  The operator $-\Delta$ from $W_{2,p}(\mathbb R^d)$ to $L_p(\mathbb R^d)$ generates a contractive semigroup.
\endtheo
\bigskip

\indent\underline{Proof}:
  The Laplacian is trivially bounded on $W_{2,p}(\mathbb R^d)$, so it is closed.
  Furthermore, by Theorem\-~\ref{theo:yukawa}, for $\lambda>0$, $(-\Delta+\lambda)$ is invertible and, for $\psi\in L_p(\mathbb R^d)$,
  \begin{equation}
    (-\Delta+\lambda)^{-1}\psi=Y_{\sqrt\lambda}\ast\psi
  \end{equation}
  so, by Young's inequality (Theorem\-~\ref{theo:young}),
  \nopagebreakaftereq
  \begin{equation}
    \|(-\Delta+\lambda)^{-1}\|\leqslant\|Y_{\sqrt\lambda}\|_1=\frac1\lambda
    .
  \end{equation}
\qed
\restorepagebreakaftereq

\subsection{Fourier analysis}
\theoname{Definition}{Fourier transform}
  The Fourier transform of a function $f$ is
  \begin{equation}
    \hat f(k):=\int dx\ e^{ikx}f(x)
    \label{fourier}
  \end{equation}
  which is defined for any $f$ such that this integral is finite.
\endtheo
\bigskip

\theoname{Theorem}{{\rm\cite[Theorems 5.3, 5.5]{LL01}}}\label{theo:fourierinv}
  If $f\in L_2(\mathbb R^d)$, then $\hat f$ is well defined, and
  \begin{equation}
    f(x)=\int\frac{dk}{(2\pi)^d}e^{-ikx}\hat f(k)
    .
    \label{fourierinv}
  \end{equation}
  In addition, $\hat f\in L_2(\mathbb R^d)$ and
  \nopagebreakaftereq
  \begin{equation}
    \|\hat f\|_2=\|f\|_2
    .
  \end{equation}
\endtheo
\restorepagebreakaftereq

%
%

\subsection{Positivity preserving operators on Banach spaces}\label{app:positivity_preserving}
\theoname{Definition}{Positivity preserving operators}\label{def:positivity_preserving}
  An operator $A$ from a Banach space of real-valued functions $\mathcal B_1$ to another $\mathcal B_2$ is said to be {\it positivity preserving} if, for any $f\in\mathcal B_1$ such that $f(x)\geqslant0$, $Af(x)\geqslant 0$.
\endtheo
\bigskip

\theo{Theorem}\label{theo:add_inv}
  If $A$, $B$, and $A+B$ generate contraction semigroups, and $e^{-tA}$ and $e^{-t B}$ are positivity preserving for all $t\geqslant 0$, then for all $t\geqslant 0$, $e^{-t(A+B)}$ and $(A+B+\lambda)^{-1}$ are positivity preserving for all $\lambda>0$.
\endtheo
\bigskip

\indent\underline{Proof}:
  By the Hille-Yosida Theorem\-~\ref{theo:hille_yosida},
  \begin{equation}
    (A+B+\lambda)^{-1}
    =
    \int_0^\infty dt\ e^{-t\lambda}e^{-t(A+B)}
  \end{equation}
  and by the Trotter product formula (Theorem\-~\ref{theo:trotter}),
  \begin{equation}
    e^{-t(A+B)}
    =
    \lim_{N\to\infty}(e^{-\frac tNA}e^{-\frac tNB})^N
    .
  \end{equation}
  We conclude the proof by noting that the product of positivity preserving operators is positivity preserving, and that the set of positivity preserving operators is closed.
\qed
\bigskip

\theoname{Lemma}{Positivity of the heat kernel on $L_{2}(\mathbb R^d)$}\label{lemma:heat_L2}
  Given $t>0$, the operator $e^{t\Delta}$ on $L_2(\mathbb R^d)$ is positivity preserving.
\endtheo
\bigskip

\indent\underline{Proof}:
  Given $\psi\in L_2(\mathbb R^d)$, by Theorem\-~\ref{theo:fourierinv}, the Fourier transform of $\psi$ is $\hat\psi\in L_2(\mathbb R^d)$, and
  \begin{equation}
    \int dx\ e^{ikx}e^{t\Delta}\psi
    =
    e^{-tk^2}\hat\psi(k)
  \end{equation}
  which is in $L_2(\mathbb R^d)$, so we can take an inverse Fourier transform:
  \begin{equation}
    e^{t\Delta}\psi(x)=\left(\int\frac{dk}{(2\pi)^d}\ e^{-ikx}e^{-tk^2}\right)\ast\psi(x)
  \end{equation}
  (to see that the convolution appears, one can do a simple explicit computation, or use\-~\cite[Theorem\-~5.8]{LL01})
  and
  \nopagebreakaftereq
  \begin{equation}
    \int\frac{dk}{(2\pi)^d}\ e^{-ikx}e^{-tk^2}
    =\prod_{i=1}^d\int_{-\infty}^\infty\frac{dk_i}{2\pi}\ e^{-ik_ix_i}e^{-tk_i^2}
    =\prod_{i=1}^d\frac1{2\sqrt{t\pi}}e^{-\frac1{4t}x_i^2}
    =\frac1{(4\pi t)^{\frac d2}}e^{-\frac1{4t}|x|^2}
    \geqslant 0
    .
  \end{equation}
\qed
\restorepagebreakaftereq
\bigskip

\theoname{Theorem}{Positivity of the heat kernel on $W_{2,p}(\mathbb R^d)$}\label{theo:heat}
  Given $t>0$, the operator $e^{t\Delta}$ on $W_{2,p}(\mathbb R^d)$ with $p\geqslant \min\{2,\frac d2\}$ is positivity preserving.
\endtheo
\bigskip

\indent\underline{Proof}:
  We take $\psi\in W_{2,p}(\mathbb R^d)$.
  If $p\leqslant 2$, then,
  by the Sobolev Embedding Theorem\-~\ref{theo:sobolev} (more precisely, by Lemma\-~\ref{lemma:Lq_from_Wmp}), we have $\psi\in L_2(\mathbb R^d)$.
  Therefore, if $\psi\geqslant 0$, then $e^{t\Delta}\psi\geqslant 0$ by Lemma\-~\ref{lemma:heat_L2}.
  \bigskip

  \indent
  Let us now turn to $p>2$.
  We introduce a smooth cutoff function $\chi_N$ which is $\mathcal C_\infty$ (infinitely continuously differentiable), non-negative, and is equal to $1$ for $|x|\leqslant N$ and equal to $0$ for $|x|\geqslant N+1$.
  For every $N\geqslant 0$, $\chi_N\psi\in W_{2,p}(\mathbb R^d)\cap L_2(\mathbb R^d)$ (see, e.g. \cite[Section\-~6.4]{LL01}).
  Therefore, if $\psi\geqslant 0$, then by Lemma\-~\ref{lemma:heat_L2}, $e^{t\Delta}\chi_N\psi\geqslant 0$.
  Furthermore, since $e^{t\Delta}$ is bounded (by Theorem\-~\ref{theo:heat_contractive}), it is continuous, and
  \begin{equation}
    \lim_{N\to\infty}\|\psi-\chi_N\psi\|_{W_{2,p}(\mathbb R^d)}=0
  \end{equation}
  so $e^{t\Delta}\psi\geqslant 0$.
\qed
\bigskip

\subsection{The Yukawa potential}
\theoname{Theorem}{{\rm\cite[Theorem 6.23]{LL01}}}\label{theo:yukawa}
  For $d\geqslant 1$, $\mu>0$, we define the Yukawa potential
  \begin{equation}
    Y_\mu(x):=\int_0^\infty\frac{dt}{(4\pi t)^{\frac n2}}\ e^{-\frac{|x|^2}{4t}-\mu^2t}
  \end{equation}
  satisfies
  \begin{equation}
    \|Y_\mu\|_1=\frac1{\mu^2}
  \end{equation}
  is in $L_q(\mathbb R^d)$ for
  \begin{equation}
    \left\{\begin{array}{ll}
      q\in[1,\infty]&\mathrm{if\ }d=1\\
      q\in[1,\infty)&\mathrm{if\ }d=2\\
      q\in[1,\frac d{d-2}]&\mathrm{if\ }d\geqslant 3
    \end{array}\right.
  \end{equation}
  and, for any $f\in L_p(\mathbb F^d)$, $p\in[1,\infty]$,
  \nopagebreakaftereq
  \begin{equation}
    (-\Delta+\mu^2)(Y_\mu\ast f)=f
    .
  \end{equation}
\endtheo
\restorepagebreakaftereq

\subsection{The Perron-Frobenius theorem}
\theoname{Theorem}{(Generalized) Perron-Frobenius theorem {\rm\cite[Theorem 1.1]{Du06}}}\label{theo:perron_frobenius}
  Let $A$ be a compact operator from a Banach space of real-valued functions $\mathcal B_1$ to another $\mathcal B_2$.
  If $A$ is positivity preserving (see Definition\-~\ref{def:positivity_preserving}) and has a finite and positive spectral radius $r(A)$ ($r(A):=\sup_{\lambda\in\mathrm{spec}(A)}|\lambda|$), then $r(A)$ is a non-degenerate eigenvalue whose corresponding eigenvector $f$ is non-negative:
  \nopagebreakaftereq
  \begin{equation}
    Af=r(A)f
    ,\quad
    f(x)\geqslant 0
    .
  \end{equation}
\endtheo
\restorepagebreakaftereq

(The original Perron-Frobenius theorem applies to matrices; this generalized version is also called the Krein-Rutman theorem.)

\section{Elements of harmonic analysis}\label{app:harmonic}
\indent
In this appendix, we state some useful results from harmonic analysis.
\bigskip

\theoname{Definition}{Harmonic, subharmonic and superharmonic functions}
  A function $f\in \mathcal C^2(\mathbb R^d)$ is said to be {\it harmonic} on $A\subset\mathbb R^d$ if $\forall x\in A$,
  \begin{equation}
    \Delta f(x)=0
  \end{equation}
  it is {\it subharmonic} if
  \begin{equation}
    \Delta f(x)\geqslant 0
  \end{equation}
  and {\it superharmonic} if
  \nopagebreakaftereq
  \begin{equation}
    \Delta f(x)\leqslant 0
    .
  \end{equation}
\endtheo
\restorepagebreakaftereq
\bigskip

\theoname{Theorem}{{\rm\cite[Theorem 9.4]{LL01}}}\label{theo:harmonic}
  A function that is subharmonic on $A$ achieves its maximum on the boundary of $A$.
  A function that is superharmonic on $A$ achieves its minimum on the boundary of $A$.
\endtheo

\section{The operator $K_e$}\label{app:Ke}
\indent
In this appendix we prove some properties on the operator $K_e$, formally defined as
\begin{equation}
  K_e:=\left(-\frac1m\Delta+v+4e\right)^{-1}
\end{equation}
from $L_p(\mathbb R^d)$ to the Sobolev space $W_{2,p}(\mathbb R^d)$, see Section\-~\ref{app:sobolev}.
In particular, we will prove that this operator is well-defined and positivity-preserving.
The first step is to prove that $v$ is relatively bounded with respect to $-\frac1m\Delta+4e$.
\bigskip

\theo{Lemma}\label{lemma:v_relative_bound}
  For $p\geqslant \min\{\frac d2,1\}$ and $f\in W_{2,p}(\mathbb R^d)$, if $v\in L_r(\mathbb R^d)$ with $r\geqslant p$, and $e,m>0$, there exists $b>0$ and $a\in[0,1)$ such that
  \begin{equation}
    \|vf\|_p\leqslant a\|(-{\textstyle \frac1m}\Delta+4e)f\|_p+b\|f\|_p
  \end{equation}
  in other words, $v$ is relatively bounded with respect to $-\frac1m\Delta+4e$.
\endtheo
\bigskip

\indent\underline{Proof}:
  First of all, we split $v$ into the sum of a bounded function and a function with small $L_r$ norm:
  \begin{equation}
    v=v_\infty+v_r
    ,\quad
    v_\infty:=v\mathds 1_{v\leqslant b}
    ,\quad
    v_r:=v\mathds 1_{v>b}
  \end{equation}
  where $\mathds 1_{v\leqslant b}\in\{0,1\}$ is equal to 1 if and only if $v\leqslant b$ and similarly for $\mathds 1_{v>b}$.
  We then bound
  \begin{equation}
    \|vf\|_p\leqslant\|v_\infty f\|_p+\|v_rf\|_p
    .
  \end{equation}
  Since $|v_\infty|\leqslant b$,
  \begin{equation}
    \|v_\infty f\|_p\leqslant b\|f\|_p
    .
  \end{equation}
  Next, by the Sobolev Embedding Theorem\-~\ref{theo:sobolev} (or rather by Lemma\-~\ref{lemma:Lq_from_Wmp}), since $p\geqslant\frac d2$, $f\in L_q(\mathbb R^d)$ for any $q\geqslant p$.
  Therefore, by the H\"older inequality (Theorem\-~\ref{theo:holder}),
  \begin{equation}
    \|v_rf\|_p\leqslant\|v_r\|_{r}\|f\|_q
    ,\quad
    \frac1q=\frac1p-\frac1r
  \end{equation}
  (note that $q\in[1,\infty]$).
  Furthermore, by Theorem\-~\ref{theo:yukawa},
  \begin{equation}
    f=(-\Delta+4em)(Y_{2\sqrt{em}}\ast f)=Y_{2\sqrt{em}}\ast(-\Delta+4em)f
  \end{equation}
  and by Young's inequality (Theorem\-~\ref{theo:young}),
  \begin{equation}
    \|f\|_q\leqslant \|Y_{2\sqrt{em}}\|_{\frac r{r-1}}\|(-\Delta+4em)f\|_p
    .
  \end{equation}
  Note that, if $d<3$ by Theorem\-~\ref{theo:yukawa}, $Y_{2\sqrt{em}}\in L_{\frac r{r-1}}(\mathbb R^d)$ for any $r$, and if $d\geqslant 3$, $Y_{2\sqrt{em}}\in L_l(\mathbb R^d)$ for $l\leqslant\frac d{d-2}$, and $\frac r{r-1}\leqslant \frac d{d-2}$ because $r\geqslant \frac d2$, so $Y_{2\sqrt{em}}\in L_{\frac r{r-1}}(\mathbb R^d)$.
  Thus,
  \begin{equation}
    \|vf\|_p\leqslant b\|f\|_p+m\|Y_{2\sqrt{em}}\|_{\frac r{r-1}}\|v_r\|_r\|(-{\textstyle\frac1m}\Delta+4e)f\|_p
    .
  \end{equation}
  Finally, by dominated convergence,
  \begin{equation}
    \lim_{b\to\infty}\|v_r\|_r=0
  \end{equation}
  so, taking $b$ to be large enough,
  \nopagebreakaftereq
  \begin{equation}
    m\|Y_{2\sqrt{em}}\|_{\frac r{r-1}}\|v_r\|_r<1
    .
  \end{equation}
\qed
\restorepagebreakaftereq
\bigskip

\theo{Lemma}\label{lemma:Ke_pos}
  If $v\geqslant 0$, $v\in L_p(\mathbb R^d)$ with $p\geqslant 1$ and $p>\frac d2$, the operator $-\frac1m\Delta+v+4e$ from $W_{2,p}(\mathbb R^d)$ to $L_p(\mathbb R^d)$ is invertible and its inverse is bounded and positivity preserving.
\endtheo
\bigskip

\indent\underline{Proof}:
  We check the assumptions of the Kato-Rellich Theorem\-~\ref{theo:kato_rellich}.
  By Theorem\-~\ref{theo:heat_contractive}, $-\frac1m\Delta$, defined on $W_{2,p}(\mathbb R^d)$, generates a contractive semigroup.
  In addition, the domain of $v$ is $L_\infty(\mathbb R^d)$: indeed, by the H\"older inequality (Theorem\-~\ref{theo:holder}), for $\psi\in L_\infty(\mathbb R^d)$,
  \begin{equation}
    \|v\psi\|_p\leqslant\|v\|_p\|\psi\|_\infty
    .
  \end{equation}
  Furthermore, by the Sobolev Embedding Theorem\-~\ref{theo:sobolev}, since $p>\frac d2$, $W_{2,p}(\mathbb R^d)\subset L_\infty(\mathbb R^d)$.
  In addition, since $v\geqslant 0$, it generates the contraction semigroup $e^{-tv(x)}$.
  Thus, by Lemma\-~\ref{lemma:v_relative_bound} and the Kato-Rellich Theorem\-~\ref{theo:kato_rellich}, $(-\frac 1m\Delta+v)$ generates a contraction semigroup.
  By Definition\-~\ref{def:generate_semigroup}, $(-\frac1m\Delta+v+4e)$ is invertible and its inverse is bounded.
  \bigskip

  \indent
  By Theorem\-~\ref{theo:heat}, $e^{\frac1mt\Delta}$ is positivity preserving, as $p>\frac d2$.
  In addition, $e^{-tv}$ is positivity preserving because $v\geqslant 0$.
  Therefore, by Theorem\-~\ref{theo:add_inv}, $(-\frac1m\Delta+v+4e)^{-1}$ is positivity preserving.
\qed
\bigskip

\indent
Thus, $K_e$ is defined as an operator from $W_{2,p}(\mathbb R^d)$ to $L_p(\mathbb R^d)$.
As we will now prove, $K_e\psi$ is actually also defined for $\psi\in L_1(\mathbb R^d)$.
\bigskip

\theo{Lemma}\label{lemma:Ke_extend}
  $K_e$ extends to a bounded operator from $L_q(\mathbb R^d)$ to $L_q(\mathbb R^d)$ for any $q\geqslant 1$.
\endtheo
\bigskip

\indent\underline{Proof}:
  We have
  \begin{equation}
    (-{\textstyle\frac 1m}\Delta+4e+v)^{-1}
    =
    (-{\textstyle\frac 1m}\Delta+4e)^{-1}
    -
    (-{\textstyle\frac 1m}\Delta+4e)^{-1}
    v
    (-{\textstyle\frac 1m}\Delta+4e+v)^{-1}
  \end{equation}
  (this identity is called the ``second resolvent identity'', and follows from an elementary computation), which we rewrite using Theorem\-~\ref{theo:yukawa}:
  \begin{equation}
    K_e\psi
    =\frac1mY_{2\sqrt{me}}\ast\psi
    -\frac1m Y_{2\sqrt{me}}\ast(vK_e\psi)
    .
  \end{equation}
  Writing
  \begin{equation}
    \psi(x)=\psi_+(x)-\psi_-(x)
    ,\quad
    \psi_+(x):=\mathds 1_{\psi(x)\geqslant 0}\psi(x)\geqslant 0
    ,\quad
    \psi_-(x):=-\mathds 1_{\psi(x)<0}\psi(x)\geqslant 0
  \end{equation}
  we have, formally, for $q\geqslant 1$,
  \begin{equation}
    \|K_e\psi\|_q\leqslant \|K_e\psi_+\|_q+\|K_e\psi_-\|_q
    .
  \end{equation}
  Now, since $Y_{2\sqrt{me}}\geqslant 0$, $v\geqslant 0$ and $K_e$ is positivity preserving and $\psi_\pm(x)\geqslant 0$,
  \begin{equation}
    0\leqslant K_e\psi_\pm
    \leqslant\frac1mY_{2\sqrt{me}}\ast\psi_\pm
  \end{equation}
  so, by Young's inequality (Theorem\-~\ref{theo:young}),
  \nopagebreakaftereq
  \begin{equation}
    \|K_e\psi\|_q
    \leqslant \frac 1m\|Y_{2{\sqrt{me}}}\|_1(\|\psi_+\|_q+\|\psi_-\|_q)
    \leqslant \frac 2{m^2e}\|\psi\|_q
    .
  \end{equation}
\qed
\restorepagebreakaftereq

\section{Proof of Theorem \expandonce{\ref{theo:compleq}}}\label{app:proof_factorization}

\subsection{Factorization}
\indent
We will first compute $u_3,u_4$.
We need to compute these up to order $V^{-2}$, because one of the terms in the equation is of order $V$ (see below).
\bigskip

\subsubsection{Factorization of $g_3$}
\theo{Lemma}\label{lemma:g3}
  Assumption\-~\ref{assum:factorization} with $i=3$ and\-~(\ref{cd_g3g4}) imply that
  \begin{equation}
    g_3(x,y,z)=(1-u_3(x-y))(1-u_3(x-z))(1-u_3(y-z))
  \end{equation}
  with
  \begin{equation}
    u_3(x-y)=\left(u(x-y)+\frac{w_3(x-y)}V\right)(1+O(V^{-2}))
    \label{u3}
  \end{equation}
  \nopagebreakaftereq
  \begin{equation}
    w_3(x-y):=(1-u(x-y))\int dz\ u(x-z)u(y-z)
    .
    \label{w3}
  \end{equation}
\endtheo
\restorepagebreakaftereq
\bigskip

\indent\underline{Proof}:
  Using\-~(\ref{cd_g3g4}) in\-~(\ref{g_factorized}),
  \begin{equation}
    g_2(x_1-x_2)=(1-u_3(x_1-x_2))
    \int \frac{dx_3}V\ (1-u_3(x_1-x_3))(1-u_3(x_2-x_3))
    .
    \label{g2_factor_inproof}
  \end{equation}
  \bigskip

  \point
  We first expand to order $V^{-1}$.
  By\-~(\ref{assum_bound}),
  \begin{equation}
    \int\frac{dz}V\ u_3(x-z)=O(V^{-1})
    \label{f3V1}
  \end{equation}
  so,
  \begin{equation}
    g_2(x-y)\equiv 1-u(x-y)=(1-u_3(x-y))
    \left(1
    +O(V^{-1})\right)
    .
  \end{equation}
  Therefore,
  \begin{equation}
    u_3(x-y)=u(x-y)(1+O(V^{-1}))
    .
    \label{3V}
  \end{equation}
  \bigskip

  \point
  We push the expansion to order $V^{-2}$: (\ref{g2_factor_inproof}) is
  \begin{equation}
    g_2(x-y)=(1-u_3(x-y))\int\frac{dz}{V}\ 
    \left(
      1
      -u_3(x-z)-u_3(y-z)
      +u_3(x-z)u_3(y-z)
    \right)
    .
  \end{equation}
  Dividing by the integral, we find that, by\-~(\ref{3V}),
  \begin{equation}
    \begin{largearray}
      (1-u_3(x-y))
      =(1-u(x-y))
      \cdot\\[0.3cm]\hfill\cdot
      \left(1+\int\frac{dz}{V}\ (u(x-z)+u(y-z)-u(x-z)u(y-z))+O(V^{-2})\right)
      .
    \end{largearray}
  \end{equation}
  Furthermore, by\-~(\ref{cd_g3g4}),
  \begin{equation}
    \int\frac{dx}V\ (1-u(x))=1
  \end{equation}
  so
  \begin{equation}
    \int dx\ u(x)=0
    .
    \label{intu0}
  \end{equation}
  Thus,
  \nopagebreakaftereq
  \begin{equation}
    1-u_3(x-y)=(1-u(x-y))
    \left(1-\int\frac{dz}{V}\ u(x-z)u(y-z)+O(V^{-2})\right)
    .
  \end{equation}
\qed
\restorepagebreakaftereq

\subsubsection{Factorization of $g_4$}

\theo{Lemma}\label{lemma:g4}
  Assumption\-~\ref{assum:factorization} and\-~(\ref{cd_g3g4}) imply that
  \begin{equation}
    g_4(x_1,x_2,x_3,x_2)=
    \left(\prod_{i<j}\left(1-u_4(x_i-x_j)\right)\right)
  \end{equation}
  with
  \begin{equation}
    u_4(x-y)=\left(u(x-y)+\frac{2w_3(x-y)}V\right)
    (1+O(V^{-2}))
    \label{u4}
  \end{equation}
  where $w_3$ is the same as in Lemma\-~\ref{lemma:g3}.
\endtheo
\bigskip

\indent\underline{Proof}:
  Using\-~(\ref{cd_g3g4}) in\-~(\ref{g_factorized}),
  \begin{equation}
    \begin{largearray}
      g_2(x_1-x_2)=(1-u_4(x_1-x_2))\int\frac{dx_3dx_4}{V^2}\ 
      (1-u_4(x_1-x_3))
      (1-u_4(x_1-x_4))
      \cdot\\\hfill\cdot
      (1-u_4(x_2-x_3))
      (1-u_4(x_2-x_4))
      (1-u_4(x_3-x_4))
      .
    \end{largearray}
  \end{equation}
  \bigskip

  \point
  We expand to order $V^{-1}$.
  By\-~(\ref{assum_bound}),
  \begin{equation}
    \int\frac{dz}V\ u_4(x-z)=O(V^{-1})
    \label{f4V1}
  \end{equation}
  so
  \begin{equation}
    g_2(x-y)\equiv1-u(x-y)=(1-u_4(x-y))\left(1+O(V^{-1})\right)
  \end{equation}
  so
  \begin{equation}
    u_4(x-y)=u(x-y)(1+O(V^{-1}))
    .
    \label{4V}
  \end{equation}
  \bigskip

  \point
  We push the expansion to order $V^{-2}$:
  by\-~(\ref{assum_bound}),
  \begin{equation}
    \int \frac{dzdt}{V^2}u_4(x-z)u_4(y-t)=O(V^{-2})
    ,\quad
    \int \frac{dzdt}{V^2}u_4(x-z)u_4(z-t)=O(V^{-2})
  \end{equation}
  \begin{equation}
    \int \frac{dzdt}{V^2}u_4(x-z)u_4(x-t)=O(V^{-2})
  \end{equation}
  so
  \begin{equation}
    \begin{largearray}
      1-u(x-y)=(1-u_4(x-y))
	\cdot\\[0.5cm]\hfill\cdot
	\left(
	1
	+\int\frac{dzdt}{V^2}
	(-2u_4(x-z)-2u_4(y-z)-u_4(z-t)+2u_4(x-z)u_4(y-z))
	+O(V^{-2})
      \right)
      .
    \end{largearray}
  \end{equation}
  By\-~(\ref{4V}),
  \begin{equation}
    \begin{largearray}
      1-u_4(x-y)
      =
      (1-u(x-y))
      \cdot\\[0.5cm]\hfill\cdot
      \left(1+
	\int\frac{dzdt}{V^2}\ (2u(x-z)+2u(y-z)+u(z-t)-2u(x-z)u(y-z))
	+O(V^{-2})
      \right)
      .
    \end{largearray}
  \end{equation}
  By~\-(\ref{intu0}),
  \nopagebreakaftereq
  \begin{equation}
    1-u_4(x-y)
    =
    (1-u(x-y))\left(1-2\int\frac{dz}{V}\ u(x-z)u(y-z)+O(V^{-2})\right)
    .
  \end{equation}
\qed
\restorepagebreakaftereq

\subsection{Proof of Theorem \expandonce{\ref{theo:compleq}}}
\point
First of all, by\-~(\ref{E0}),
\begin{equation}
  E_0=\frac{N(N-1)}{2V}\int dx\ v(x)(1-u(x))
  .
  \label{E0u}
\end{equation}
\bigskip

\point
We rewrite\-~(\ref{g2_hierarchy}) using Lemmas, \ref{lemma:g3} and\-~\ref{lemma:g4}.
Let
\begin{equation}
  G_3:=
  \frac{N-2}V \int dx_3\ (v(x_1-x_3)+v(x_2-x_3))g_3(x_1,x_2,x_3)
\end{equation}
and
\begin{equation}
  G_4:=
  \frac{(N-2)(N-3)}{2V^2}
  \int dx_3dx_4\ v(x_3-x_4)
  g_4(x_1,x_2,x_3,x_4)
  .
\end{equation}
By Lemma\-~\ref{lemma:g3},
\begin{equation}
  \begin{largearray}
    G_3(x-y)=
    \frac NV(1-u(x-y))
    \cdot\\\hfill\cdot
    \left(\left(\int dz\ (v(x-z)+v(y-z))(1-u(x-z))(1-u(y-z))\right)+O(V^{-1})\right)
  \end{largearray}
\end{equation}
and by Lemma\-~\ref{lemma:g4},
\begin{equation}
  \begin{largearray}
    G_4(x-y)=
    \left(\frac{N^2}{2V^2}(1-u_4(x-y))\int dzdt\ v(z-t)(1-u_4(z-t))\Pi(x,y,z,t)
    -\right.\\\hfill\left.-
    \frac{5N}{2V^2}(1-u(x-y))\int dzdt\ v(z-t)(1-u(z-t))
    +O(V^{-1})
    \right)
  \end{largearray}
\end{equation}
\begin{equation}
  \Pi(x,y,z,t):=
  (1-u(x-z))(1-u(x-t))(1-u(y-z))(1-u(y-t))
  .
  \label{Pi}
\end{equation}
(Note that, in this line and the following, the term $O(V^{-1})$ is outside the integrals.)
The first term in $G_4$ is of order $V$: by\-~(\ref{u4}),
\begin{equation}
  \begin{array}{>\displaystyle l}
    \frac{N^2}{2V^2}(1-u_4(x-y))\int dzdt\ v(z-t)(1-u_4(z-t))
    \Pi(x-y-z-y)
    =\\[0.5cm]\indent=
    \frac{N^2}{2V^2}(1-u(x-y))\int dzdt\ v(z-t)(1-u(z-t))
    -\\[0.5cm]\indent-
    \frac{N^2}{V^3}w_3(x-y)\int dzdt\ v(z-t)(1-u(z-t))
    -\\[0.5cm]\indent-
    \frac{N^2}{V^3}(1-u(x-y))\int dzdt\ v(z-t)w_3(z-t)
    +\\[0.5cm]\indent+
    \frac{N^2}{2V^2}(1-u(x-y))\int dzdt\ v(z-t)(1-u(z-t))
    \left(\Pi(x,y,z,t)-1\right)
    +O(V^{-1})
  \end{array}
\end{equation}
in which the only term of order $V$ is the first one, and is equal to the term of order $V$ in $E_0g_2$:
\begin{equation}
  (1-u(x-y))\frac{N^2}{2V}\int dz\ v(z)(1-u(z))
\end{equation}
and thus cancels out.
Therefore, recalling that $\rho=\frac NV$,
\begin{equation}
  \left(
    -\frac1m\Delta
    +v(x-y)
    +\bar G_3(x-y)
    +\bar G_4(x-y)
    -\bar E_0
  \right)
  (1-u(x-y))
  =
  O(V^{-1})
  \label{g2bar}
\end{equation}
with
\begin{equation}
  \bar G_3(x-y):=
  \rho\int dz\ (v(x-z)+v(y-z))(1-u(x-z))(1-u(y-z))
\end{equation}
and by\-~(\ref{w3}),

\begin{equation}
  \begin{array}{r@{\ }>\displaystyle l}
    \bar G_4(x-y)
    :=&
    -\frac\rho2\left(5+2\rho \int dr\ u(x-r)u(y-r)\right)\int \frac{dzdt}V\ v(z-t)(1-u(z-t))
    -\\[0.3cm]&-
    \rho^2\int \frac{dzdt}V\ v(z-t)(1-u(z-t))\int dr\ u(z-r)u(t-r)
    +\\&+
    \frac{\rho^2}2\int dzdt\ v(z-t)(1-u(z-t))
    \left(\Pi(x,y,z,t)-1\right)
  \end{array}
\end{equation}
\begin{equation}
  \bar E_0=
  \frac\rho2\int \frac{dxdy}V\ v(x-y)(1-u(x-y))
  .
\end{equation}
\bigskip

\point Thus, in the thermodynamic limit, using translation invariance,
\begin{equation}
  \bar E_0=
  -\frac\rho2\int dz\ v(z)(1-u(z))
  .
\end{equation}
Furthermore,
\begin{equation}
  \bar G_3(x)=
  -4\bar E_0
  -2\rho u\ast S(x)
  .
\end{equation}
In addition,
\begin{equation}
  -\frac{5\rho}2\int \frac{dzdt}V\ v(z-t)(1-u(z-t))
  =
  5\bar E_0
\end{equation}
\begin{equation}
  -\rho^2\int dr\ u(x-r)u(y-r)\int\frac{dzdt}V\ v(z-t)(1-u(z-t))
  =
  2\rho\bar E_0 u\ast u(x-y)
\end{equation}
\begin{equation}
  -\rho^2\int\frac{dzdt}V\ v(z-t)(1-u(z-t))\int dr\ u(z-r)u(t-r)
  =
  -\rho^2\int dz\ S(z)u\ast u(z)
\end{equation}
and, by\-~(\ref{intu0}) and\-~(\ref{Pi}),
\begin{equation}
  \begin{largearray}
    \frac{\rho^2}2\int dzdt\ v(z-t)(1-u(z-t))(\Pi(x,y,z,t)-1)
    =
    \\[0.5cm]\indent
    =\rho^2\left(
      \int dz\ S(z)u\ast u(z)
      -\frac2\rho u\ast u(x-y)\bar E_0
      +u\ast u\ast S(x-y)
    \right.\\[0.5cm]\hfill\left.
      -2u\ast(u(u\ast S))(x-y)
      +\frac12\int dzdt\ u(y-z)u(y-t)u(x-z)u(x-t)S(z-t)
    \right)
    .
  \end{largearray}
\end{equation}
Inserting these into\-~(\ref{g2bar}), we find\-~(\ref{compleq}).
\qed

\section{The momentum distribution in Bogolyubov theory}\label{app:bog_Nk}
\indent
In this appendix we recover the prediction of Bogolyubov theory for the momentum distribution\-~(\ref{momentum_distribution_def}) from\-~\cite[Appendix\-~A]{LSe05}.
\bigskip

\indent
The expression of the Bogolyubov Hamiltonian in second quantized form is\-~\cite[(A.19)-(A.20)]{LSe05}
\begin{equation}
  H_B=\frac12N\rho\hat v(0)
  +\sum_{k\neq0}\left(
    (\epsilon(k)+\rho\hat v(k))a_k^\dagger a_k+\frac12\rho \hat v(k)(a_ka_{-k}+a_{-k}^\dagger a_k^\dagger
  \right)
\end{equation}
where $\epsilon(k)$ is given below\-~\cite[(A.6)]{LSe05}:
\begin{equation}
  \epsilon(k)=\frac{k^2}{2m}
  \label{epsilon}
\end{equation}
(recall that we are using units where $\hbar=1$).
Without going through the details of the second quantized formalism, suffice it to say that $a_k^\dagger a_k$ is an operator that counts the number of particles in the momentum $k$ state.
Therefore, the average number of particles in the momentum $k$ state is
\begin{equation}
  N_k=\left<\psi_0\right|a_k^\dagger a_k\left|\psi_0\right>
  =\partial_{\epsilon(k)}E_0
  .
\end{equation}
The momentum distribution is the probability of finding a particle in the momentum state $k$, and so,
\begin{equation}
  \mathcal M_0^{(\mathrm{Bogolyubov})}(k)=\frac1\rho N_k=\frac1\rho\partial_{\epsilon(k)}E_0
\end{equation}
The ground state energy $E_0$ can be computed exactly\-~\cite[(A.26)]{LSe05}:
\begin{equation}
  E_0=\frac12N\rho\hat v(0)
  -\frac12\sum_k \left(\epsilon(k)+\rho\hat v(k)-\sqrt{\epsilon(k)^2+2\epsilon(k)\rho\hat v(k)}\right)
\end{equation}
(we obtain this from\-~\cite[(A.26)]{LSe05} by first writing the integral over $|k|$ as one over $k$: $\int dk=4\pi\int_0^\infty d|k||k|^2$, and then undoing the ``bulk limit'' by replacing $\int dk$ by $\frac{8\pi^3}V\sum_k$).
Therefore,
\begin{equation}
  \mathcal M_0^{(\mathrm{Bogolyubov})}(k)=\frac1\rho\partial_{\epsilon(k)}E_0
  =-\frac1{2\rho}\left(
    1-\frac{\epsilon(k)+\rho\hat v(k)}{\sqrt{\epsilon(k)^2+2\epsilon(k)\rho\hat v(k)}}
  \right)
\end{equation}
and so, by\-~(\ref{epsilon}),
\begin{equation}
  \mathcal M_0^{(\mathrm{Bogolyubov})}(k)
  =\frac1{2\rho}\left(
    \frac{k^2+2m\rho\hat v(k)}{\sqrt{k^4+4mk^2\rho\hat v(k)}}-1
  \right)
  .
\end{equation}
This is not quite\-~(\ref{Nk_conj}), which we may rewrite using\-~(\ref{kappak_conj}) as
\begin{equation}
  \frac1{2\rho}\left(\frac{k^2+8m\pi\rho a}{\sqrt{k^4+16m\pi k^2\rho a}}-1\right)
  .
\end{equation}
To get this expression, we must do one more operation, which is required to get correct low-density predictions in Bogolyubov theory, which is to replace $v$ with a ``pseudo-potential''\-~\cite{LHY57}, which is defined as $4\pi a\delta(x)$.
In other, words, we replace $\hat v(k)$ with the constant $4\pi a$.
Thus, $\mathcal M_0^{(\mathrm{Bogolyubov})}$ becomes\-~(\ref{Nk_conj}).

\vfill
\eject

\vfill
\eject

\exercises

\problem\label{ex:BE_distr} (solution on p.\-~\ref{sol:BE_distr})\par
\smallskip

Prove\-~(\ref{BE}) starting from\-~(\ref{BE_pre}).
\bigskip

\problem\label{ex:bec} (solution on p.\-~\ref{sol:bec})\par
\smallskip

Check\-~(\ref{rhomu}) explicitly using the Riemann theorem.
Prove that $\mu\mapsto\rho(\mu)$ is strictly increasing using the dominated convergence theorem.
Prove\-~(\ref{lim_rhomu_infty}) and\-~(\ref{lim_rhomu_0}).
\bigskip

\problem\label{ex:FD_distr} (solution on p.\-~\ref{sol:FD_distr})\par
Compute the analogue of\-~(\ref{BE}) for an ideal gas of {\it Fermions}:
\begin{equation}
  \left<\mathcal B_q\right>=
  \frac1{e^{\beta(\frac{q^2}{2m}-\mu)}+1}
  \label{FD}
\end{equation}
{\it Hint}: Proceed in the same way as the ideal gas of Bosons by parametrizing the space using the occupation number of plane waves $e^{ikx}$.
Except that now, because the wavefunctions are antisymmetric, there can only be 0 or 1 particle in each state, so the sums over $N_k$ are over $\{0,1\}$.
Can there be Bose-Einstein condensation for the ideal gas of Fermions?
\bigskip

\problem\label{ex:scattering_softcore} (solution on p.\-~\ref{sol:scattering_softcore})\par
\smallskip
Prove that the solution of\-~(\ref{scat_softcore_spherical}) is\-~(\ref{sol_softcore}).\par
{\it Hint}: use the fact that $\psi\to1$ as $|x|\to\infty$, as well as the fact that $\psi$ is continuous at $0$ and at $R$.
\bigskip

\problem\label{ex:perron_frobenius} (solution on p.\-~\ref{sol:perron_frobenius})\par
\smallskip
Prove that the ground state $\psi_0$ of\-~(\ref{Ham}) is unique and $\psi_0\geqslant 0$.\par
{\it Hint}: 
Instead of the Hamiltonian $H_N$, consider the operator $e^{-tH_N}$ for $t\geqslant0$.
Use Theorem\-~\ref{theo:add_inv} to show that $e^{-tH_N}$ is positivity preserving (see Definition\-~\ref{def:positivity_preserving}).
Use the Perron-Frobenius theorem (Theorem\-~\ref{theo:perron_frobenius}) to conclude.
\bigskip

\problem\label{ex:nonneg} (solution on p.\-~\ref{sol:nonneg})\par
\smallskip
In this exercise, we will derive an alternate proof that $\psi_0\geqslant 0$ under the extra assumption that $v$ is continuous.
To do so, consider the energy of $\psi_0$:
\begin{equation}
  \mathcal E(\psi_0):=\left<\psi_0\right|H_N\left|\psi_0\right>
  =\int_{(\mathbb R/(L\mathbb Z))^{3N}} dx\ 
  \left(
    -\frac1{2m}\psi_0^*(x)\Delta\psi(x)
    +V(x)|\psi_0(x)|^2
  \right)
\end{equation}
where $\Delta$ is the Laplacian on $\mathbb R^{3N}$ and $V(x)\equiv\sum_{i<j}v(x_i-x_j)$.
Prove that
\begin{equation}
  \mathcal E(|\psi_0|)=\mathcal E(\psi_0)
\end{equation}
and use that to prove that $\psi_0\geqslant 0$.
\bigskip


\vfill
\eject

\solution{BE_distr}
To compute
\begin{equation}
  \sum_{N=0}^\infty Ne^{-tN}
\end{equation}
we differentiate
\begin{equation}
  \sum_{N=0}^\infty e^{-tN}
  =\frac1{1-e^{-t}}
\end{equation}
with respect to $-t$:
\begin{equation}
  \sum_{N=0}^\infty Ne^{-tN}
  =\frac{e^{-t}}{(1-e^{-t})^2}
  .
\end{equation}
Therefore, by\-~(\ref{Xi_ideal}) and\-~(\ref{BE_pre}),
\begin{equation}
  \left<\mathcal B_q\right>=
  \frac{e^{-\beta(\frac{k^2}{2m}-\mu)}}{1-e^{-\beta(\frac{k^2}{2m}-\mu)}}
  \frac{
    \prod_{k\in\frac{2\pi}L\mathbb Z^3}
    \frac1{1-e^{-\beta(\frac{k^2}{2m}-\mu)}}
  }
    {\prod_{k\in\frac{2\pi}L\mathbb Z^3}
    \frac1{1-e^{-\beta(\frac{k^2}{2m}-\mu)}}
  }
  =
  \frac{e^{-\beta(\frac{k^2}{2m}-\mu)}}{1-e^{-\beta(\frac{k^2}{2m}-\mu)}}
\end{equation}
from which\-~(\ref{BE}) follows.
\bigskip

\solution{bec}
\point
(\ref{rhomu_pre}) is of the form
\begin{equation}
  \rho_L(\mu)
  =
  \frac1{(2\pi)^3}
  \frac1D^3\sum_{k\mathbb Z^3}f(Dk)
\end{equation}
with
\begin{equation}
  D\equiv\frac L{2\pi}
  ,\quad
  f(k):=\frac1{e^{\beta(\frac{k^2}{2m}-\mu)-1}}
\end{equation}
so\-~(\ref{rhomu}) follows from the Riemann theorem.
\bigskip

\point
For $\mu<0$,
\begin{equation}
  \left|\frac1{e^{\beta(\frac{q^2}{2m}-\mu)}-1}\right|
\end{equation}
is integrable and, for $\mu_0<\mu<0$,
\begin{equation}
  \left|\partial_\mu\frac1{e^{\beta(\frac{q^2}{2m}-\mu)}-1}\right|
  =
  \beta\frac{e^{\beta(\frac{q^2}{2m}-\mu)}}{(e^{\beta(\frac{q^2}{2m}-\mu)}-1)^2}
  \leqslant
  \beta\frac{e^{\beta(\frac{q^2}{2m}-\mu_0)}}{(e^{\beta\frac{q^2}{2m}}-1)^2}
  .
\end{equation}
Now, since
\begin{equation}
  e^{\beta\frac{q^2}{2m}}-1
  \sim_{q\to0}\frac{\beta q^2}{2m}
\end{equation}
this is integrable in dimensions three or more.
\bigskip

\point
Since
\begin{equation}
  \left|\frac1{e^{\beta(\frac{q^2}{2m}-\mu)}-1}\right|
  \leqslant
  \frac1{e^{\beta(\frac{q^2}{2m})}-1}
\end{equation}
is integrable, (\ref{lim_rhomu_infty}) and\-~(\ref{lim_rhomu_0}) follow from dominated convergence.
\begin{equation}
  \lim_{\mu\to-\infty}\rho(\mu)=0
\end{equation}

\solution{FD_distr}
In the case of Fermions, each state can be occupied by at most one particle, so
\begin{equation}
  \Xi=
  \prod_{k\in\frac{2\pi}L\mathbb Z^3}
  \sum_{N_k=0}^1
  e^{-\beta(\frac{k^2}{2m}-\mu)N_k}
  \equiv
  \prod_{k\in\frac{2\pi}L\mathbb Z^3}
  \left(
    1+
    e^{-\beta(\frac{k^2}{2m}-\mu)}
  \right)
\end{equation}
and
\begin{equation}
  \left<\mathcal B_q\right>=
  \frac1\Xi
  e^{-\beta(\frac{q^2}{2m}-\mu)}
  \prod_{k\in\frac{2\pi}L\mathbb Z^3\setminus\{q\}}
  \left(
    1+
    e^{-\beta(\frac{k^2}{2m}-\mu)}
  \right)
\end{equation}
so
\begin{equation}
  \left<\mathcal B_q\right>=
  \frac{e^{-\beta(\frac{q^2}{2m}-\mu)}}
  {1+e^{-\beta(\frac{q^2}{2m}-\mu)}}
  =
  \frac{1}
  {1+e^{\beta(\frac{q^2}{2m}-\mu)}}
  .
\end{equation}
There cannot be Bose-Einstein condensation for Fermions, because each state can only be occupied by at most one particle, so it is not possible for a state to have an occupation number that is proportional to $N$.
\bigskip

\solution{scattering_softcore}
For $r\leqslant R$,
\begin{equation}
  \partial_r^2(r\psi)=mUr\psi
\end{equation}
so
\begin{equation}
  r\psi=A\cosh(\sqrt{mU}r)+B\sinh(\sqrt{mU}r)
  .
\end{equation}
For $r>R$,
\begin{equation}
  \partial_r^2(r\psi)=0
\end{equation}
so, since $\psi\to1$ as $r\to\infty$,
\begin{equation}
  r\psi=r-a
  .
\end{equation}
Since $r\psi=0$ at $r=0$,
\begin{equation}
  r\psi=B\sinh(\sqrt{mU}r)
\end{equation}
and since $r\psi$ is continuous at $R$,
\begin{equation}
  B=\frac{R-a}{\sinh(\sqrt{mU}R)}
  .
\end{equation}
This implies\-~(\ref{sol_softcore}).
\bigskip

\solution{perron_frobenius}
By Lemma\-~\ref{lemma:heat_L2}, $e^{\frac tm\sum_i\Delta_i}$ is positivity preserving, and because $v\geqslant 0$, so is $e^{t\sum_{i<j}v(|x_i-x_j|)}$.
Therefore, by Theorem\-~\ref{theo:add_inv}, $e^{-tH_N}$ is positivity preserving.
In addition, $e^{-tH_N}$ is compact by Theorem\-~\ref{theo:compact_schrodinger}, and the spectrum of $e^{-tH_N}$ is $e^{-t\ \mathrm{spec}(H_N)}$, and the largest eigenvector of $e^{-tH_N}$ with the largest eigenvalue is the ground-state of $H_N$.
Finally, since $\mathrm{spec}(H_N)\geqslant 0$ (because $-\Delta\geqslant 0$ and $v\geqslant 0$, we can apply the Perron-Frobenius theorem, which implies that $\psi_0$ is unique and $\geqslant 0$.
\bigskip

\solution{nonneg}
The potential term $\int dx\ V(x)|\psi_0|^2$ is obviously the same for $\psi_0$ and $|\psi_0|$, so we only need to worry about the kinetic term.
Let
\begin{equation}
  P:=\{x\in\mathbb R^{3N}:\ \psi_0(x)\neq0\}
  .
\end{equation}
Now, $\psi_0$ is twice continuously differentiable, since it is an eigenfunction and $v$ is continous (see\-~\cite[Theorem 11.7(vi)]{LL01}), so $|\psi_0|$ is twice continuously differentiable on $P$,  and so
\begin{equation}
  -\int dx\ |\psi_0|\Delta|\psi_0|
  \equiv
  -\int_P dx\ |\psi_0|\Delta|\psi_0|
  =
  \int_P dx\ (\nabla|\psi_0|)^2
  +\int_{\partial P} dx\ |\psi_0(x)|(n(x)\cdot\nabla|\psi_0|)
\end{equation}
where $n(x)$ is the normal vector at $x$ ($\partial P$ is a differentiable manifold because $\psi_0$ is differentiable), but since $\psi_0(x)=0$ on $\partial P$,
\begin{equation}
  -\int dx\ |\psi_0|\Delta|\psi_0|
  =
  \int_P dx\ (\nabla|\psi_0|)^2
  =
  \int_P dx\ \left(\nabla\psi_0\frac{\psi_0}{|\psi_0|}\right)^2
  =
  \int_P dx\ (\nabla\psi_0)^2
  =
  -\int_P dx\ \psi_0\Delta\psi_0
  .
\end{equation}
Thus, $\mathcal E(|\psi_0|)=\mathcal E(\psi_0)$.
Since $\psi_0$ is the minimizer of $\mathcal E(\psi_0)$, so is $|\psi_0|$, but since the ground state is unique,
\begin{equation}
  \psi_0=|\psi_0|
\end{equation}
so $\psi_0\geqslant 0$.
\bigskip


\end{document}